\documentclass[aps,prd,onecolumn,amsmath,11pt,superscriptaddress,floatfix,nofootinbib,preprintnumbers]{revtex4}

\usepackage{mathrsfs}
\usepackage{amsfonts}
\usepackage{amsmath,amssymb,bm}
\usepackage{array}
\usepackage{verbatim}
\usepackage{epsfig}
\usepackage{graphicx}
\usepackage{hyperref}
\hypersetup{colorlinks, linkcolor = [rgb]{0, 0, 0.5}, citecolor = [rgb]{0,0.0,0.5}, urlcolor = [rgb]{0,0.0,0.5}}
\usepackage[normalem]{ulem}
\usepackage{xcolor}
\usepackage{ulem}
\usepackage{multirow}

\usepackage{graphicx}
\graphicspath{{./figs/}}
\usepackage{dcolumn}
\usepackage{bm}
\usepackage{slashed}
\usepackage{siunitx}
\usepackage{ulem,xpatch}
\usepackage{hyperref}
\usepackage[mathlines]{lineno}
\usepackage{comment}
\usepackage{soul}
\usepackage{xcolor}
\usepackage{xfrac}

\allowdisplaybreaks[4]

\begin{document}

\title{Polarization analysis of two baryons with various spin combinations produced in electron-positron annihilation}

\newcommand*{\SDU}{Key Laboratory of Particle Physics and Particle Irradiation (MOE), Institute of Frontier and Interdisciplinary Science, Shandong University, Qingdao, Shandong 266237, China}\affiliation{\SDU}
\newcommand*{\ihep}{Institute of High Energy Physics, Chinese Academy of Sciences, Beijing 100049, China}\affiliation{\ihep}
\newcommand*{\ucas}{University of Chinese Academy of Sciences, Beijing 100049, China}\affiliation{\ucas}
\newcommand*{\SCNT}{Southern Center for Nuclear-Science Theory (SCNT), Institute of Modern Physics, Chinese Academy of Sciences, Huizhou 516000, China}\affiliation{\SCNT}
\newcommand*{\HTU}{School of Physics, He'nan Normal University, Xinxiang, Henan 453007, China}\affiliation{\HTU}
\newcommand*{\ytu}{College of Nuclear Equipment and Nuclear Engineering, Yantai University, Yantai, Shandong 264005, China}\affiliation{\ytu}

\author{Zhe Zhang}\email{zhangzhe@mail.sdu.edu.cn}\affiliation{\SDU}
\author{Rong-Gang Ping}\email{pingrg@mail.ihep.ac.cn}\affiliation{\ihep}\affiliation{\ucas}
\author{Tianbo Liu}\email{liutb@sdu.edu.cn}\affiliation{\SDU}\affiliation{\SCNT}
\author{Jiao Jiao Song}\email{songjiaojiao@ihep.ac.cn}\affiliation{\HTU}
\author{Weihua Yang}\email{yangwh@ytu.edu.cn}\affiliation{\ytu}
\author{Ya-jin Zhou}\email{zhouyj@sdu.edu.cn}\affiliation{\SDU}

\begin{abstract}
We develop a systematic approach to analyze polarization correlations of two baryons $B_{1}\bar{B}_{2}$ produced in the electron-positron annihilation process. With spin density matrices for arbitrary spin particles established in the standard, the Cartesian, and the helicity forms, we provide analyses of polarization correlations for two baryons with various spin combinations. This framework can be applied to determine the spin and the parity of excited baryons, and therefore offers opportunities for the investigation of baryon spectrum and transition form factors in present and future electron-positron annihilation experiments.

\end{abstract}

\maketitle

\newpage

\section{Introduction\label{sec:Intro}}

Quantum chromodynamics (QCD) as the accepted theory for strong interactions has been widely tested at high-energy scales, where the strong coupling constant is small and one can apply perturbation theory. However, nonperturbative properties of QCD at low-energy scales continue challenging our understanding~\cite{Gross:2022hyw}. The phenomenon of color confinement obstructs the direct observation of elementary degrees of freedom, {\it i.e.} quarks and gluons, making the exploration of the spectrum and interactions of hadrons a crucial and formidable frontier in modern particle physics.

The quark model is proven successful in describing ordinary mesons and baryons as $q\bar q$ and $qqq$ states. However, other configurations, including multiquarks, glueballs, and hybrids, are not excluded by QCD. Recently, some exotic mesons, which exhibit quantum numbers that cannot be constructed in a pure $q\bar q$ state, were observed in experiments. These states have been widely pursued and survive as candidates of glueballs or hybrid mesons ~\cite{Bugg:2000zy, Klempt:2007cp, Mathieu:2008me, Crede:2008vw, Meyer:2010ku, Meyer:2015eta, Ochs:2013gi, Sonnenschein:2016pim, Bass:2018xmz, Ketzer:2019wmd, Roberts:2021nhw, Fang:2021wes, Jin:2021vct, Chen:2022asf, ParticleDataGroup:2022pth}. On the other hand, hybrid baryon candidates\cite{Kisslinger:1995yw, Capstick:2002wm, Barnes:1982fj, Golowich:1982kx, Carlson:1983xr, Duck:1983ju, Chow:1998tq, Dudek:2012ag} received much less attention because of the lack of exotic properties for their spin and parity, $J^P$. These baryons can be identified by investigating their unique strong decay amplitudes or distinct production patterns in processes like $J/\psi$ hadronic decays~\cite{Carlson:1991tg, Zou:1999wd, Barnes:2000vn, Page:2002mt, Lanza:2017qel}.


Currently, there is a gap between experimental measurements and theoretical predictions of the baryon spectrum~\cite{Hey:1982aj, Capstick:2000qj,Krusche:2003ik, Klempt:2009pi,Crede:2013kia, Ireland:2019uwn,Thiel:2022xtb, Eichmann:2022zxn}, particularly concerning hyperons with increased strangeness, such as the $\Xi$ baryons. 
The identified $\Xi$ baryons are markedly fewer than their $\Sigma$ and $\Lambda$ counterparts. Furthermore, for most discovered excited $\Xi^*$ states, the precise determination of spin and parity are missing~\cite{ParticleDataGroup:2022pth}. This is due to lower production rates of $\Xi^{*}$ and prior measurements focusing on individual $\Xi^{*}$, hindering simultaneous determinations of spin and parity~\cite{Schlein:1963zza, Smith:1965zze, Button-Shafer:1966buh, Merrill:1968zz, Amsterdam-CERN-Nijmegen-Oxford:1976ezm, Teodoro:1978bu, Biagi:1986vs,BaBar:2008myc}. Electron-positron annihilation into two baryons with opposite strangeness ensures strangeness conservation and generates adequate production cross sections. Analyzing the polarization correlations of the two-baryon system allows for concurrent assessment of both spin and parity.  Investigations into these processes at facilities like $BABAR$~\cite{BaBar:2001yhh}, BESIII~\cite{BESIII:2009fln, BESIII:2020nme}, Belle II~\cite{Belle-II:2018jsg} and the proposed STCF~\cite{Achasov:2023gey} offers valuable opportunities to study properties of excited baryons.

Previous studies on polarization correlations have focused on baryons with spins of 1/2 or 3/2 and established parity~\cite{Donohue:1969fu, Dubnickova:1992ii, Tomasi-Gustafsson:2005svz, Czyz:2007wi, Chen:2007zzf,Faldt:2017kgy, Perotti:2018wxm, He:2022jjc, Salone:2022lpt, Batozskaya:2023rek, Zhang:2023box}. Building on these, the BESIII Collaboration has reported a series of measurements on ground-state octet and decuplet baryons~\cite{BESIII:2018cnd, BESIII:2021ypr, BESIII:2022qax, BESIII:2023drj, BESIII:2020fqg, BESIII:2023cvk, BESIII:2022lsz, BESIII:2023jhj, BESIII:2020lkm}. For other excited baryons, determining both parity and whether spin exceeds 3/2 is essential.  Spin density matrices, where we decompose the polarization information, are developed in two primary formats: the standard form~\cite{Minnaert:1966, Byers:1967tw, Chung:1971ri, Doncel:1972ez} and the Cartesian form~\cite{Biedenharn:1958, Bourrely:1980mr, Leader:2011vwq, Bacchetta:2000jk, Song:1967,Zhao:2022lbw,Zhang:2023box}. The standard form allows for the straightforward decomposition of matrices for high spins but lacks clarity in interpreting spin components (parameters). Conversely, the Cartesian form provides clearer physical interpretations of spin components but lacks a general method for decomposing matrices for high spins. Although these forms are intrinsically consistent, a direct method to associate their expressions remains absent.

Besides the spin and parity quantum numbers of individual baryons, more information can be obtained from the study of effective transition form factors, which describe the production of various baryons via a virtual photon or some resonance states~\cite{Korner:1976hv, Capstick:2000qj, Burkert:2004sk, Pascalutsa:2006up, Aznauryan:2011qj, Aznauryan:2012ba, Eichmann:2016yit, Ramalho:2023hqd}. Within the helicity formalism,  transition form factors are captured in the helicity transition amplitudes and reflected in the polarization correlation coefficients of two-baryon systems~\cite{Jacob:1959at}. The electron-positron annihilation process, with its rich two-baryon production channels, offers a fertile ground for studying these form factors. Furthermore, it is possible to search for $CP$ violation by comparing the production processes of $B_{1}\bar{B}_{2}$ and its conjugate $\bar{B}_{1}B_{2}$, which is directly linked to the asymmetry between matter and antimatter.

In this paper, we explore the polarization dynamics of the $e^{+}e^{-} \rightarrow B_{1}\bar{B}_{2}$ process, where $B_{1}\bar{B}_{2}$ are two baryons with various potential spin combinations and parity combinations. We decompose density matrices for any spin particles in both standard and Cartesian formats and connect these two forms by linking their expressions.  We then give the application of the spin density matrix in the helicity formalism and represent the polarization correlations of a two-baryon system with the polarization correlation matrix. We outline the method to derive the polarization correlations for baryons produced in electron-positron annihilation. We introduce a parameter as the product of the parities of the two baryons and detail the polarization correlation matrices for the spin combinations of (1/2, 1/2), (1/2, 3/2), and (1/2, 5/2). In this connection, it is also interesting to note that the polarization correlation of baryons can be used to test the Bell's inequality~\cite{Tornqvist:1980af, Tornqvist:1986pe, Lednicky:2001yw, Lednicky:2004gz, Baranov:2008zzb, Shi:2019kjf, Qian:2020ini, Gong:2021bcp, Wu:2024asu}. This paper will spur studies of the Bell's inequality for high-spin baryons.

The polarization of baryons is usually deduced from decay processes. The representations of baryon decays have been widely discussed e.g. in Refs.~\cite{ Jacob:1959at, Doncel:1972ez, Chung:1971ri,Chen:2007zzf, Perotti:2018wxm, Zhang:2023box,  Rose:1957, Lee:1957qs, Kim:1992az, Byers:1963zz, Button-Shafer:1965,Donohue:1969fu, Lednicky:1975ry, Lednicky:1985zx, Ademollo:1964, Berman:1965, Xia:2019fjf, Cao:2024tvz}. In this paper, we present a systematic approach for expressing the decay of any $J^P$ baryon with polarization transfer matrices. For two-baryon systems with established spins and parities, specific polarization correlation matrices and decay formulas enable the measurement of polarization correlations and transition form factors. Our analysis extends to excited baryons with unknown spins and parities, particularly the ${\Xi}^{*}$. Taking the $e^{+}e^{-} \rightarrow \Xi^{-}\bar{\Xi}^{*+}$ process as an example, we introduce a technique to identify the spin and parity of the $\bar{\Xi}^{*+}$($\Xi^{*-}$).

In Sec.~\ref{sec:sdm}, we introduce the methodology for decomposing spin density matrices for high spins, focusing on the spin-5/2 case. In Sec.~\ref{sec:Production}, we present the production density matrix of two baryons and detail the polarization correlation matrices for the spin combinations of (1/2, 1/2), (1/2, 3/2), and (1/2, 5/2). In Sec.~\ref{sec:Decay_chains}, we outline the general steps for calculating polarization transfer matrices for baryon decays and demonstrate how to determine the spin and parity of the $\Xi^{*-}$ using the $e^{+}e^{-}\rightarrow\Xi^{-}\bar{\Xi}^{*+}$ process as an example. In Sec.~\ref{sec:Transition_Form_factors}, we establish the correspondences between helicity amplitudes and transition form factors for the processes under consideration. In Sec.~\ref{sec:Summary}, we give a brief summary.

\section{spin density matrix\label{sec:sdm}}

In this section, we provide the general expressions for polarization correlations of a two-baryon system. We decompose the spin density matrix for any spin particles in both standard and Cartesian forms, focusing on constructing orthogonal and complete spin projection matrices and the complete set of spin components (parameters). We establish the association between these forms by linking the matrices and their components.  As an example, we present the spin density matrix for spin-5/2 particles. Moving to the application of the spin density matrix within the helicity formalism, we express it with polarization expansion coefficients and polarization projection matrices. We refine the selection of polarization projection matrices to directly connect the polarization expansion coefficients in the helicity formalism with spin components in the Cartesian form. Finally, we present the spin density matrix for a two-baryon system, capturing polarization information within the polarization correlation matrix.

\subsection{Decomposition of the spin density matrix\label{subsec:decomposition_of_SDM}}

First, we review the decomposition of the spin density matrix in the standard form. Using spherical tensor operators, the spin density matrix for any spin particles can be represented as~\cite{Doncel:1972ez},
\begin{align}
\rho_{s}= & \frac{1}{2s+1}\left(I+2s\sum_{L=1}^{2s}\sum_{M=-L}^{L}r_{M}^{s,L}Q_{M}^{s,L}\right),\label{eq:standard_form}
\end{align}
$r_{M}^{s,L}$ are real multipole parameters and $Q_{M}^{s,L}$ denote hermitian basis matrices. These matrices are defined as follows,
\begin{align}
M\geq1, & \quad Q_{M}^{s,L}=\frac{\left(-1\right)^{M}}{2}\sqrt{\frac{2L+1}{s}}\left(T_{M}^{s,L}+T_{M}^{s,L\dagger}\right),\nonumber \\
M=0, & \quad Q_{0}^{s,L}=\sqrt{\frac{2L+1}{2s}}T_{0}^{s,L},\nonumber \\
M\leq-1, & \quad Q_{M}^{s,L}=\frac{\left(-1\right)^{M}}{2i}\sqrt{\frac{2L+1}{s}}\left(T_{-M}^{s,L}-T_{-M}^{s,L\dagger}\right).\label{eq:QLM}
\end{align}
Here, $\left(T_{M}^{s,L}\right)_{mm^{\prime}}=\left\langle sm|sm^{\prime};LM\right\rangle $ represent the Clebsch-Gordan coefficients. Then, we have $\text{Tr}[Q_{M}^{s,L}Q_{M}^{s,L}]=(2s+1)/(2s)$. This approach enables the straightforward decomposition of the spin density matrix for any spin particles, where polarization information of particles is contained in the multipole parameters $r_{M}^{s,L}$. In this form, we can easily obtain the algebraic structure of the spin projection matrices $Q_{M}^{s,L}$. For $M=0$, the projection matrices are diagonal. As the absolute value of $M$ increases, the non-zero components of the matrices move further away from the diagonal elements. Moreover, the value of $M$ is directly related to the rotation properties of $Q_{M}^{s,L}$, which will be discussed in more detail in the next section. Nonetheless, the physical interpretations of these spin components are not intuitive in this form.

We decompose the spin density matrix in the Cartesian form to provide physical interpretations for spin components. In Refs.~\cite{Biedenharn:1958,Bourrely:1980mr,Leader:2011vwq,Bacchetta:2000jk, Song:1967,Zhao:2022lbw,Zhang:2023box}, the decomposition of spin density matrices for spin 1/2, 1, and 3/2 particles has been established. We introduce a universal method for decomposing the density matrix for any spin particles.

For the spin density matrix of spin-$s$ particles, a $\left(2s+1\right)\times\left(2s+1\right)$ matrix satisfying the hermiticity condition $\rho=\rho^{\dagger}$, its general expression can be written as
\begin{align}
\rho_{s}= & a_{0}I+a_{1}S^{i_{1}}\Sigma^{i_{1}}+a_{2}T^{i_{1}i_{2}}\Sigma^{i_{1}i_{2}}+\cdots+a_{2s}T^{i_{1}i_{2}\cdots i_{2s}}\Sigma^{i_{1}i_{2}\cdots i_{2s}},\label{eq:Cartesian_form}
\end{align}
where $\Sigma^{i_{1}i_{2}\cdots i_{n}}$ are traceless matrices and symmetric under index permutation, and $a_{0}$, ..., $a_{2s}$ denote normalization coefficients. We note that a summation is implied over repeated indices. This convention applies to all formulas in this paper. In the $z$-representation, $\Sigma^{i_{1}}$ can be expressed as
\begin{align}
\left(\Sigma^{x}\right)_{mm^{\prime}}= & \frac{1}{2}\left(\delta_{m,m^{\prime}+1}+\delta_{m+1,m^{\prime}}\right)\sqrt{(s+1)(m+m^{\prime}-1)-mm^{\prime}},\nonumber \\
\left(\Sigma^{y}\right)_{mm^{\prime}}= & \frac{i}{2}\left(\delta_{m,m^{\prime}+1}-\delta_{m+1,m^{\prime}}\right)\sqrt{(s+1)(m+m^{\prime}-1)-mm^{\prime}},\nonumber \\
\left(\Sigma^{z}\right)_{mm^{\prime}}= & (s+1-m)\delta_{m,m^{\prime}},\label{eq:Sigma_xyz}
\end{align}
where $m$ and $m^{\prime}$ run from 1 to $2s+1$. These matrices satisfy algebraic relations $\Sigma^{i}\Sigma^{j}-\Sigma^{j}\Sigma^{i}=i\varepsilon_{ijk}\Sigma^{k}$ and $\Sigma^{i}\Sigma^{j}\delta^{i,j}=s\left(s+1\right)$. The matrices $\Sigma^{i_{1} i_{2} \cdots i_{n}}$ with different numbers of indices are designed to be orthogonal to each other. They are constructed from $\Sigma^{i_{1}}$, given by,
\begin{align}
\Sigma^{i_{1}i_{2}\cdots i_{2n}}= & \frac{1}{\left(2n\right)!}\left(\Sigma^{\{i_{1}}\Sigma^{i_{2}}\cdots\Sigma^{i_{2n}\}}+b_{1}\delta^{\{i_{1},i_{2}}\Sigma^{i_{3}}\Sigma^{i_{4}}\cdots\Sigma^{i_{2n}\}}\right.\nonumber \\
&\left.+\cdots+b_{n}\delta^{\{i_{1},i_{2}}\delta^{i_{3},i_{4}}\cdots\delta^{i_{2n-1},i_{2n}\}}\right),\nonumber \\
\Sigma^{i_{1}i_{2}\cdots i_{2n+1}}= & \frac{1}{\left(2n+1\right)!}\left(\Sigma^{\{i_{1}}\Sigma^{i_{2}}\cdots\Sigma^{i_{2n+1}\}}+c_{1}\delta^{\{i_{1},i_{2}}\Sigma^{i_{3}}\Sigma^{i_{4}}\cdots\Sigma^{i_{2n+1}\}}\right.\nonumber \\
&\left.+\cdots+c_{n}\delta^{\{i_{1},i_{2}}\delta^{i_{3},i_{4}}\cdots\delta^{i_{2n-1},i_{2n}}\Sigma^{i_{2n+1}\}}\right),\label{eq:Sigma_ijk}
\end{align}
where $\{\cdots\}$ denotes the symmetrization of the indices, and the coefficients $b_i$ and $c_i$ are determined by the following relations,
\begin{align}
& \text{Tr}[\Sigma^{i_{1}i_{2}\cdots i_{2n}}]=0,\,\text{Tr}[\Sigma^{i_{1}i_{2}\cdots i_{2n}}\Sigma^{j_{1}}\Sigma^{j_{2}}]=0,\,\cdots,\,\text{Tr}[\Sigma^{i_{1}i_{2}\cdots i_{2n}}\Sigma^{j_{1}}\Sigma^{j_{2}}\cdots\Sigma^{j_{2n-2}}]=0,\nonumber \\
& \text{Tr}[\Sigma^{i_{1}i_{2}\cdots i_{2n+1}}\Sigma^{j_{1}}]=0,\,\text{Tr}[\Sigma^{i_{1}i_{2}\cdots i_{2n+1}}\Sigma^{j_{1}}\Sigma^{j_{2}}\Sigma^{j_{3}}]=0,\,\cdots,\,\text{Tr}[\Sigma^{i_{1}i_{2}\cdots i_{2n+1}}\Sigma^{j_{1}}\Sigma^{j_{2}}\cdots\Sigma^{j_{2n-1}}]=0.\label{eq:coefficient_Sigma_ijk}
\end{align}
All indices in the above formulas can be taken as $z$ in calculations for simplicity and without loss of generality. Moreover, these matrices adhere to the relation $\delta^{i_{1},i_{2}}\Sigma^{i_{1}i_{2}\cdots i_{n}}=0$.

The rank-$n$ spin tensor $T^{i_{1}\cdots i_{n}}$ is constructed from $2n+1$ independent spin components,
\begin{align}
S_{L\cdots LL},\,S_{L\cdots LT}^{x},\,S_{L\cdots LT}^{y},\,\cdots,\,S_{T\cdots TT}^{x\cdots xx},\,S_{T\cdots TT}^{x\cdots xy}.\label{eq:spin_components}
\end{align}
We use the convention that the $y$ index appears at most once. For any component, the total number of the subscripts $L$ and $T$ is $n$, and the total number of the superscripts $x$ and $y$ corresponds to the number of $T$ subscripts. This notation system indicates the symmetry of specific spin components, directly reflected in the associated project matrices.

The spin components in Eq.~\eqref{eq:spin_components} are defined using a set of orthogonal spin projection matrices\footnote{To establish a more intuitive connection, we use the same subscript and superscript notation for the projection matrix $\Sigma_{L\cdots T}^{x\cdots xx}$ and the spin component $S_{L\cdots T}^{x\cdots xx}$. To avoid confusion between the subscript $L$ in $\Sigma_{L\cdots T}^{x\cdots xx}$ and the superscript $L$ in $Q_M^{s,L}$, one can alternatively use $\Sigma_{Z\cdots X}$ to denote the projection matrix.},
\begin{align}
S_{L\cdots T}^{x\cdots xx}  =\left\langle \Sigma_{L\cdots T}^{x\cdots xx}\right\rangle,\quad
S_{L\cdots T}^{x\cdots xy}  =\left\langle \Sigma_{L\cdots T}^{x\cdots xy}\right\rangle ,\label{eq:defination_of_spin_components}
\end{align}
where $\left\langle \Sigma_{L\cdots T}^{i_{1}\cdots i_{m}}\right\rangle \equiv \text{Tr}[\rho_s \Sigma_{L\cdots T}^{i_{1}\cdots i_{m}}]$ and the matrices $\Sigma_{L\cdots T}^{i_{1}\cdots i_{m}}$ are linear combinations of $\Sigma^{i_{1}\cdots i_{n}}$. For $\Sigma^{i_{1}\cdots i_{n}}$ related to $\Sigma_{L\cdots T}^{x\cdots xx}$, the $z$ superscripts match the $L$ subscripts, while the $x$ and $y$ superscripts correspond to the $T$ subscripts, with $y$ appearing an even number of times. For $\Sigma^{i_{1}\cdots i_{n}}$ linked to $\Sigma_{L\cdots T}^{x\cdots xy}$, the conditions are similar, but $y$ occurs an odd number of times. This notation system highlights the correlation between the indices of spin components and those of the corresponding matrices, indicating the dependencies of the physical interpretations of the spin components on the orientation of coordinate axes. For instance, the axis-dependent physical interpretations of the spin-3/2 components are detailed in Ref.~\cite{Zhang:2023box}, revealing potential non-zero polarization correlations for the spin-3/2 baryon pairs producing in annihilation. 

We then detail the specific expressions of the project matrices. For longitudinal spin components, the representation is direct,
\begin{align}
S_{L\cdots LL}= & \left\langle \Sigma_{L\cdots LL}\right\rangle =\left\langle \Sigma^{z\cdots zz}\right\rangle.\label{eq:defination_SL}
\end{align}
For spin components with $2t-1$ transverse indices, the project matrices are combinations of the $\Sigma^{i_{1}\cdots i_{n}}$ matrices with $2t-1$ transverse indices,
\begin{align}
S_{L\cdots T}^{x\cdots xx}= & \left\langle \Sigma_{L\cdots T}^{x\cdots xx}\right\rangle=\left\langle \Sigma^{x\cdots xz\cdots z}+d_{1}\Sigma^{x\cdots xyyz\cdots z}+\cdots+d_{t-1}\Sigma^{xy\cdots yz\cdots z}\right\rangle, \nonumber \\
S_{L\cdots T}^{x\cdots xy}= & \left\langle \Sigma_{L\cdots T}^{x\cdots xy}\right\rangle=\left\langle \left(-1\right)^{t-1}\Sigma^{y\cdots yz\cdots z}+e_{1}\Sigma^{xxy\cdots yz\cdots z}+\cdots+e_{t-1}\Sigma^{x\cdots xyz\cdots z}\right\rangle. \label{eq:defination_SLT_1}
\end{align}
The coefficients for these expressions are determined by ensuring orthogonality with matrices $\Sigma^{i_{1}\cdots i_{n}}$ that potentially have $\{1,3,\cdots,2t-3\}$ transverse indices,
\begin{align}
& \text{Tr}\left[\Sigma_{L\cdots T}^{x\cdots xx}\Sigma^{zzx\cdots xz\cdots z}\right]=0,\,\text{Tr}\left[\Sigma_{L\cdots T}^{x\cdots xx}\Sigma^{zzzzx\cdots xz\cdots z}\right]=0,\,\cdots,\,\text{Tr}\left[\Sigma_{L\cdots T}^{x\cdots xx}\Sigma^{z\cdots zxz\cdots z}\right]=0,\nonumber \\
& \text{Tr}\left[\Sigma_{L\cdots T}^{x\cdots xy}\Sigma^{zzy\cdots yz\cdots z}\right]=0,\,\text{Tr}\left[\Sigma_{L\cdots T}^{x\cdots xy}\Sigma^{zzzzy\cdots yz\cdots z}\right]=0,\,\cdots,\,\text{Tr}\left[\Sigma_{L\cdots T}^{x\cdots xy}\Sigma^{z\cdots zyz\cdots z}\right]=0.\label{eq:coefficients_SLT_1}
\end{align}
For spin components with $2t$ transverse indices, the project matrices derive from combinations of the $\Sigma^{i_{1}\cdots i_{n}}$ matrices with $2t$ transverse indices,
\begin{align}
S_{L\cdots T}^{x\cdots xx}= & \left\langle \Sigma_{L\cdots T}^{x\cdots xx}\right\rangle
=\left\langle \Sigma^{x\cdots xz\cdots z}+d_{1}\Sigma^{x\cdots xyyz\cdots z}+\cdots+d_{t}\Sigma^{y\cdots yz\cdots z}\right\rangle, \nonumber \\
S_{L\cdots T}^{x\cdots xy}= & \left\langle \Sigma_{L\cdots T}^{x\cdots xy}\right\rangle
=\left\langle \left(-1\right)^{t-1}\Sigma^{xy\cdots yz\cdots z}+e_{1}\Sigma^{xxxy\cdots yz\cdots z}+\cdots+e_{t-1}\Sigma^{x\cdots xyz\cdots z}\right\rangle. \label{eq:defination_SLT_2}
\end{align}
The coefficients for these expressions are determined by ensuring orthogonality with matrices $\Sigma^{i_{1}\cdots i_{n}}$ that potentially have $\left\{ 0,2,\cdots,2t-2\right\} $ or $\left\{ 2,4,\cdots,2t-2\right\} $ transverse indices,
\begin{align}
& \text{Tr}\left[\Sigma_{L\cdots T}^{x\cdots xx}\Sigma^{zzx\cdots xz\cdots z}\right]=0,\,\text{Tr}\left[\Sigma_{L\cdots T}^{x\cdots xx}\Sigma^{zzzzx\cdots xz\cdots z}\right]=0,\,\cdots,\,\text{Tr}\left[\Sigma_{L\cdots T}^{x\cdots xx}\Sigma^{z\cdots zz\cdots z}\right]=0,\nonumber \\
& \text{Tr}\left[\Sigma_{L\cdots T}^{x\cdots xy}\Sigma^{xzzy\cdots yz\cdots z}\right]=0,\,\text{Tr}\left[\Sigma_{L\cdots T}^{x\cdots xy}\Sigma^{xzzzzy\cdots yz\cdots z}\right]=0,\,\cdots,\,\text{Tr}\left[\Sigma_{L\cdots T}^{x\cdots xy}\Sigma^{xz\cdots zyz\cdots z}\right]=0.\label{eq:coefficients_SLT_2}
\end{align}
In the last step, we derive the normalization coefficients $a_i$ in Eq.~\eqref{eq:Cartesian_form}. The spin density matrix is typically normalized as $\text{Tr}[\rho_s]$=1, resulting in $a_{0}=1/\left(2s+1\right)$. The determination of the normalization coefficient $a_{n}$ relies on the parametrization of $T^{i_{1}\cdots i_{n}}$. We suggest a scheme as follows,
\begin{align}
T^{i_{1}\cdots i_{n}} & =\sum_{j_{1},\cdots,j_{m}}\frac{\text{Tr}[\Sigma^{i_{1}\cdots i_{n}}\Sigma_{L\cdots T}^{j_{1}\cdots j_{m}}]}{\text{Tr}[\Sigma_{L\cdots T}^{j_{1}\cdots j_{m}}\Sigma_{L\cdots T}^{j_{1}\cdots j_{m}}]}S_{L\cdots T}^{j_{1}\cdots j_{m}},\label{eq:Tijk}
\end{align}
where the summation ensures that $S_{L\cdots T}^{i_{1}\cdots i_{m}}$ can cover all possible spin components as detailed in Eq.~\eqref{eq:spin_components}. Then, the normalization coefficient before $T^{i_{1}\cdots i_{n}}$ is given by
\begin{align}
a_{n}= & \frac{\text{Tr}[\Sigma^{z\cdots zz}\Sigma^{z\cdots zz}]}{\underset{i_{1},\cdots,i_{n}}{\sum}\left(\text{Tr}\left[\Sigma^{i_{1}\cdots i_{n}}\Sigma^{z\cdots zz}\right]\right)^{2}}.\label{eq:coefficients_Tijk}
\end{align}
The total degree of polarization from the spin density matrix is given by,
\begin{align}
d=&\frac{1}{\sqrt{2s}}\sqrt{(2s+1)\text{Tr}[\rho_s^2]-1}\nonumber\\
=&\frac{\sqrt{2s+1}}{\sqrt{2s}}\left(a_1 S^{i_1}S^{i_1}+a_2 T^{i_1 i_2}T^{i_1 i_2}+\cdots+a_{2s} T^{i_1 i_2\cdots i_{2s}}T^{i_1 i_2\cdots i_{2s}}\right)^{1/2},\label{eq:degree}
\end{align}
where $d \in[0,1]$.

The methodology above provides a generalized approach for decomposing the spin density matrix of any spin particles in the Cartesian form. To bridge it with the standard form, we establish connections for projection matrices between these forms,
\begin{align}
M\geq1, & \quad a_{M}^{s,L}Q_{M}^{s,L}=\Sigma_{L\cdots T}^{x\cdots xx},\nonumber \\
M=0, & \quad a_{0}^{s,L}Q_{0}^{s,L}=\Sigma_{L\cdots L},\nonumber \\
M\leq-1, & \quad a_{M}^{s,L}Q_{M}^{s,L}=\Sigma_{L\cdots T}^{x\cdots xy}.\label{eq:relation_QLM_SigmaLT}
\end{align}
The total number and the transverse ($T$) number of the subscripts of matrices $\Sigma_{L\cdots T}^{x\cdots xx}$ and $\Sigma_{L\cdots T}^{x\cdots xy}$ match the indices $L$, $ M$ of $Q_{M}^{s,L}$. The coefficients $a_{M}^{s,L}$ are determined by
\begin{align}
M\geq1, & \quad a_{M}^{s,L}=\frac{2s}{2s+1}\text{Tr}\left[\Sigma_{L\cdots T}^{x\cdots xx}Q_{M}^{s,L}\right],\nonumber \\
M=0, & \quad a_{0}^{s,L}=\frac{2s}{2s+1}\text{Tr}\left[\Sigma_{L\cdots L}Q_{0}^{s,L}\right],\nonumber \\
M\leq-1, & \quad a_{M}^{s,L}=\frac{2s}{2s+1}\text{Tr}\left[\Sigma_{L\cdots T}^{x\cdots xy}Q_{M}^{s,L}\right].\label{eq:coefficients_QLM_SigmaLT}
\end{align}
Then, we obtain the relations of the spin components (parameters) between these two forms, given by
\begin{align}
M\geq1, & \quad a_{M}^{s,L}t_{M}^{s,L}=S_{L\cdots T}^{x\cdots xx},\nonumber \\
M=0, & \quad a_{0}^{s,L}t_{0}^{s,L}=S_{L\cdots L},\nonumber \\
M\leq-1, & \quad a_{M}^{s,L}t_{M}^{s,L}=S_{L\cdots T}^{x\cdots xy}.\label{eq:relation_tLM_SLT}
\end{align}

\subsection{Spin density matrix of spin-5/2 particles}
 
As an example and for later use, we present the spin density matrix for spin-5/2 particles. The standard form expression, directly derived from Eq.~\eqref{eq:standard_form}, is omitted for brevity. We focus on the expressions in the Cartesian form and link them with counterparts in the standard form.

In the Cartesian form, the spin-5/2 density matrix is represented as
\begin{align}
\rho_{5/2}= & \frac{1}{5} I+\frac{2}{35}S^{i}\Sigma^{i}+\frac{1}{56}T^{ij}\Sigma^{ij}+\frac{1}{162}T^{ijk}\Sigma^{ijk}+\frac{1}{360}T^{ijkl}\Sigma^{ijkl}+\frac{1}{450}T^{ijklm}\Sigma^{ijklm},\label{eq:spin_5h_g}
\end{align}
where $\Sigma^{i}$, $\Sigma^{ij}$, $\Sigma^{ijk}$, $\Sigma^{ijkl}$, and $\Sigma^{ijklm}$ are complete, orthonormal, and hermitian basis matrices. the matrices $\Sigma^{i}$ in the $S_{z}$ representation are determined by Eq.~\eqref{eq:Sigma_xyz}, detailed in Appendix~\ref{sec:spin_5h}. Following Eq.~\eqref{eq:Sigma_ijk}, the remaining basis matrices are constructed from $\Sigma^{i}$,
\begin{align}
\Sigma^{ij}  = & \frac{1}{2}\Sigma^{\{i}\Sigma^{j\}}-\frac{35}{24}\delta^{\{i,j\}}I,\\
\Sigma^{ijk} = & \frac{1}{6}\Sigma^{\{i}\Sigma^{j}\Sigma^{k\}}-\frac{101}{120}\delta^{\{i,j}\Sigma^{k\}},\\
\Sigma^{ijkl}  = & \frac{1}{24}\Sigma^{\{i}\Sigma^{j}\Sigma^{k}\Sigma^{l\}}-\frac{95}{336}\delta^{\{i,j}\Sigma^{k}\Sigma^{l\}}+\frac{27}{128}\delta^{\{i,j}\delta^{k,l\}}I,\\
\Sigma^{ijklm}  = & \frac{1}{120}\Sigma^{\{i}\Sigma^{j}\Sigma^{k}\Sigma^{l}\Sigma^{m\}}-\frac{29}{432}\delta^{\{i,j}\Sigma^{k}\Sigma^{l}\Sigma^{m\}}+\frac{11567}{120960}\delta^{\{i,j}\delta^{k,l}\Sigma^{m\}}.\label{eq:spin_5h_m2}
\end{align}
Coefficients in these formulas are derived according to Eq.~\eqref{eq:coefficient_Sigma_ijk}. The polarization information is defined through the spin vector $S^{i}$ and the spin tensors $T^{ij}$, $T^{ijk}$, $T^{ijkl}$, and $T^{ijklm}$, constituting 35 independent spin components. These spin components include:
\begin{align}
S^{i}: &  \,S_{L}, \,S_{T}^{x},\, S_{T}^{y},\label{eq:spin_5h_Si}\\
T^{ij}: & \, S_{LL},\, S_{LT}^{x},\, S_{LT}^{y},\, S_{TT}^{xx},\, S_{TT}^{xy},\\
T^{ijk}: & \, S_{LLL},\, S_{LLT}^{x},\, S_{LLT}^{y},\, S_{LTT}^{xx},\,S_{LTT}^{xy},\, S_{TTT}^{xxx},\, S_{TTT}^{xxy},\\
T^{ijkl}: &\,  S_{LLLL},\, S_{LLLT}^{x},\, S_{LLLT}^{y},\,S_{LLTT}^{xx},\, S_{LLTT}^{xy},\nonumber\\
&S_{LTTT}^{xxx},\,S_{LTTT}^{xxy},\, S_{TTTT}^{xxxx},\, S_{TTTT}^{xxxy},\\
T^{ijklm}: &\,  S_{LLLLL},\, S_{LLLLT}^{x},\, S_{LLLLT}^{y},\,S_{LLLTT}^{xx},\, S_{LLLTT}^{xy},\,S_{LLTTT}^{xxx},\nonumber\\
& S_{LLTTT}^{xxy},\, S_{LTTTT}^{xxxx},\, S_{LTTTT}^{xxxy},\,S_{TTTTT}^{xxxxx},\, S_{TTTTT}^{xxxxy}.\label{eq:Spin_5h_Tijklm}
\end{align}
The indices of these spin components indicate the relation between their physical interpretations and the orientations of the coordinate axes. Detailed definitions and physical interpretations of these spin components are provided in Appendix~\ref{sec:spin_5h}. The coefficients in Eq.~\eqref{eq:spin_5h_g} are determined by Eq.~\eqref{eq:coefficients_Tijk}.

The relations between the spin projection matrices in the standard form and those in Cartesian form are determined by Eq.~\eqref{eq:relation_QLM_SigmaLT}, and the correlations of spin components are inferred by Eq.~\eqref{eq:relation_tLM_SLT}. The coefficients $a_{M}^{s,L}$ are presented in Appendix~\ref{sec:spin_5h}.

\subsection{Application of the spin density matrices in helicity formalism}

This paper focuses on analyzing the polarization of particles within the helicity formalism. Here, the spin density matrix is generally expressed as
\begin{align}
\rho_{s}= & \sum_{\mu=0}^{4s\left(s+1\right)}S_{\mu}\Sigma_{\mu},\label{eq:SDM_helicity}
\end{align}
where $S_{\mu}$ denote the polarization expansion coefficients, and $\Sigma_{\mu}$ are the polarization projection matrices. In Sec.~\ref{subsec:decomposition_of_SDM}, we note that spin projection matrices are more explicitly defined in the standard form, whereas the Cartesian form offers clearer physical interpretations for spin components. To deepen our understanding of the spin density matrix within the helicity formalism, we aim to establish connections between $\Sigma_{\mu}$ and $Q_{M}^{s,L}$. Additionally, we present the correlations between $S_{\mu}$ and the Cartesian form spin components.

For a particle with spin 1/2, the formulation in Eq.~\eqref{eq:SDM_helicity} includes,
\begin{align}
S_{\mu}= & \left\{ S_{0},\,S_{1},\,S_{2},\,S_{3}\right\} ,\label{eq:helicity_1h_components}\\
\Sigma_{\mu}= & \frac{1}{2}\left\{ I,\,\sigma_{x},\,\sigma_{y},\,\sigma_{z}\right\} ,\label{eq:helicity_1h_matrix}
\end{align}
where $\sigma_{x}$, $\sigma_{y}$, and $\sigma_{z}$ are the Pauli matrices. Then, the relation between $\Sigma_{\mu}$ and $Q_{M}^{1/2, L}$ is given by
\begin{align}
\Sigma_{\mu}=  \frac{1}{2}\left\{ I,\,Q_{1}^{1/2, 1},\,Q_{-1}^{1/2, 1},\,Q_{0}^{1/2, 1}\right\} .\label{eq:helicity_1h_matrix_2}
\end{align}
The trace of the density matrix, $\text{Tr}[\rho_{s}]$, equals $S_{0}$ and corresponds to the cross section, usually not normalized to 1. To align with conventional spin components in the Cartesian form, the polarization expansion coefficients are divided by $S_{0}$, resulting in
\begin{align}
\{S_{1}, \,S_{2},\, S_{3}\}/S_{0}=\{S_{T}^{x},\, S_{T}^{y},\, S_{L}\}.\label{eq:helicity_1h_components_2}
\end{align}

For spin-3/2 particles, $\Sigma_{\mu}$ can be chosen from the basis matrices $Q_{M}^{3/2, L}$~\cite{Perotti:2018wxm}. However, we prefer the refined selection scheme~\cite{Zhang:2023wmd, Zhang:2023box},
\begin{align}
\left\{\begin{array}{cccc}
\Sigma_{0}, & \Sigma_{1}, & \Sigma_{2}, & \Sigma_{3},\\
\Sigma_{4}, & \Sigma_{5}, & \Sigma_{6}, & \Sigma_{7},\\
\Sigma_{8}, & \Sigma_{9}, & \Sigma_{10}, & \Sigma_{11},\\
\Sigma_{12}, & \Sigma_{13}, & \Sigma_{14}, & \Sigma_{15}
\end{array}\right\}=&
\frac{1}{4}\left\{\begin{array}{cccc}
I, & \frac{2\sqrt{15}}{5}Q_{0}^{3/2,1}, & \frac{2\sqrt{15}}{5}Q_{1}^{3/2,1}, & \frac{2\sqrt{15}}{5}Q_{-1}^{3/2,1},\\
\sqrt{3}Q_{0}^{3/2,2}, & Q_{1}^{3/2,2}, & Q_{-1}^{3/2,2}, & Q_{2}^{3/2,2},\\
Q_{-2}^{3/2,2}, & \frac{2\sqrt{15}}{3}Q_{0}^{3/2,3}, & \sqrt{10}Q_{1}^{3/2,3}, & \sqrt{10}Q_{-1}^{3/2,3},\\
Q_{2}^{3/2,3}, & Q_{-2}^{3/2,3}, & \frac{\sqrt{6}}{3}Q_{3}^{3/2,3}, & \frac{\sqrt{6}}{3}Q_{-3}^{3/2,3}
\end{array}\right\},\label{eq:helicity_3h_matrix}
\end{align}
where $Q_{M}^{3/2, L}$ are defined in Eq.~\eqref{eq:QLM}. The specific expressions of $\Sigma_\mu$ are provided in Appendix~\ref{sec:basis matrices}. In this scheme, $S_{\mu}$ correspond directly to the Cartesian form spin components, expressed as
\begin{align}
\frac{1}{S_0}
\left\{
\begin{array}{cccc}
S_{0},&S_{1},&S_{2},&S_{3},\\
S_{4},&S_{5},&S_{6},&S_{7},\\
S_{8},&S_{9},&S_{10},&S_{11},\\
S_{12},&S_{13},&S_{14},&S_{15}
\end{array}\right\}
=\left\{
\begin{array}{cccc}
1,&S_{L},&S_{T}^{x},&S_{T}^{y},\\
S_{LL},&S_{LT}^{x},&S_{LT}^{y},&S_{TT}^{xx},\\
S_{TT}^{xy},&S_{LLL},&S_{LLT}^{x},&S_{LLT}^{y},\\
S_{LTT}^{xx},&S_{LTT}^{xy},&S_{TTT}^{xxx},&S_{TTT}^{yxx}
\end{array}\right\}.\label{eq:helicity_3h_components}
\end{align}

For higher spin states, we also aim to establish direct correspondences between $S_{\mu}$ and the spin components in the Cartesian form. For this purpose, the projection matrices can be chosen as:
\begin{align}
\Sigma_{L\left(L+1\right)+M} & =\frac{2s}{\left(2s+1\right)a_{M}^{s,L}}Q_{M}^{s,L},\label{eq:helicity_higher_spin_matrix}
\end{align}
where $a_{M}^{s,L}$ is derived from Eq.~\eqref{eq:coefficients_QLM_SigmaLT}. The correspondences between $S_{\mu}$ and the Cartesian form spin components are then given by:
\begin{align}
\frac{1}{S_{0}}
\left\{\begin{array}{c}
 S_{L\left(L+1\right)-L}, \cdots, S_{L\left(L+1\right)}, \cdots, S_{L\left(L+1\right)+L}
\end{array} \right\}=  \left\{\begin{array}{c} S_{T\cdots T}^{x\cdots xy}, \cdots, S_{L\cdots L}, \cdots, S_{T\cdots T}^{x\cdots xx}\end{array}\right\} .\label{eq:helicity_higher_spin_components}
\end{align}
The total degree of polarization is
\begin{align}
d=&\frac{1}{\sqrt{2s}}\sqrt{(2s+1)\text{Tr}[\rho_s^2]-1}=\sqrt{\sum_{L=1}^{2s}\sum_{M=-L}^{L}\left(\frac{S_{L\left(L+1\right)+M}}{a_{M}^{s,L}S_0}\right)^{2}}.
\end{align}

In the method above, we introduce coefficients between $\Sigma^{\mu}$ and $Q_{M}^{s,L}$, which allows us to establish a direct correspondence between $S_{\mu}$ and the spin components in the Cartesian form. Additionally, an alternative approach aligns $\Sigma^{\mu}$ with $Q_{M}^{s,L}$, as demonstrated in the analysis for spin-3/2 particles detailed in Ref.~\cite{Perotti:2018wxm}. This method introduces specific coefficients to bridge $S_{\mu}$ with the Cartesian form spin components. Our selection scheme simplifies their application across different research areas, as polarization expansion coefficients or spin components often emerge as outcomes in various studies.

Using the coefficients $a_{M}^{5/2, L}$ in Appendix~\ref{sec:spin_5h}, one can detail all polarization projection matrices for spin-5/2 particles, achieving a full decomposition of the spin density matrix. The polarization description of spin-5/2 particles is much more complex than that of spin-1/2 particles with only 3 polarization coefficients or spin-3/2 particles with 15 polarization coefficients. We will focus on a selected subset of these spin components.

In this study, we investigate the decay of a spin-5/2 excited baryon into a spin-1/2 baryon and a pseudoscalar meson, a common decay mode for these baryons. We average over the azimuthal angle $\phi$ of the daughter baryon to evaluate the influence of the parent baryon polarization on the polar angle $\theta$ distribution of the daughter baryon. In this case, only the longitudinal spin components $S_{L}$, $S_{LL}$, $S_{LLL}$, $S_{LLLL}$, and $S_{LLLLL}$ are relevant. Consequently, for spin-5/2 baryons, we concentrate primarily on these components. According to Eq.~\eqref{eq:helicity_higher_spin_matrix}, the corresponding polarization projection matrices are defined as:
\begin{align}
\Sigma_{0}= & \frac{1}{6}I,\, \Sigma_{1}=\frac{\sqrt{21}}{21}Q_{0}^{5/2,1},\,
\Sigma_{2}=\frac{\sqrt{70}}{56}Q_{0}^{5/2,2},\nonumber\\
\Sigma_{3}=&  \frac{5\sqrt{6}}{108}Q_{0}^{5/2,3},\,
\Sigma_{4}=\frac{\sqrt{210}}{144}Q_{0}^{5/2,4},\, \Sigma_{5}=\frac{\sqrt{210}}{120}Q_{0}^{5/2,5},\label{eq:helicity_5h_matrix}
\end{align}
where $Q_{0}^{5/2,L}$ is defined in Eq.~\eqref{eq:QLM}. The specific expressions of these matrices are presented in Appendix \ref{sec:basis matrices}. The associated polarization expansion coefficients are expressed as:
\begin{align}
\frac{1}{S_{0}}
\left\{\begin{array}{c}
S_{1},\, S_{2},\, S_{3},\,S_{4},\, S_{5}\end{array}\right\} =
\left\{ \begin{array}{c}
S_{L},\, S_{LL},\, S_{LLL},\,
S_{LLLL},\, S_{LLLLL}\end{array}\right\} .\label{eq:helicity_5h_component}
\end{align}

In the preceding analysis, we have already outlined the polarization representation for individual particles. We concentrates on the polarization description for the production of two baryons, $B_{1}\bar{B}_{2}$. The spin density matrix of a two-baryon system is represented by,
\begin{align}
\rho_{B_{1}\bar{B}_{2}}= & \sum_{\mu,\nu=0}^{ }S_{\mu\nu}\Sigma_{\mu}\otimes\Sigma_{\nu},\label{eq:polarization_correlation_matrix}
\end{align}
where $S_{\mu\nu}$ is the polarization correlation matrix,  $\Sigma_{\mu}$ and $\Sigma_{\nu}$ are the polarization projection matrices for the baryon $B_{1}$ and antibaryon $\bar{B}_{2}$, respectively.

\section{Production process\label{sec:Production}}

In this section, we provide the method to calculate  $S_{\mu\nu}$ for $B_{1}\bar{B}_{2}$ produced in $e^+e^-$ annihilation. Within the helicity formalism, we detail the production density matrix of $B_{1}\bar{B}_{2}$. This matrix depends on the helicity amplitudes $A_{i,j}$ and corresponds to the spin density matrix of the two-baryon system. We identify non-zero polarization correlation components based on their physical properties. For the conjugate process $\bar{B}_{1}B_{2}$, we derive the polarization correlations by detailing the behavior of helicity amplitudes under $CP$ transformation. This enables the research for the $CP$ violation signals by comparing helicity amplitudes between these conjugate processes. In this paper, we only discuss the two-body baryon production processes. For three-body production processes involving two baryons, the polarization correlations between two baryons are more complicated~\cite{Wick:1962zz, Berman:1965gi}. We leave the analysis of three-body or multi-body production processes involving high-spin baryons to future studies.

In experimental searches for excited baryons, the process typically involves detecting one ground state baryon while reconstructing the invariant mass spectrum of the recoiled baryon. Therefore, we conduct detailed analyses for spin combinations of (1/2, 1/2), (1/2, 3/2), and (1/2, 5/2) in two-body productions. We derive specific expressions of $A_{i,j}$, introduce parametrization schemes for them, and detail the corresponding polarization correlation matrices. Then we establish numerical boundaries for the polarization correlation coefficients. 

\subsection{General framework\label{subsec:general_framework}}

\begin{figure}
\begin{centering}
\includegraphics[width=0.8\textwidth]{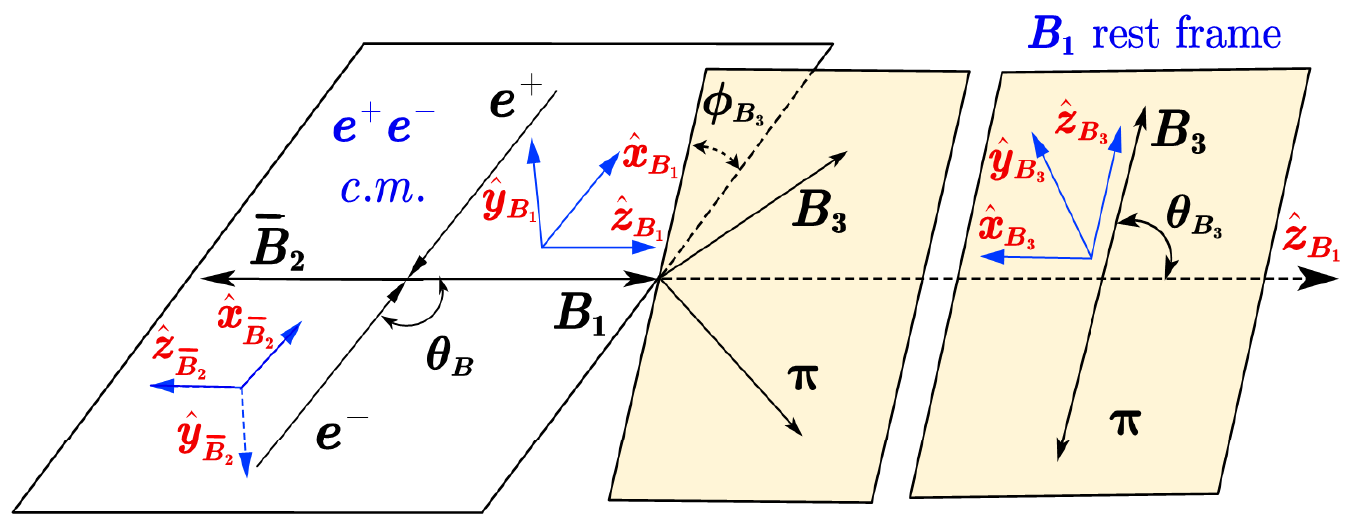}
\par\end{centering}
\caption{\label{fig: production}Definition of coordinate systems and angles in the helicity formalism. The helicity angle $\theta_{B}$ for the $B_{1}\bar{B}_{2}$ production process is the angle between $B_{1}$ and the positron. The decay angles $\theta_{B_{3}}$ and $\phi_{B_{3}}$ for the decay process, $B_{1}\rightarrow B_{3}\pi$, are the polar and azimuthal angles of $B_{3}$ in $B_{1}$ rest frame.}
\end{figure}

Firstly, we introduce the coordinate systems and related angles in the helicity formalism, as shown in Fig.~\ref{fig: production}. We consider the production of $B_{1}\bar{B}_{2}$ in the center-of-mass (c.m.) frame of the initial electron and positron. The positron moving direction is taken as the $z$-axis of the c.m. frame. The helicity angle of $B_{1}$, $\theta_{B}$, is defined as the angle between the $B_{1}$ and the positron. For the $B_{1}$ coordinate system, where we project its polarization components, $z_{B_{1}}$ aligns with its moving direction, $y_{B_{1}}$ is defined by the cross product of the moving directions of the positron and $B_{1}$, and $x_{B_{1}}$ is determined by the right-hand rule. For the $\bar{B}_{2}$ coordinate system, we have $\left\{ \hat{x}_{\bar{B}_{2}},\hat{y}_{\bar{B}_{2}},\hat{z}_{\bar{B}_{2}}\right\} =\left\{ \hat{x}_{B_{1}},-\hat{y}_{B_{1}},-\hat{z}_{B_{1}}\right\}$.

The coordinate systems and decay angles for baryon decay processes are defined in the rest frame of the parent particle. Taking the decay process $B_{1}\rightarrow B_{3}\pi$ as an example, the decay angles $\theta_{B_{3}}$ and $\phi_{B_{3}}$ represent the polar and azimuthal angles of  $B_{3}$ in the rest frame of $B_{1}$. The coordinate system for $B_{3}$ is defined as follows: $\hat{z}_{B_{3}}$ is aligned with the direction of its motion, $\hat{y}_{B_{3}}$ is the cross product of $\hat{z}_{B_{1}}$ and $\hat{z}_{B_{3}}$, and $x_{B_{3}}$ is determined using the right-hand rule. This approach to define helicity systems and angles is similarly applied to multi-step decays or the decays of $\bar{B}_{2}$, see for instance in~\cite{Zhang:2023box}.

Under the one-photon exchange approximation, the production density matrix is given by~\cite{Perotti:2018wxm}
\begin{align}
\rho_{B_{1}\bar{B}_{2}}^{\lambda_{1},\lambda_{2};\lambda_{1}^{\prime},\lambda_{2}^{\prime}}\propto & A_{\lambda_{1},\lambda_{2}}A_{\lambda_{1}^{\prime},\lambda_{2}^{\prime}}^{*} \rho_{1}^{\lambda_{1}-\lambda_{2},\lambda_{1}^{\prime}-\lambda_{2}^{\prime}},\label{eq:production_matrix}
\end{align}
where $A_{\lambda_{1},\lambda_{2}}$ and $A_{\lambda_{1}^{\prime},\lambda_{2}^{\prime}}$ represent the transition amplitudes with helicities $\lambda_{1},\lambda_{1}^{\prime}$ for $B_{1}$ and $\lambda_{2},\lambda_{2}^{\prime}$ for $\bar{B}_{2}$. The matrix $\rho_{1}$ is given by
\begin{align}
\rho_{1}^{i,j}\left(\theta_{B}\right)= & \sum_{\kappa=\pm1}\mathcal{D}_{\kappa,i}^{1*}\left(0,\theta_{B},0\right)\mathcal{D}_{\kappa,j}^{1}\left(0,\theta_{B},0\right),
\end{align}
where $\mathcal{D}_{\kappa,i}^{J}(0,\theta_{B},0)$ denotes the Wigner $\mathcal{D}$-matrix. $\kappa=\pm 1$ is the helicity difference between the positron and electron. For unpolarized electron-positron beams, a summation over $\kappa$ ensures the helicity conservation in the $e^-e^+$ annihilation  into a virtual photon.

We focus on the production of $B_{1}\bar{B}_{2}$ via intermediate states with $J^{P_m}=1^-$, such as $\gamma^{*}$ or vector mesons like $\psi$. Considering parity conservation in the decay process $\gamma^{*}(\psi)\rightarrow B_{1}\bar{B}_{2}$, the transition amplitudes $A_{i,j}$, as defined in Eq.~\eqref{eq:production_matrix}, follow the relationship,
\begin{align}
A_{\lambda_{1},\lambda_{2}} & =P_{B_{1}}P_{\bar{B}_{2}}P_m\left(-1\right)^{J-s_{1}-s_{2}}A_{-\lambda_{1},-\lambda_{2}},\label{eq:parity_conservation}
\end{align}
where $J$, $s_{1}$, and $s_{2}$ denote the spins of $\gamma^{*}(\psi)$, $B_{1}$ and $\bar{B}_{2}$ respectively. 

The production matrix corresponds
to the two-baryon spin density matrix detailed in Eq.~\eqref{eq:polarization_coefficients}. So that, we can obtain the polarization correlation matrix as follows
\begin{align}
S_{\mu\nu}= & \frac{\text{Tr}\left[\left(\rho_{B_{1}\bar{B}_{2}}^{\lambda_{1},\lambda_{2};\lambda_{1}^{\prime},\lambda_{2}^{\prime}}\right)\left(\Sigma_{\mu}\otimes\Sigma_{\nu}\right)\right]}{\text{Tr}\left[\left(\Sigma_{\mu}\otimes\Sigma_{\nu}\right)\left(\Sigma_{\mu}\otimes\Sigma_{\nu}\right)\right]},\label{eq:polarization_coefficients}
\end{align}
where $\Sigma_{\mu}$ and $\Sigma_{\nu}$ are detailed in Eqs.~\eqref{eq:helicity_1h_matrix_2},~\eqref{eq:helicity_3h_matrix}, and~\eqref{eq:helicity_higher_spin_matrix}.

We formulate the selection rule for the possible components of $S_{\mu\nu}$ based on their properties. This is an extended analysis for the method provided in Ref.~\cite{Zhang:2023box}.  When examining the properties of projection matrices, the subscript $M$ of $Q_M^{s,L}$ represents the rotation properties, as mentioned in Sec.~\ref{subsec:decomposition_of_SDM}. The production of $B_1 \bar{B}_2$ can be treated as the Wick helicity rotation or Wigner rotation from the intermediate states into $B_1 \bar{B}_2$~\cite{Jacob:1959at, Doncel:1972ez,Leader:2011vwq}. In our analysis, we consider the spin-1
intermediate states. The rotation difference is limited to a maximum of 2. So certain components are forbidden when the associated matrices $Q_M^{s,L}$ satisfy the condition $\left||M_{B_1}| -| M_{\bar{B}_2}|\right| > 2$. Using Eqs.~\eqref{eq:helicity_1h_matrix_2}, \eqref{eq:helicity_3h_matrix}, and \eqref{eq:helicity_higher_spin_matrix},  we can systematically exclude these prohibited components. Furthermore, according to the established coordinate systems in Fig.~\ref{fig: production}, parity conservation aligns with the $y$-axis, whereas the $x$- and $z$-axes are associated with parity violation~\cite{Zhang:2023box}. Based on the correspondence between the polarization coefficients and the Cartesian form spin components, detailed in Eqs.~\eqref{eq:helicity_1h_components_2}, \eqref{eq:helicity_3h_components}, and \eqref{eq:helicity_higher_spin_components}, when the total number of the longitudinal and $x$ indices is even, the components satisfy parity conservation and are allowed; otherwise, they are prohibited.

The properties of $S_{\mu\nu}$ could be understood from the perspective of the spins of $B_{1}\bar{B}_{2}$. As spins increase, we obtain a greater variety of $A_{i,j}$ and an enriched collection of $\Sigma_{\mu}$ and $\Sigma_{\nu}$. According to Eqs.~\eqref{eq:production_matrix} and~\eqref{eq:polarization_coefficients}, it results in a broader array of $S_{\mu\nu}$ with more complex dependencies on the production angle. This mechanism is crucial for identifying the spin and parity of excited baryons through polarization correlations.

For the conjugate process $e^{+}e^{-}\rightarrow\bar{B}_{1}B_{2}$, similar results can be obtained by applying a $CP$ transformation to the previously analyzed process. The behavior of the helicity amplitudes under the $CP$ transformation is represented by
\begin{align}
A_{i,j}^{B_{1}\bar{B}_{2}} & \xrightarrow{\text{$CP$ transform}}A_{-i,-j}^{\bar{B}_{1}B_{2}}.\label{eq:CP_transform}
\end{align}
For processes where $CP$ is conserved, this formula becomes an equality. $CP$ violation tests can be conducted by comparing helicity amplitudes between these conjugate production processes.

\subsection{Baryons with spin combination of (1/2, 1/2)\label{subsec:1h_1h}}

We consider the production process that the $J^P$ of $B_{1}$ and $\bar{B}_{2}$ are taken as $(\frac{1}{2})^{P_{B_1}}$ and $(\frac{1}{2})^{P_{\bar B_2}}$ respectively. The antibaryon $\bar{B}_{2}$ has an opposite parity to its corresponding baryon $B_{2}$, $P_{\bar{B}_{2}}=-P_{B_{2}}$. The product of parities of $B_{1}$ and $B_{2}$ is defined as
\begin{align}
P_{B_{1}B_{2}}= & P_{B_{1}}P_{B_{2}}=-P_{B_{1}}P_{\bar{B}_{2}}.
\end{align}
Considering  parity  conservation, as detailed in~\eqref{eq:parity_conservation}, out of the four possible helicity transition amplitudes, two are independent,
\begin{align}
h_{1}= & A_{1/2,1/2}=P_{B_{1}B_{2}}A_{-1/2,-1/2},\nonumber\\
h_{2}= & A_{1/2,-1/2}=P_{B_{1}B_{2}}A_{-1/2,1/2}.
\end{align}
The transition amplitude matrix is given by,
\begin{align}
A_{i,j} =
\left(\begin{array}{cc}
h_{1} & h_{2}\\
P_{B_{1}B_{2}}h_{2} & P_{B_{1}B_{2}}h_{1}
\end{array}\right).\label{eq:Aij_1h_1h}
\end{align}
When $\bar B_{2}$ is the corresponding antibaryon of ${B}_{1}$, $P_{B_{1} B_{2}}=1$ and Eq.~\eqref{eq:Aij_1h_1h} is reduced to the form presented in Ref.~\cite{Perotti:2018wxm}.

By substituting Eq.~\eqref{eq:Aij_1h_1h} into Eqs.~\eqref{eq:production_matrix} and~\eqref{eq:polarization_coefficients}, one can identify the polarization correlations.  With the parametrization $h_{1}=\sqrt{1-\alpha_{\psi}}/2$ and $h_{2}=\sqrt{1+\alpha_{\psi}}\text{exp}[-i\phi_{1}]/\sqrt{2}$, we obtain
\begin{align}
S_{\mu\nu}= & \begin{cases}
C_{\mu\nu} & \text{for }\mu=0,3,\\
P_{B_{1}B_{2}}C_{\mu\nu} & \text{for }\mu=1,2,
\end{cases}\label{eq:Smunu_1h_1h}
\end{align}
where the coefficients $C_{\mu\nu}$ are functions of the angle $\theta_{B}$ and parameters $\alpha_{\psi}$ and $\phi_{1}$. The non-zero components are given by,
\begin{align}
C_{0,0}= & 1+\alpha_{\psi}\cos^{2}\theta_{B},\nonumber\\
C_{0,2}= & \sqrt{1-\alpha_{\psi}^{2}}\cos\theta_{B}\sin\theta_{B}\sin\phi_{1},\nonumber\\
C_{1,1}= & \sin^{2}\theta_{B},\nonumber\\
C_{1,3}= & \sqrt{1-\alpha_{\psi}^{2}}\cos\theta_{B}\sin\theta_{B}\cos\phi_{1},\nonumber\\
C_{2,0}= & -C_{0,2},\nonumber\\
C_{2,2}= & \alpha_{\psi}C_{1,1},\nonumber\\
C_{3,1}= & -C_{1,3},\nonumber\\
C_{3,3}= & -\alpha_{\psi}-\cos^{2}\theta_{B}.\label{eq:Cij_1h_1h_g}
\end{align}

For the conjugate process $e^{+}e^{-}\rightarrow\bar{B}_{1}B_{2}$, similar results can be obtained by applying a $CP$ transformation to the previously analyzed process. Considering both $CP$ and parity  conservation, as detailed in Eqs.~\eqref{eq:CP_transform} and~\eqref{eq:parity_conservation}, we obtain the helicity amplitude matrix for $\bar{B}_{1}B_{2}$ production as $A_{i,j}^{\bar{B}_{1}B_{2}}=P_{B_{1}B_{2}}A_{i,j}^{B_{1}\bar{B}_{2}}$. This allows us to derive the polarization correlation matrix for $\bar{B}_{1}B_{2}$, where coefficients have the similar expressions as those for $B_{1}\bar{B}_{2}$, except for substituting $\theta_{B}$ with $\theta_{\bar{B}}$. Here, $\theta_{\bar{B}}$ is the angle between the antibaryon $\bar{B}_{1}$ and the positron.  Under $CP$ conservation, the parameters $\alpha_{\psi}$ and $\phi_{1}$ should remain consistent for these conjugate production processes. By comparing these parameters in $B_{1}\bar{B}_{2}$ and $\bar{B}_{1}B_{2}$ production processes, one can precisely test $CP$ violation. For example, the BESIII Collaboration reported the first measurement of $CP$ violation by comparing the processes $e^{+}e^{-}\rightarrow\Lambda\bar{\Sigma}$ and $e^{+}e^{-}\rightarrow\bar{\Lambda}\Sigma$~\cite{BESIII:2023cvk}.

It is worth mentioning that we apply a uniform parametrization scheme for the helicity amplitudes in both $B_{1}\bar{B}_{2}$ and $\bar{B}_{1}B_{2}$ production processes. This method ensures consistency in parameters $\alpha_{\psi}$ and $\phi_{1}$ across them. This differs from Ref.~\cite{BESIII:2023cvk}, where helicity amplitudes for $\bar{\Lambda}\Sigma^0$ and $\Lambda\bar{\Sigma}^0$ were parametrized differently. Specifically,  $h_2^{\bar{\Lambda}\Sigma^0} = \sqrt{1+\alpha_{\psi}}\text{exp}[-i\phi_{1}^{\bar{\Lambda}\Sigma^0}]$ and $h_2^{\Lambda\bar{\Sigma}^0} = \sqrt{1+\alpha_{\psi}}\text{exp}[-i(\pi-\phi_{1}^{\Lambda\bar{\Sigma}^0})]$, leading to a  phase relation $\phi_{1}^{\bar{\Lambda}\Sigma^0} + \phi_{1}^{\Lambda\bar{\Sigma}^0} = \pi$.

\begin{figure}[ht]
\centering
\includegraphics[width=0.3\textwidth]{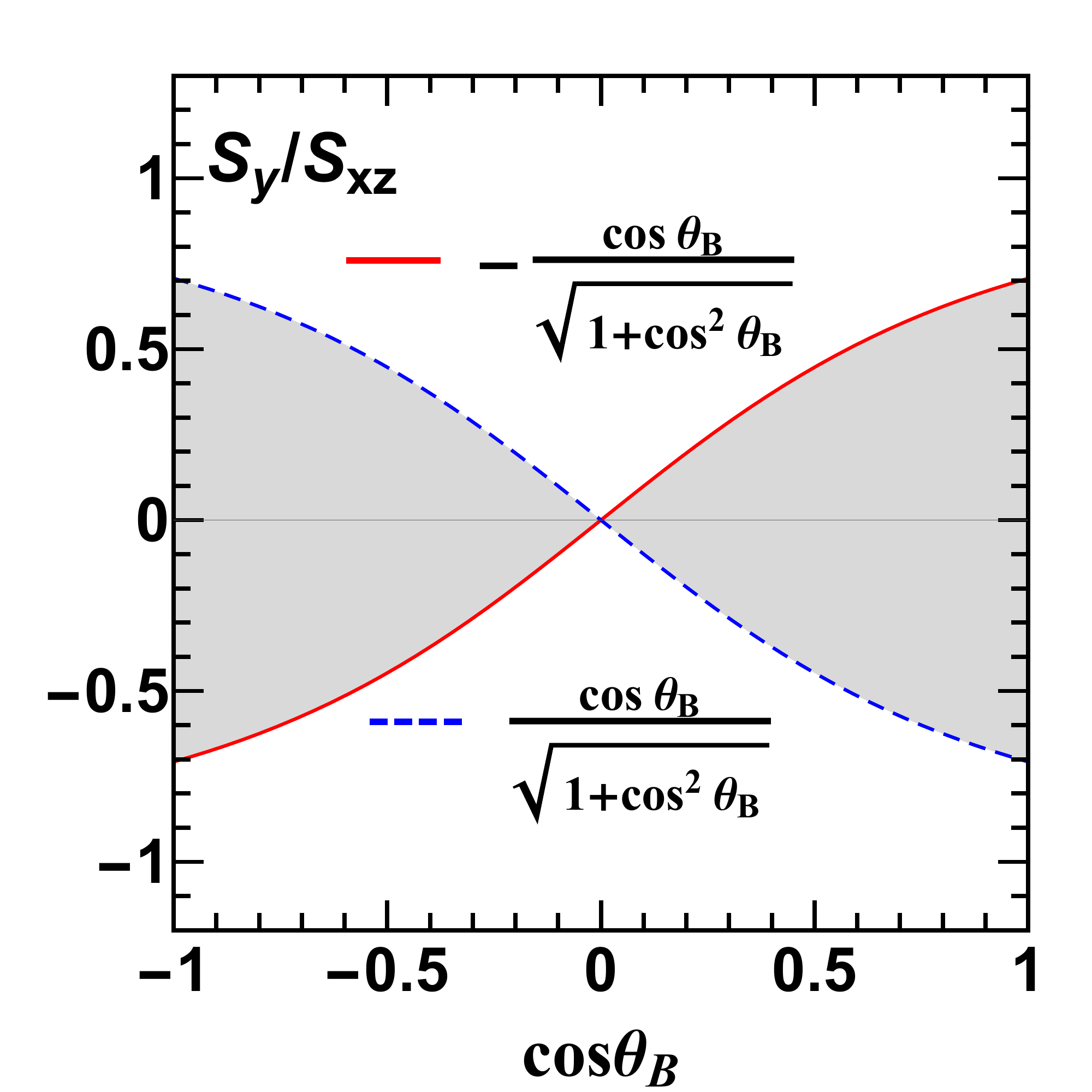}
\quad
\includegraphics[width=0.3\textwidth]{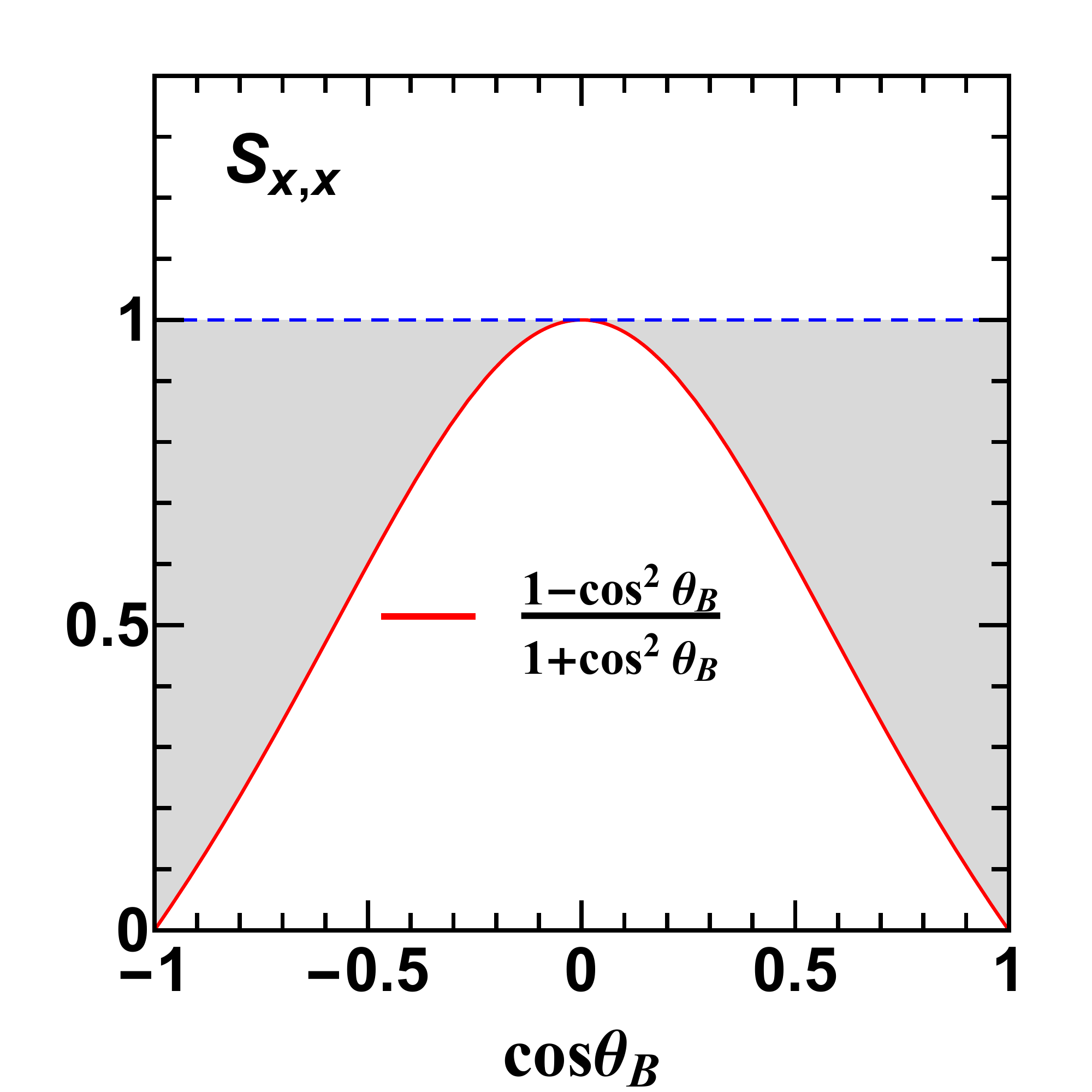}\\
\includegraphics[width=0.3\textwidth]{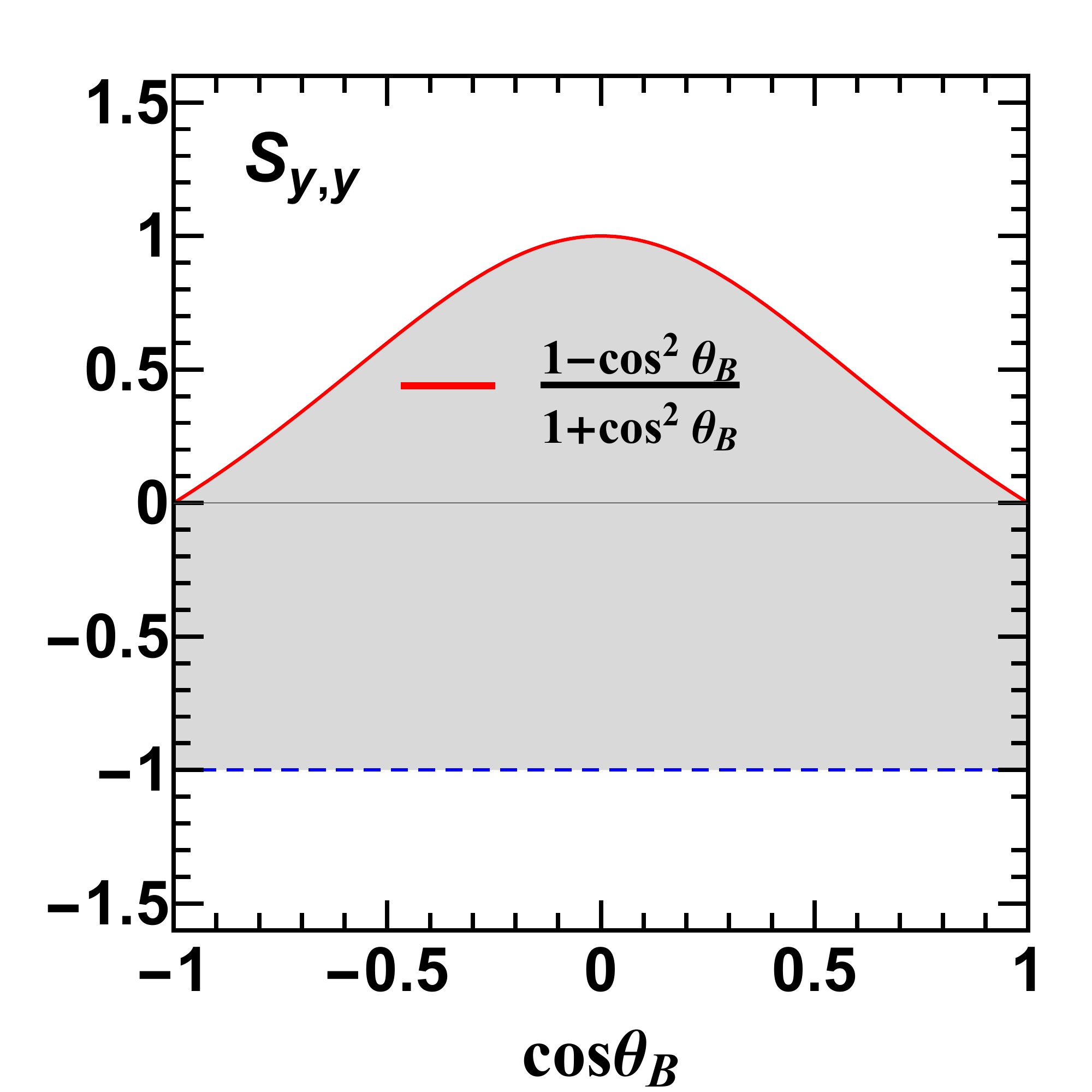}
\quad
\includegraphics[width=0.3\textwidth]{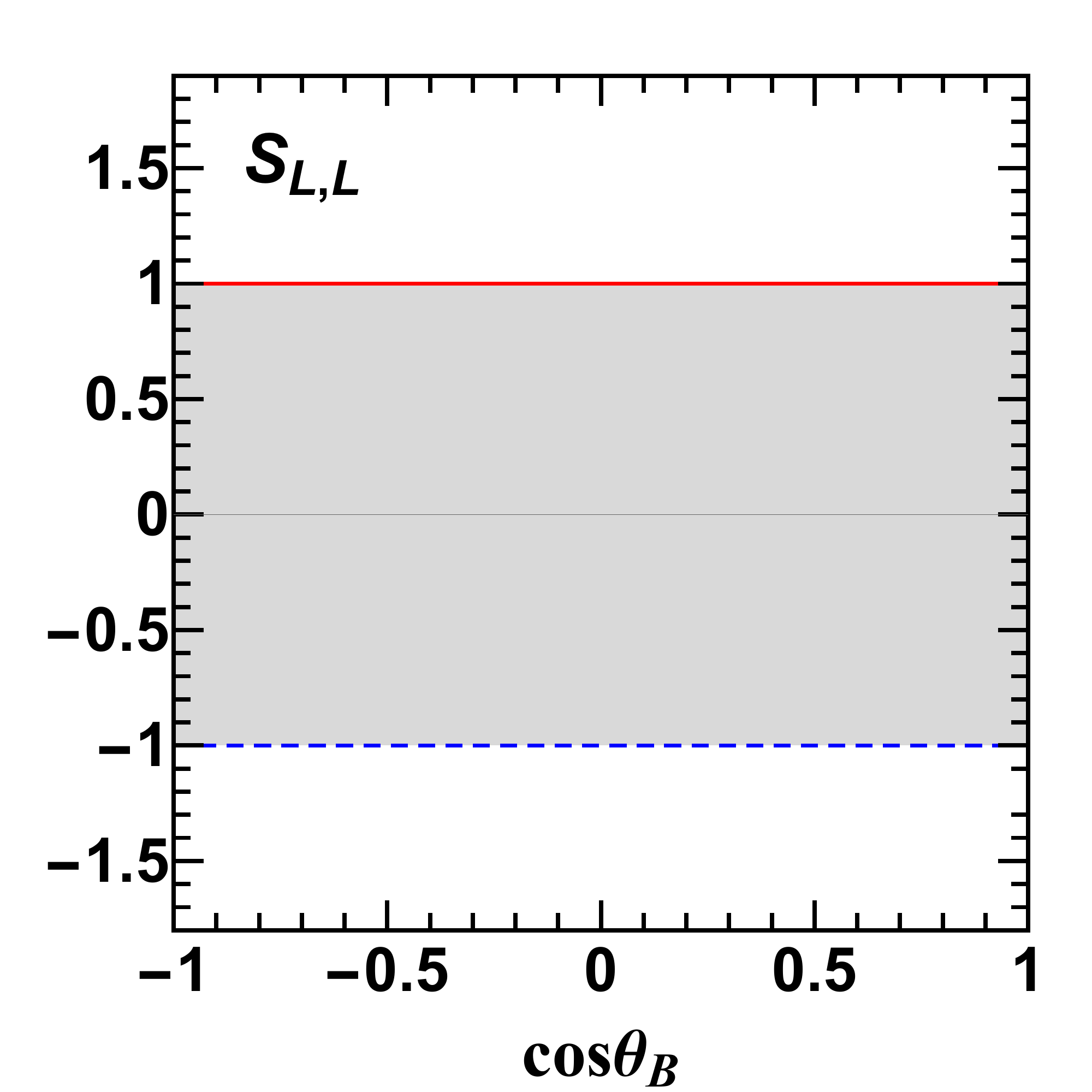}
\caption{\label{fig:1h_1h}Boundaries of the normalization polarization coefficients for the spin combination of (1/2, 1/2).}
\end{figure}

Considering the range of $\alpha_{\psi}$, which is $-1\leq\alpha_{\psi}\leq1$, one can determine the boundaries for the polarization correlations in Eq.~\eqref{eq:Smunu_1h_1h}. We define the following normalized polarization coefficients,
\begin{align}
S_{y}= & S_{2,0}/S_{0,0}=-S_{0,2}/S_{0,0},\nonumber\\
S_{xz}= & S_{1,3}/S_{0,0}=-S_{3,1}/S_{0,0},\nonumber\\
S_{xx}= & S_{1,1}/S_{0,0},\nonumber\\
S_{yy}= & S_{2,2}/S_{0,0},\nonumber\\
S_{zz}= & S_{3,3}/S_{0,0}.
\end{align}
The boundaries for these normalized components are illustrated in Fig.~\ref{fig:1h_1h}. These boundaries show the natural constraints for the polarization correlations and would broaden our understanding of the polarization-related physical mechanisms in this process.

\subsection{Baryons with spin combination of (1/2, 3/2)\label{subsec:1h_3h}}

For the case where $B_1$ is spin 1/2 and $\bar{B}_{2}$ is spin 3/2, the one-photon approximation constrain the helicity transitions $\left|\lambda_{1}-\lambda_{2}\right|\leq1$. Considering  parity  conservation, as detailed in~\eqref{eq:parity_conservation}, only three independent transition amplitudes are obtained,
\begin{align}
h_{1}= & A_{1/2,1/2}=-P_{B_{1}B_{2}}A_{-1/2,-1/2},\nonumber\\
h_{2}= & A_{1/2,-1/2}=-P_{B_{1}B_{2}}A_{-1/2,1/2},\nonumber\\
h_{3}= & A_{1/2,3/2}=-P_{B_{1}B_{2}}A_{-1/2,-3/2}.
\end{align}
In terms of the transition amplitude matrix form, we have
\begin{align}
A_{i,j}=
\left(\begin{array}{cccc}
h_{3} & h_{1} & h_{2} & 0\\
0 & -P_{B_{1}B_{2}}h_{2} & -P_{B_{1}B_{2}}h_{1} & -P_{B_{1}B_{2}}h_{3}
\end{array}\right).\label{eq:Aij_1h_3h}
\end{align}

The polarization correlation matrix can be obtained by substituting Eq.~\eqref{eq:Aij_1h_3h} into Eqs.~\eqref{eq:production_matrix} and~\eqref{eq:polarization_coefficients}. With the following parametrizations,
\begin{eqnarray}
h_{1} & = & \frac{1}{2}\sqrt{1-\alpha_{\psi}}\text{exp}\left[i\phi_{1}\right],\nonumber\\
h_{2} & = & \frac{\sqrt{2}}{2}\sqrt{1+\alpha_{\psi}-\alpha_{1}},\nonumber\\
h_{3} & = & \frac{\sqrt{2}}{2}\sqrt{\alpha_{1}}\text{exp}\left[i\phi_{3}\right],\label{eq:parametrization_g}
\end{eqnarray}
where $ -1\leq\alpha_{\psi}\leq1$ and $0\leq\alpha_{1}\leq1+\alpha_{\psi}$, we have
\begin{align}
S_{\mu\nu}= & \begin{cases}
C_{\mu\nu} & \qquad\text{for }\mu=0,3,\\
P_{B_{1}B_{2}}C_{\mu\nu} & \qquad\text{for }\mu=1,2,
\end{cases}\label{eq:Smunu_1h_3h}
\end{align}
where coefficients $C_{\mu\nu}$ are functions of the angle $\theta_{B}$ and parameters $\alpha_{\psi}$, $\alpha_{1}$, $\phi_{1}$, $\phi_{2}$,
\begin{align}
C_{0,0}= & 1+\alpha_{\psi}\cos^{2}\theta_{B},\nonumber\\
C_{0,3}= & \frac{1}{4}\left(2D_{1}^{s}+\sqrt{3}D_{2}^{s}\right)\sin2\theta_{B},\nonumber\\
C_{0,4}= & -\left(1-\alpha_{1}\right)+\left(\alpha_{1}-\alpha_{\psi}\right)\cos^{2}\theta_{B},\nonumber\\
C_{0,5}= & \frac{\sqrt{3}}{2}D_{2}^{c}\sin2\theta_{B},\nonumber\\
C_{0,7}= & \sqrt{3}D_{3}^{c}\sin^{2}\theta_{B},\nonumber\\
C_{0,11}= & -\frac{1}{10}\left(3D_{1}^{s}-\sqrt{3}D_{2}^{s}\right)\sin2\theta_{B},\nonumber\\
C_{0,13}= & -\sqrt{3}D_{3}^{s}\sin^{2}\theta_{B},\nonumber\\
C_{1,1}= & -\frac{1}{4}D_{1}^{c}\sin2\theta_{B},\nonumber\\
C_{1,2}= & -\frac{1}{2}\left(2-\alpha_{1}\right)\sin^{2}\theta_{B}-\frac{\sqrt{3}}{4}D_{3}^{c}\left(3+\cos2\theta_{B}\right),\nonumber\\
C_{1,6}= & \frac{\sqrt{3}}{2}D_{3}^{s}\left(3+\cos2\theta_{B}\right),\nonumber\\
C_{1,8}= & -\frac{\sqrt{3}}{2}D_{2}^{s}\sin2\theta_{B},\nonumber\\
C_{1,9}= & \frac{9}{20}D_{1}^{c}\sin2\theta_{B},\nonumber\\
C_{1,10}= & \frac{3}{10}\left(2-\alpha_{1}\right)\sin^{2}\theta_{B}-\frac{\sqrt{3}}{10}D_{3}^{c}\left(3+\cos2\theta_{B}\right),\nonumber\\
C_{1,12}= & -\frac{\sqrt{3}}{2}D_{2}^{c}\sin2\theta_{B},\nonumber\\
C_{1,14}= & -\frac{3}{2}\alpha_{1}\sin^{2}\theta_{B},\nonumber\\
C_{2,0}= & \frac{1}{2}D_{1}^{s}\sin2\theta_{B},\nonumber\\
C_{2,3}= & \frac{1}{2}\left(\alpha_{1}-2\alpha_{\psi}\right)\sin^{2}\theta_{B}+\frac{\sqrt{3}}{4}D_{3}^{c}\left(3+\cos2\theta_{B}\right),\nonumber\\
C_{2,4}= & -\frac{1}{2}D_{1}^{s}\sin2\theta_{B},\nonumber\\
C_{2,5}= & \frac{\sqrt{3}}{2}D_{3}^{s}\left(3+\cos2\theta_{B}\right),\nonumber\\
C_{2,7}= & -\frac{\sqrt{3}}{2}D_{2}^{s}\sin2\theta_{B},\nonumber\\
C_{2,11}= & \frac{3}{10}\left(2\alpha_{\psi}-\alpha_{1}\right)\sin^{2}\theta_{B}+\frac{\sqrt{3}}{10}D_{3}^{c}\left(3+\cos2\theta_{B}\right),\nonumber\\
C_{2,13}= & \frac{\sqrt{3}}{2}D_{2}^{c}\sin2\theta_{B},\nonumber\\
C_{2,15}= & \frac{3}{2}\alpha_{1}\sin^{2}\theta_{B},\nonumber\\
C_{3,1}= & -\frac{1}{2}\left(\alpha_{\psi}-2\alpha_{1}\right)-\frac{1}{2}\left(1-2\alpha_{1}\right)\cos^{2}\theta_{B},\nonumber\\
C_{3,2}= & -\frac{1}{4}\left(2D_{1}^{c}-\sqrt{3}D_{2}^{c}\right)\sin2\theta_{B},\nonumber\\
C_{3,6}= & \frac{\sqrt{3}}{2}D_{2}^{s}\sin2\theta_{B},\nonumber\\
C_{3,8}= & -\sqrt{3}D_{3}^{s}\sin^{2}\theta_{B},\nonumber\\
C_{3,9}= & \frac{3}{10}\left(3-\alpha_{1}\right)\cos^{2}\theta_{B}+\frac{3}{10}\left(3\alpha_{\psi}-\alpha_{1}\right),\nonumber\\
C_{3,10}= & \frac{1}{10}\left(3D_{1}^{c}+\sqrt{3}D_{2}^{c}\right)\sin2\theta_{B},\nonumber\\
C_{3,12}= & \sqrt{3}D_{3}^{c}\sin^{2}\theta_{B}.\label{eq:Cij_1h_3h_g}
\end{align}
The $D$-type functions above are defined as
\begin{align}
D_{1}^{s} & =\sqrt{\left(1-\alpha_{\psi}\right)\left(1+\alpha_{\psi}-\alpha_{1}\right)}\sin\phi_{1},\nonumber\\
D_{1}^{c} & =\sqrt{\left(1-\alpha_{\psi}\right)\left(1+\alpha_{\psi}-\alpha_{1}\right)}\cos\phi_{1},\nonumber\\
D_{2}^{s} & =\sqrt{\alpha_{1}\left(1-\alpha_{\psi}\right)}\sin\left(\phi_{1}-\phi_{3}\right),\nonumber\\
D_{2}^{c} & =\sqrt{\alpha_{1}\left(1-\alpha_{\psi}\right)}\cos\left(\phi_{1}-\phi_{3}\right),\nonumber\\
D_{3}^{s} & =\sqrt{\alpha_{1}\left(1+\alpha_{\psi}-\alpha_{1}\right)}\sin\phi_{3},\nonumber\\
D_{3}^{c} & =\sqrt{\alpha_{1}\left(1+\alpha_{\psi}-\alpha_{1}\right)}\cos\phi_{3}.\label{eq:D_function_g}
\end{align}

\begin{figure}[ht]
\centering
\includegraphics[width=0.3\textwidth]{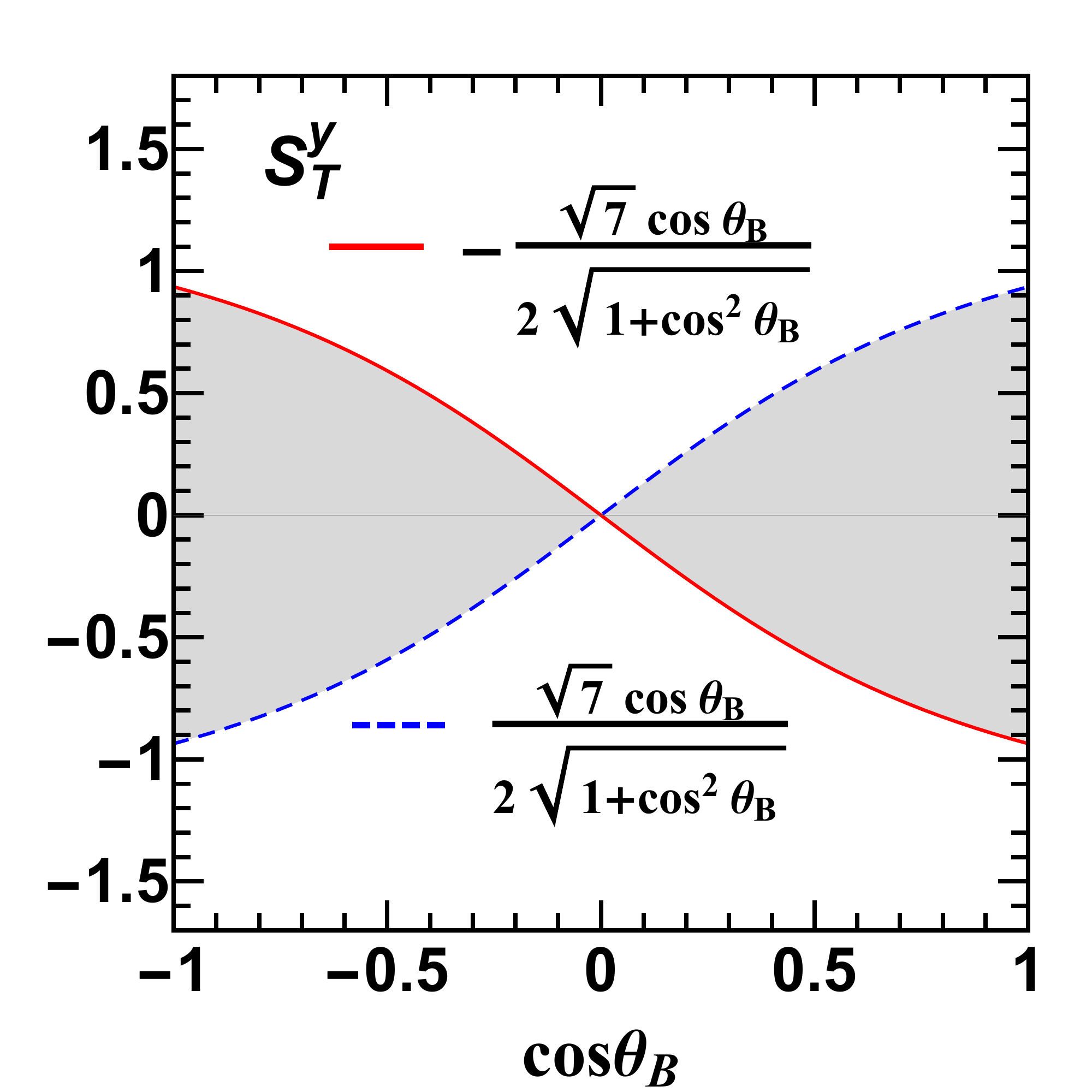}
\quad
\includegraphics[width=0.3\textwidth]{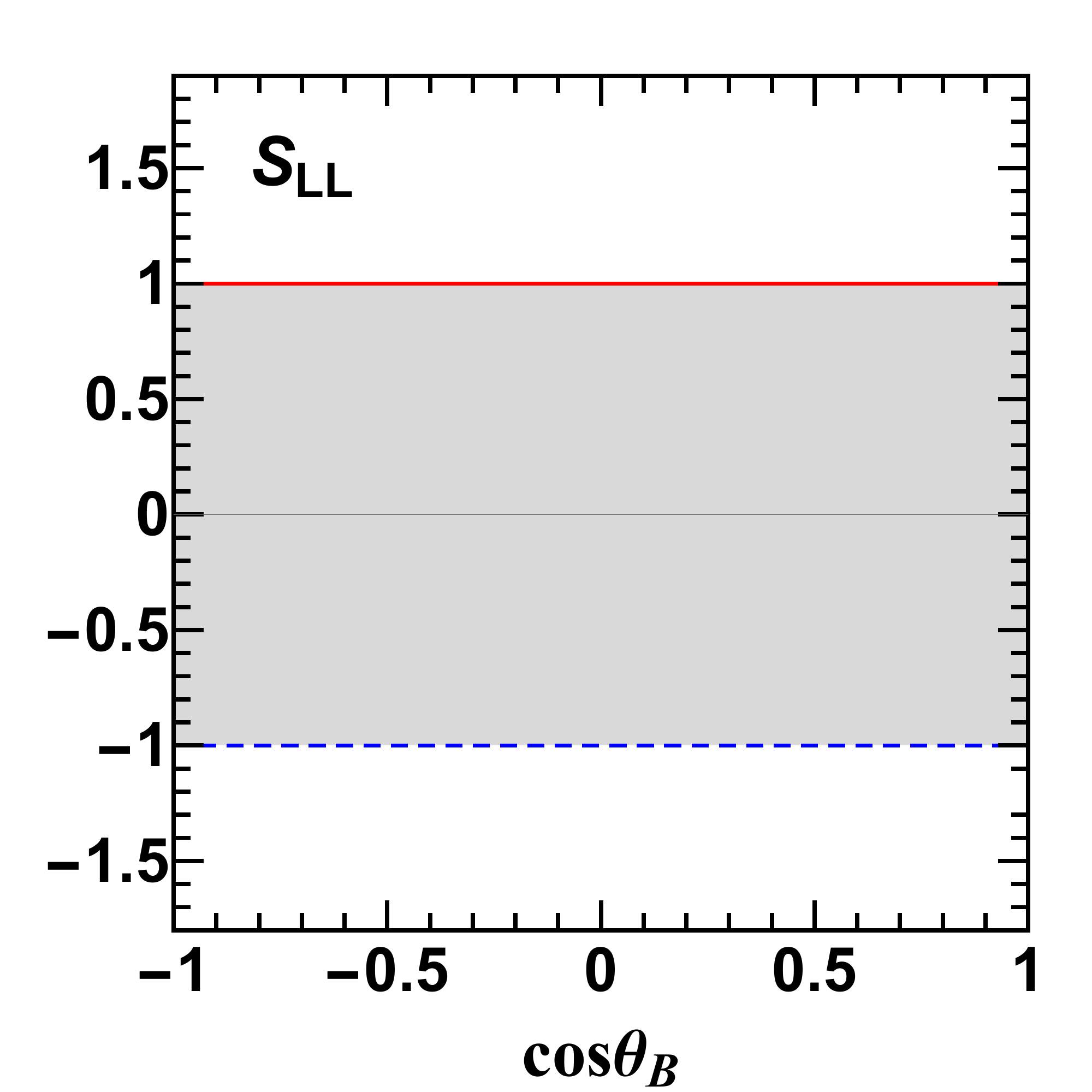}
\quad
\includegraphics[width=0.3\textwidth]{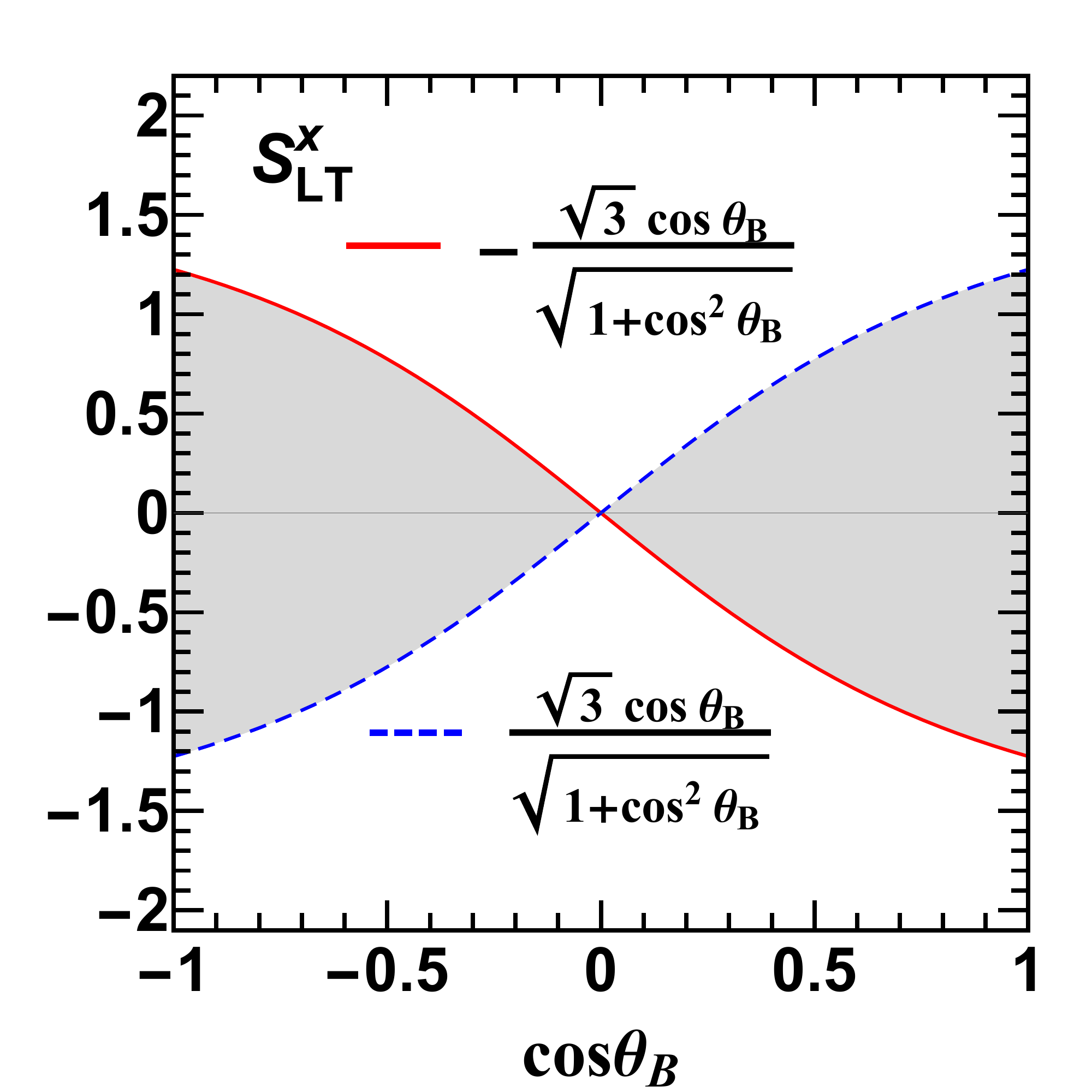}\\
\includegraphics[width=0.3\textwidth]{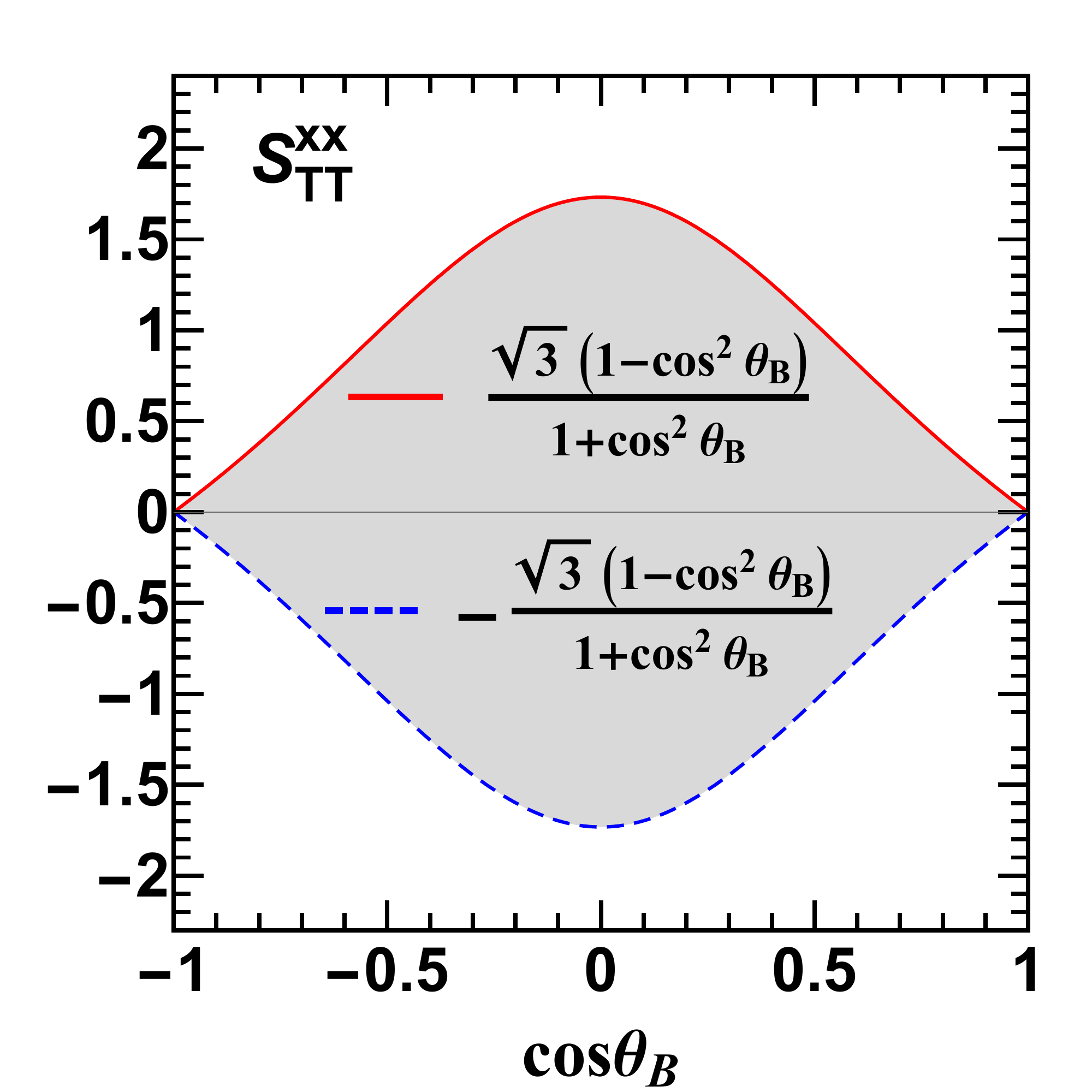}
\quad
\includegraphics[width=0.3\textwidth]{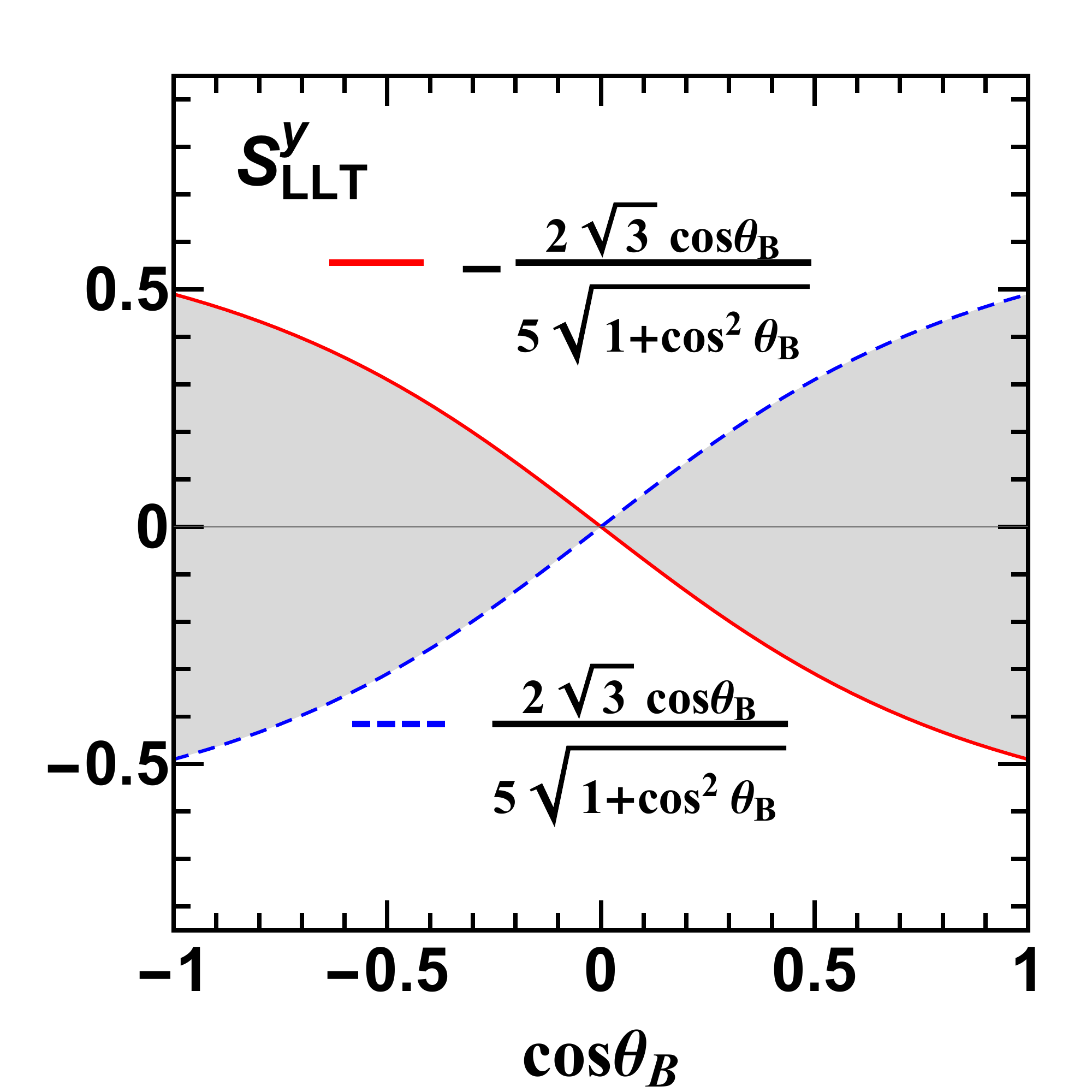}
\quad
\includegraphics[width=0.3\textwidth]{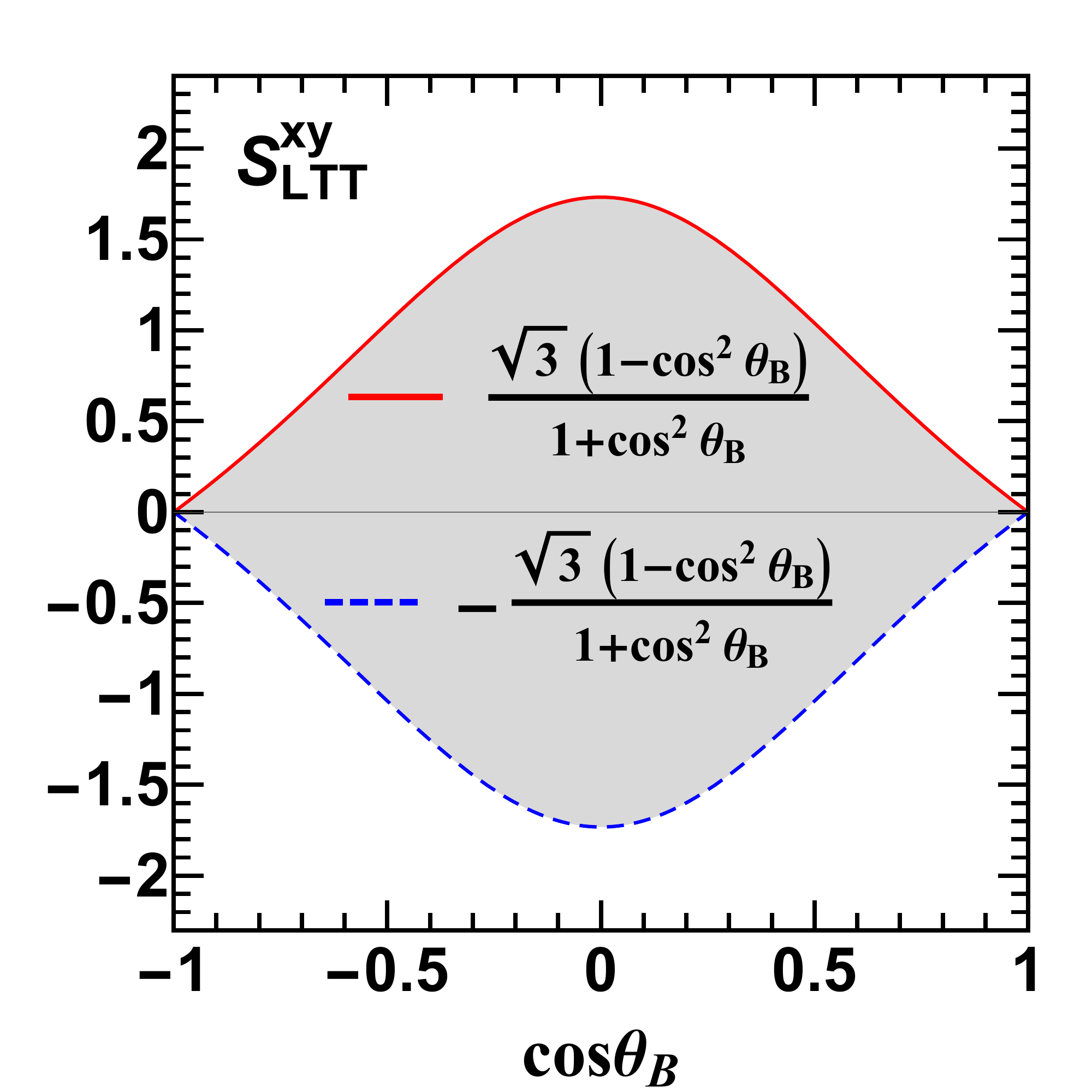}
\caption{\label{fig:1h_3h}Boundaries of the normalization polarization coefficients only associated with the polarization of $\bar{B}_{2}$ for the spin combination of (1/2, 3/2), where $S_{T}^{y}=S_{0,3}/S_{0,0}$, $S_{LL}=S_{0,4}/S_{0,0}$, $S_{LT}^{x}=S_{0,5}/S_{0,0}$, $S_{TT}^{xx}=S_{0,7}/S_{0,0}$, $S_{LLT}^{y}=S_{0,11}/S_{0,0}$, and $S_{LTT}^{xy}=S_{0,13}/S_{0,0}$.}
\end{figure}

For the conjugate process $e^{+}e^{-}\rightarrow\bar{B}_{1}B_{2}$, the helicity amplitude matrix is obtained by using $A_{i,j}^{\bar{B}_{1}B_{2}}=-P_{B_{1}B_{2}}A_{i,j}^{B_{1}\bar{B}_{2}}$ due to the $CP$ and parity  conservation, as detailed in Eqs.~\eqref{eq:CP_transform} and~\eqref{eq:parity_conservation}. Correspondingly, the polarization correlation coefficients for $\bar{B}_{1}B_{2}$ are consistent with those for $B_{1}\bar{B}_{2}$, where $\theta_{B}$ should be replaced by $\theta_{\bar{B}}$. Parameters, $\alpha_{\psi}$, $\alpha_{1}$, $\phi_{1}$, $\phi_{2}$, for these conjugate processes should be consistent under $CP$ conservation.

Normalizing the polarization correlations in Eq.~\eqref{eq:Smunu_1h_3h} by the cross section term $S_{0,0}$, one can obtain normalized polarization correlation coefficients. By considering the ranges of $\alpha_{\psi}$ and $\alpha_{1}$ specified in Eq.~\eqref{eq:parametrization_g}, we determine the boundaries for these normalized correlations. In Fig.~\ref{fig:1h_3h}, we illustrate the boundaries for the components only associated with the polarization of $\bar{B}_{2}$. We can find that the constraints for polarization of the spin-3/2 baryons differ from those in the spin-3/2 baryon pairs production process~\cite{Zhang:2023box}. This difference shows the unique polarization transfer mechanisms for the production of the spin-3/2 baryons with different associated baryons.

\subsection{Baryons with spin combination of (1/2, 5/2)\label{subsec:1h_5h}}

For the production process of two baryons with spin 1/2 and spin 5/2, the analysis  is similar to that in the above subsections. Considering  parity  conservation, as detailed in~\eqref{eq:parity_conservation}, there are three independent helicity transition amplitudes under the constraint of helicity transitions $\left|\lambda_{1}-\lambda_{2}\right|\leq1$,
\begin{align}
h_{1}= & A_{1/2,1/2}=P_{B_{1}B_{2}}A_{-1/2,-1/2},\nonumber\\
h_{2}= & A_{1/2,-1/2}=P_{B_{1}B_{2}}A_{-1/2,1/2},\nonumber\\
h_{3}= & A_{1/2,3/2}=P_{B_{1}B_{2}}A_{-1/2,-3/2}.
\end{align}
The transition amplitude matrix can be expressed as,
\begin{align}
A_{i,j}  =
\left(\begin{array}{cccccc}
0 & h_{3} & h_{1} & h_{2} & 0 & 0\\
0 & 0 & P_{B_{1}B_{2}}h_{2} & P_{B_{1}B_{2}}h_{1} & P_{B_{1}B_{2}}h_{3} & 0
\end{array}\right).\label{eq:Aij_1h_5h}
\end{align}

For simplicity, we are focusing solely on the longitudinal polarization components of the spin-5/2 baryon. By substituting Eq.~\eqref{eq:Aij_1h_5h} into Eqs.~\eqref{eq:production_matrix} and~\eqref{eq:polarization_coefficients}, the polarization correlation matrix is given by,
\begin{align}
S_{\mu\nu}= & \begin{cases}
C_{\mu\nu} & \qquad\text{for }\mu=0,3,\\
P_{B_{1}B_{2}}C_{\mu\nu} & \qquad\text{for }\mu=1,2,
\end{cases}\label{eq:Smunu_1h_5h}
\end{align}
where the coefficients $C_{\mu\nu}$ are functions of $\theta_{B}$, $\alpha_{\psi}$, $\alpha_{1}$, and $\phi_{1}$, given by
\begin{align}
C_{0,0}= & 1+\alpha_{\psi}\cos^{2}\theta_{B},\nonumber\\
C_{0,2}= & -\frac{1}{3}\left(8-3\alpha_{1}\right)-\frac{1}{3}\left(8\alpha_{\psi}-3\alpha_{1}\right)\cos^{2}\theta_{B},\nonumber\\
C_{0,4}= & \frac{6}{7}\left(4-5\alpha_{1}\right)+\frac{6}{7}\left(4\alpha_{\psi}-5\alpha_{1}\right)\cos^{2}\theta_{B},\nonumber\\
C_{1,1}= & \frac{1}{4}D_{1}^{c}\sin2\theta_{B},\nonumber\\
C_{1,3}= & -\frac{6}{5}D_{1}^{c}\sin2\theta_{B},\nonumber\\
C_{1,5}= & \frac{50}{21}D_{1}^{c}\sin2\theta_{B},\nonumber\\
C_{2,0}= & -\frac{1}{2}D_{1}^{s}\sin2\theta_{B},\nonumber\\
C_{2,2}= & \frac{4}{3}D_{1}^{s}\sin2\theta_{B},\nonumber\\
C_{2,4}= & -\frac{12}{7}D_{1}^{s}\sin2\theta_{B},\nonumber\\
C_{3,1}= & -\frac{1}{2}\left(1-2\alpha_{1}\right)\cos^{2}\theta_{B}-\frac{1}{2}\left(\alpha_{\psi}-2\alpha_{1}\right),\nonumber\\
C_{3,3}= & \frac{3}{10}\left(8-11\alpha_{1}\right)\cos^{2}\theta_{B}+\frac{3}{10}\left(8\alpha_{\psi}-11\alpha_{1}\right),\nonumber\\
C_{3,5}= & \frac{100}{21}\left(1-\alpha_{\psi}\right)-\frac{25}{42}\left(4-\alpha_{1}\right)\left(3+\cos2\theta_{B}\right),\label{eq:Cij_1h_5h_g}
\end{align}
where the $D$-type functions are defined in Eq.~\eqref{eq:D_function_g}. For the conjugate process $e^{+}e^{-}\rightarrow\bar{B}_{1}B_{2}$,  the helicity amplitude matrix follows the relation $\ensuremath{A_{i,j}^{\bar{B}_{1}B_{2}}=P_{B_{1}B_{2}}A_{i,j}^{B_{1}\bar{B}_{2}}}$. So polarization correlation coefficients for $\bar{B}_{1}B_{2}$ align with those for $B_{1}\bar{B}_{2}$ with the replacement of $\theta_{B}$ with $\theta_{\bar{B}}$. Under the $CP$ conservation, the parameters $\alpha_{\psi}$, $\alpha_{1}$, $\phi_{1}$ and $\phi_{2}$ should remain consistent across these conjugate processes.
\begin{figure}[ht]
\centering
\includegraphics[width=0.3\textwidth]{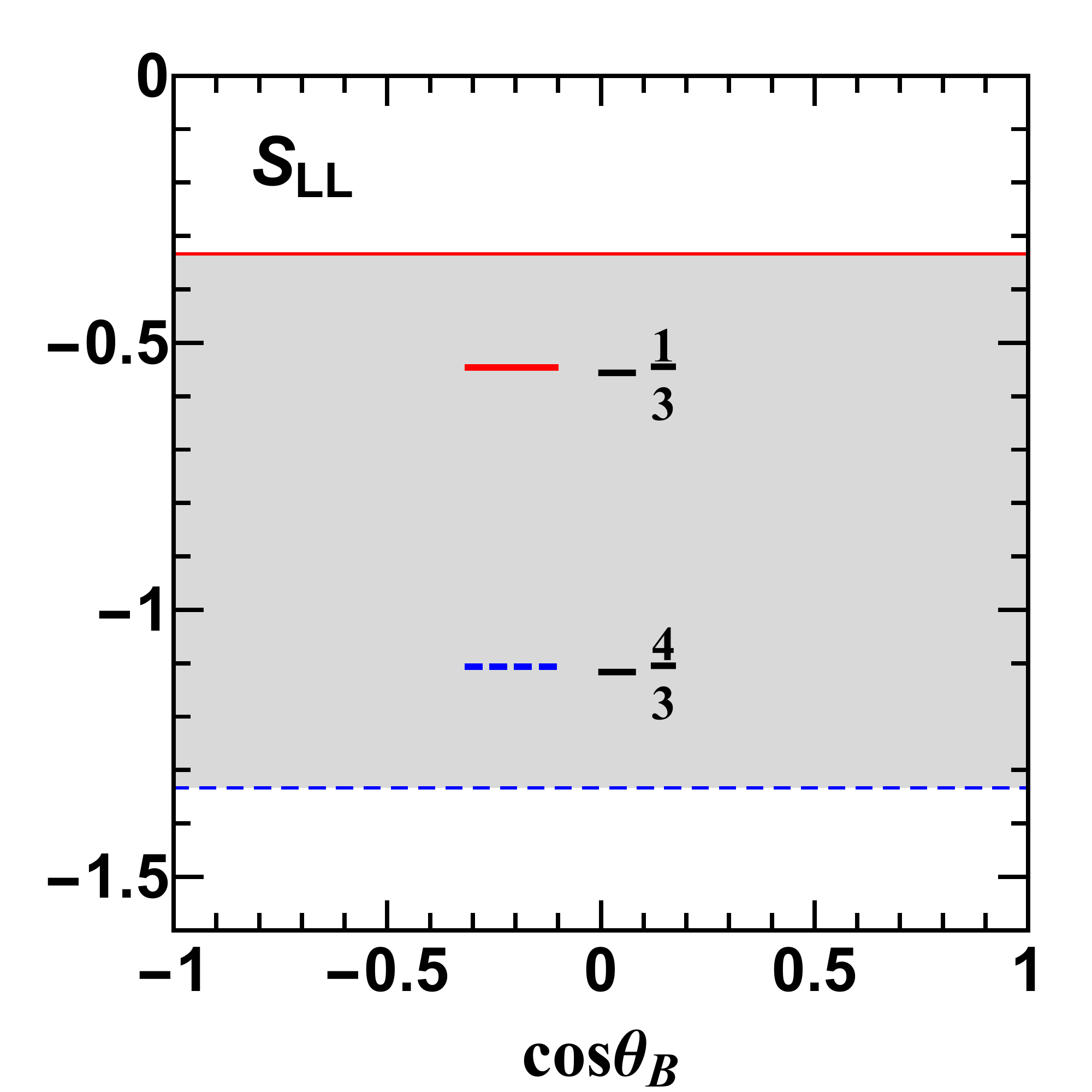}
\quad
\includegraphics[width=0.3\textwidth]{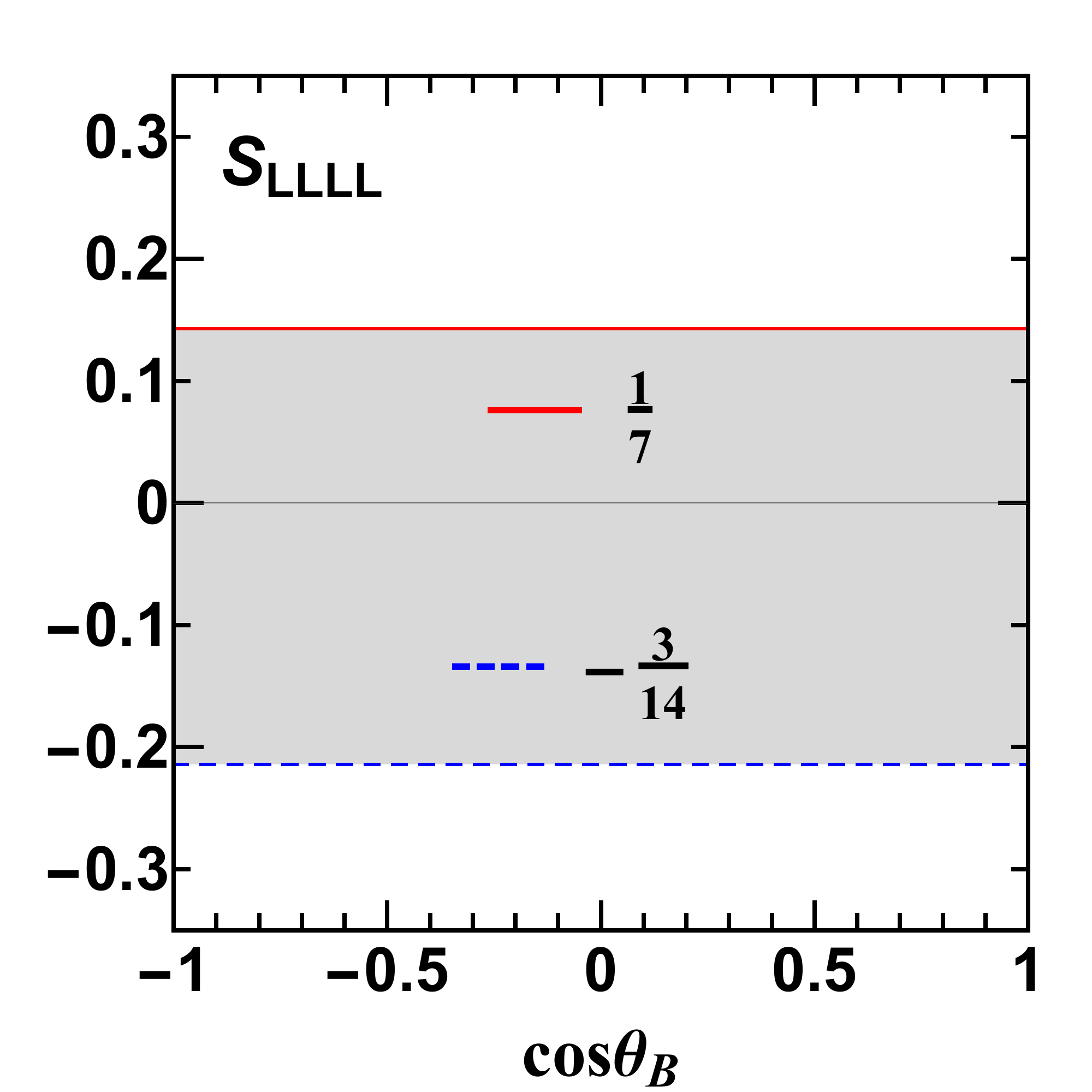}
\caption{\label{fig:1h_5h}Boundaries of the normalization polarization coefficients only associated with the polarization of $\bar{B}_{2}$ for the spin combination of (1/2, 5/2), where $S_{LL}=S_{0,2}/S_{0,0}$ and $S_{LLLL}=S_{0,4}/S_{0,0}$.}
\end{figure}

Taking into account the ranges of $\alpha_{\psi}$ and $\alpha_{1}$, one can determine the boundaries for the normalized polarization correlations in Eq.~\eqref{eq:Smunu_1h_5h}. In Fig.~\ref{fig:1h_5h}, we show the boundaries for components solely related to the polarization of $\bar{B}_{2}$.

\section{Decay chains\label{sec:Decay_chains}}

In this section, we explore the technique for experimental measurements of polarization correlations through the decay processes. Baryon decay has been widely discussed, and the expressions can be categorized into three types: Cartesian spin components~\cite{Lee:1957qs, Kim:1992az, Zhang:2023box}, multipole parameters~\cite{Byers:1963zz, Button-Shafer:1965, Donohue:1969fu, Chung:1971ri, Lednicky:1975ry, Lednicky:1985zx} or real multipole parameters~\cite{Doncel:1972ez, Perotti:2018wxm}, and spin density matrix elements~\cite{Ademollo:1964, Berman:1965, Chen:2007zzf, Xia:2019fjf, Cao:2024tvz}. We follow the first type, and one can transform the expressions into other types according to the relations given in Sec.~\ref{sec:sdm}. The analysis presented in this section is an extension of the work in Ref.~\cite{Perotti:2018wxm, Zhang:2023box}. Within the helicity formalism, baryon decay can be represented with the polarization transfer matrix~\cite{Perotti:2018wxm, Zhang:2023box}. We outline the general steps to calculate the polarization transfer matrix for the spin-$J$ parent baryon, focusing on the predominant decay pathway involving a transition to a spin-1/2 baryon and a spin-0 pseudoscalar meson. We present the helicity amplitudes for the decay process, analyze their parity, and relate them to the canonical amplitudes.  We detail decay expressions for baryons with spin 1/2, 3/2, or 5/2 within this decay mode. 

For two-baryon systems with established spins and parities, we give the joint angular distribution of all products, enabling the measurement of the associated parameters. Our analysis extends to excited baryons with unknown spins and parities, particularly the ${\Xi}^{*}$. Identifying spin and parity is crucial. Taking the $e^{+}e^{-} \rightarrow \Xi^{-}\bar{\Xi}^{*+}$ process as an example, we introduce a technique to identify the spin and parity of the $\bar{\Xi}^{*+}$($\Xi^{*-}$). Variations in spin and parity lead to distinct changes in the joint angular distribution of the final products. We specify the moments of the angular distributions sensitive to the spin and parity.

In the moment analysis, different moments are projected onto the polarization correlations of two baryons through the corresponding polarization transfer matrices. We note that there are other forms of moment analysis for single-baryon decays~\cite{Byers:1963zz, Button-Shafer:1965, Donohue:1969fu, Chung:1971ri, Lednicky:1975ry, Lednicky:1985zx}, in which the angular distribution of final-state particles is described with multipole and decay multipole parameters multiplied by the corresponding Wigner-$\mathcal{D}$ functions. These parameters thus represent the independent moments of the angular distribution, which can be projected using the corresponding Wigner-$\mathcal{D}$ functions. This method of moment analysis is widely used in spin and parity analysis via the decay of a single particle~\cite{Schlein:1963zza, Smith:1965zze, Button-Shafer:1966buh, Merrill:1968zz, Amsterdam-CERN-Nijmegen-Oxford:1976ezm, Teodoro:1978bu, Biagi:1986vs, Shafer:1964, Berge:1966zz, Apsell:1974ca, Amsterdam-CERN-Nijmegen-Oxford:1976caz, Hemingway:1984pz}. In our approach, we focus on the polarization correlations of two baryons, which allows for the projection of numerous moments and accurate determination of the spin and the parity of excited baryons. An alternative approach to the moment analysis using multipole and decay multipole parameters in the polarization correlations of two baryons is left for future studies.

\subsection{General description}

For the decay process $J^{P_{B}}\rightarrow1/2^{+}+0^{-}$, the spin density matrix of the daughter baryon is expressed as~\cite{Perotti:2018wxm, Zhang:2023box}
\begin{align}
\rho_{1/2}^{d}= & \sum_{\mu,\nu}S_{\mu}a_{\mu\nu}\Sigma_{\nu}^d,\label{eq:decay}
\end{align}
where $S_{\mu}$ denote the polarization coefficients of the parent baryon. $a_{\mu\nu}$ represents the polarization transfer matrix, which describes the polarization transfer from a spin-$J$ parent baryon to a spin-1/2 daughter baryon. Typically, different symbols are assigned to the polarization transfer matrices based on the spin of the parent particle. For instance, $b_{\mu\nu}$ is used for parent particles with spin 3/2~\cite{Zhang:2023box}, while $d_{\mu\nu}$ is introduced for those with spin 5/2. In the above formula, $a_{\mu\nu}$ is used as a universal symbol, with specific substitutions made according to the spin of the parent particle.

In general, the polarization transfer matrix $a_{\mu\nu}$ can be expressed as,
\begin{align}
a_{\mu\nu}= & \frac{2J+1}{2\pi}\sum_{\kappa,\kappa^{\prime}}\sum_{\lambda,\lambda^{\prime}}B_{\lambda}B_{\lambda^{\prime}}^{*}\left(\varSigma_{\mu}\right)^{\kappa,\kappa^{\prime}}\left(\Sigma_{\nu}\right)^{\lambda^{\prime},\lambda}\mathcal{D}_{\kappa,\lambda}^{J*}\left(\Omega\right)\mathcal{D}_{\kappa^{\prime},\lambda^{\prime}}^{J}\left(\Omega\right),\label{eq:amunu}
\end{align}
where $\mathcal{D}_{\kappa,\lambda}^{J}(\Omega)=\mathcal{D}_{\kappa,\lambda}^{J}(0,\theta,\phi)$ denotes the Wigner $\mathcal{D}$-matrix with $J$ is the spin of the parent baryon, and $B_{\lambda}$, $B_{\lambda^{\prime}}$ represent the helicity amplitudes for this decay process. The variables $\kappa,\kappa^{\prime}$ and $\lambda,\lambda^{\prime}$ denote the helicities of the parent and daughter baryons, while $\Sigma_{\mu}$ and $\Sigma_{\nu}$ are their polarization projection matrices respectively.

We explore the properties of polarization transfer matrices from the perspective of spins. In our analysis, the spins of the final products are limited to 1/2 and 0. $B_{\lambda}$ and $B_{\lambda^{\prime}}$ are distinctly constrained. However, the calculation of polarization transfer matrices involves the polarization projection matrices of the parent particle and the Wigner $\mathcal{D}$-matrices, which are dependent on the spin of the parent particle. An increase in the spin enriches the transfer matrix with more elements and more dependencies on the angles of the final products. This, combined with the effects in the production process, establishes a basis for identifying the spins and parities of excited state particles by analyzing the angular distribution of all products.

The relation between the helicity amplitudes $B_{\lambda}$ and the canonical amplitudes $A_{L}$ is given by~\cite{Jacob:1959at}
\begin{align}
B_{\lambda}= & \sum_{L}\left(\frac{2L+1}{2J+1}\right)^{1/2}\left\langle L,0;S,\lambda|J,\lambda\right\rangle A_{L},\label{eq:canonical amplitudes}
\end{align}
where $\left\langle L,0; S,\lambda|J,\lambda\right\rangle $ denotes the Clebsch-Gordan coefficients, which involve spin of the parent particle $J$, the spin of the daughter particle $S$, and the orbital angular momentum $L$. Canonical amplitudes, labeled as $A_{S}$, $A_{P}$, $A_{D}$, $A_{F}$..., represent different angular momentum states. For the decay process $J^{P_{B}}\rightarrow1/2^{+}+0^{-}$, we have,
\begin{align}
B_{-1/2}= & \frac{\sqrt{2}}{2}\left(A_{J-1/2}+A_{J+1/2}\right),\nonumber\\
B_{1/2}= & \frac{\sqrt{2}}{2}\left(A_{J-1/2}-A_{J+1/2}\right).\label{eq:B1h_1/2}
\end{align}
These amplitudes account for both parity-conserving and parity-violating effects. The conservation of parity in these decays is determined by the relationship,
\begin{align}
B_{1\text{/2}}= & P_{B} P_{B^\prime}P_{m}\left(-1\right)^{J-s_{1}-s_{2}}B_{-1/2},
\end{align}
with $P_{B}$, $ P_{B^\prime}$, and $P_{m}$ being the parities of the parent and daughter particles, and $J$, $s_{1}$, and $s_{2}$ being their spins. In our case, this means that,
\begin{align}
B_{1/2}= & \left(-1\right)^{J+1/2}P_{B}B_{-1\text{/2}}.
\end{align}
This relation indicates that $A_{J-1/2}$ aligns with the parity-conserving aspect for $(-1)^{J+1/2}P_{B}=1$ and the parity-violating aspect for $(-1)^{J+1/2}P_{B}=-1$. Meanwhile, $A_{J+1/2}$ corresponds to parity-conserving aspect for $(-1)^{J+1/2}P_{B}=-1$ and the parity-violating aspect for $(-1)^{J+1/2}P_{B}=1$.

Under the normalization condition $\left|A_{J-1/2}\right|^{2}+\left|A_{J+1/2}\right|^{2}=\left|B_{-1/2}\right|^{2}+\left|B_{1/2}\right|^{2}=1$, we parametrize these amplitudes as follows~\cite{Lee:1957qs}:
\begin{align}
\alpha_{D}= & -2\,\text{Re}\left[A_{J-1/2}^{*}A_{J+1/2}\right]=\left|B_{1/2}\right|^{2}-\left|B_{-1/2}\right|^{2},\nonumber\\
\beta_{D}= & -2\,\text{Im}\left[A_{J-1/2}^{*}A_{J+1/2}\right]=2\,\text{Im}\left[B_{1/2}B_{-1/2}^{*}\right],\nonumber\\
\gamma_{D}= & \left|A_{J-1/2}\right|^{2}-\left|A_{J+1/2}\right|^{2}=2\,\text{Re}\left[B_{1/2}B_{-1/2}^{*}\right],
\end{align}
where $\beta_{D}=\sqrt{1-\alpha_{D}^{2}}\sin\phi_{D}$, and $\gamma_{D}=\sqrt{1-\alpha_{D}^{2}}\cos\phi_{D}$. By substituting Eqs.~\eqref{eq:helicity_1h_matrix_2},~\eqref{eq:helicity_3h_matrix}, and~\eqref{eq:helicity_5h_matrix} into Eq.~\eqref{eq:amunu} respectively, we derive specific expressions for the polarization transfer matrices. The matrices $a_{\mu\nu}$ and $b_{\mu\nu}$ for parent particles with spins 1/2 and 3/2 are already provided in Refs.~\cite{Perotti:2018wxm, Zhang:2023box} and are included in Appendix \ref{sec:tranfer_matrix}. Additionally, we provide expressions for the polarization transfer matrix $d_{\mu\nu}$, which relates to the longitudinal polarization components of the spin-5/2 parent baryons, also detailed in Appendix \ref{sec:tranfer_matrix}. It is worth noting that $\alpha_{D}$, $\beta_{D}$, and $\gamma_{D}$ serve as hyperon decay parameters in weak decay processes. In strong decay processes, where parity is conserved, these parameters are set to $\alpha_{D}=0$, $\beta_{D}=0$, and $\gamma_{D}=(-1)^{J+1/2}P_{B}$.

Combining the polarization correlations with the decay transfer matrices, the joint angular distribution of the final products can be represented as,
\begin{align}
\mathcal{W} \propto &  \sum_{\mu=0}^{ }\sum_{\nu=0}^{ }S_{\mu\nu}^{B_{1}\bar{B}_{2}}a_{\mu0}^{B_{1}}a_{\nu0}^{\bar{B}_{2}},\label{eq:joint_angle_distribution_0}
\end{align}
where $S_{\mu\nu}^{B_{1}\bar{B}_{2}}$, $a_{\mu0}^{B_{1}}$, and $a_{\nu0}^{\bar{B}_{2}}$ are determined according to the spin combinations of $B_{1}\bar{B}_{2}$. When dealing with cascade decays, one can simply extend the decay chains, such as $a_{\mu0}^{B_1}\rightarrow \sum\limits _{\mu_{2}\cdots\mu_{n}} a_{\mu\mu_2}^{B_1}\cdots a_{\mu_n 0}^{B_n}$.  For processes with determined spins and parities,  applying the relevant formulas allows for the measurement of the corresponding polarization correlation coefficients, form factors, decay parameters, and so on.

In our analysis, we also encounter excited states with unknown spins and parities, especially when studying the production of the $\Xi^{*}$. By substituting the polarization correlation matrices into Eq.~\eqref{eq:joint_angle_distribution_0} and applying the maximum likelihood fit method, one can identify the specific matrix that best fits the experimental data. As the spin of the excited baryon increases, the underlying physical mechanism results in a more complex angular dependence pattern of the joint angular distribution of the final products. Intuitively, this complexity enables us to extract a broader range of moments of angular distribution. In the following subsection, we will explicitly detail the moments sensitive to the spins and parities of the excited baryons.

\subsection{Determination of the spin and parity of $\bar{\Xi}^*$\label{subsec:Xi_decay}}

Taking the process $e^{+}e^{-}\rightarrow\Xi^{-}\bar{\Xi}^{*+}$ as an example, we show how to determine the spin and parity of the $\bar{\Xi}^{*+}$($\Xi^{*-}$). The primary decay mode of a $\bar{\Xi}^{*+}$ generally falls into two cases: $\bar{\Xi}^{*+}\rightarrow\bar{\Xi}\pi$ below the $\Lambda K$ production threshold, or $\bar{\Xi}^{*+}\rightarrow\bar{\Lambda} K$ above this threshold. For simplicity, we focus on the decay mode $\bar{\Xi}^{*+}\rightarrow\bar{\Xi}^{+}\pi^{0}$. Using the possible polarization correlation matrices of $\Xi^{-}\bar{\Xi}^{+*}$ as detailed in Eqs.~\eqref{eq:Smunu_1h_1h},~\eqref{eq:Smunu_1h_3h}, and~\eqref{eq:Smunu_1h_5h}, along with the polarization transfer matrices $a_{\mu\nu}$, $b_{\mu\nu}$, and $d_{\mu\nu}$ in Appendix~\ref{sec:tranfer_matrix}, we derive the polarization correlation matrix for $\Xi^-\bar{\Xi}^+$. For spin-5/2 bartons, we focus solely on the components related to longitudinal polarization, necessitating averaging over the azimuthal angle $\phi_{\bar{\Xi}}$ of the daughter baryon $\bar{\Xi}^{+}$. To ensure comparability, we utilize the same approach for cases where the spin of the $\bar{\Xi}^{+*}$ is either 1/2 or 3/2.

When the $\bar{\Xi}^{*+}$ is spin 1/2, the polarization correlation matrix for $\Xi^-\bar{\Xi}^+$ can be expressed as
\begin{align}
S_{\mu\nu}^{\Xi\bar{\Xi}}= & \frac{1}{2\pi}\int_{0}^{2\pi}\sum_{\nu^{\prime}=0}^{}S_{\mu\nu^{\prime}}^{\Xi\bar{\Xi}^{*}}
a_{\nu^{\prime}\nu}^{\bar{\Xi}^{*}}d\phi_{\bar{\Xi}}. \label{eq:smunu}
\end{align}
For the $\bar{\Xi}^{*+}$ with spins 3/2 or 5/2, we can obtain the polarization correlation matrix of $\Xi^-\bar{\Xi}^+$ by replacing $a_{\mu\nu}$ with $b_{\mu\nu}$ or $d_{\mu\nu}$ respectively. The general expression for $S_{\mu\nu}^{\Xi\bar{\Xi}}$ can be written as
\begin{align}
S_{\mu\nu}^{\Xi\bar{\Xi}}= &
\left(\begin{array}{cccc}
S_{0,0} & 0 & 0 & 0\\
0 & S_{1,1} & 0 & S_{1,3}\\
S_{2,0} & 0 & 0 & 0\\
0 & S_{3,1} & 0 & S_{3,3}
\end{array}\right).\label{eq:S_Xi_Xibar}
\end{align}
The explicit expressions of polarization correlation coefficients $S_{0,0}$, ..., $S_{3,3}$ depend on the spin of the $\bar{\Xi}^{*+}$. If the $\bar{\Xi}^{*+}$ is spin $1/2$, the expressions for these coefficients are given by,
\begin{align}
S_{0,0}= & C_{0,0},\nonumber\\
S_{1,1}= & C_{1,3}\sin\theta_{\bar{\Xi}},\nonumber\\
S_{1,3}= & P_{\Xi^{*}}C_{1,3}\cos\theta_{\bar{\Xi}},\nonumber\\
S_{2,0}= & P_{\Xi^{*}}C_{2,0},\nonumber\\
S_{3,1}= & P_{\Xi^{*}}C_{3,3}\sin\theta_{\bar{\Xi}},\nonumber\\
S_{3,3}= & C_{3,3}\cos\theta_{\bar{\Xi}},
\end{align}
where $P_{\Xi^{*}}$ is the parity of the $\Xi^{*-}$ and these coefficients $C_{\mu\nu}$ are detailed in Eq.~\eqref{eq:Cij_1h_1h_g}.

When the $\bar{\Xi}^{*+}$ is spin $3/2$, the expressions for these coefficients are given by,
\begin{align}
S_{0,0}= & C_{0,0}-\frac{1}{4}\left(1+3\cos2\theta_{\bar{\Xi}}\right)C_{0,4}, \nonumber\\
S_{1,1}= & -\frac{4}{5}\sin\theta_{\bar{\Xi}}C_{1,1}+\frac{1}{4} \left(\sin\theta_{\bar{\Xi}}+5\sin3\theta_{\bar{\Xi}}\right)C_{1,9},\nonumber\\
S_{1,3}= & P_{\Xi^{*}}\left[\frac{2}{5}\cos\theta_{\bar{\Xi}}C_{1,1}-\frac{1}{4} \left(3\cos\theta_{\bar{\Xi}}+5\cos3\theta_{\bar{\Xi}}\right)C_{1,9}\right],\nonumber\\
S_{2,0}= & P_{\Xi^{*}}\left[C_{2,0}-\frac{1}{4}\left(1+3\cos2\theta_{\bar{\Xi}}\right)C_{2,4}\right],\nonumber\\
S_{3,1}= & P_{\Xi^{*}}\left[-\frac{4}{5}\sin\theta_{\bar{\Xi}}C_{3,1}+\frac{1}{4} \left(\sin\theta_{\bar{\Xi}}+5\sin3\theta_{\bar{\Xi}}\right)C_{3,9}\right],\nonumber\\
S_{3,3}= & \frac{2}{5}\cos\theta_{\bar{\Xi}}C_{3,1}-\frac{1}{4}\left(3\cos\theta_{\bar{\Xi}} +5\cos3\theta_{\bar{\Xi}}\right)C_{3,9}.
\end{align}
These coefficients $C_{\mu\nu}$ are shown in Eq.~\eqref{eq:Cij_1h_3h_g}.

When the $\bar{\Xi}^{*+}$ have a spin $5/2$, these coefficients are given by,
\begin{align}
S_{0,0}= &C_{0,0}-\frac{3}{14}\left(1+3\cos2\theta_{\bar{\Xi}}\right)C_{0,2}\nonumber\\
&+\frac{3}{32} \left(9+20\cos2\theta_{\bar{\Xi}}+35\cos4\theta_{\bar{\Xi}}\right)C_{0,4},\nonumber\\
S_{1,1}= & \frac{18}{35}\sin\theta_{\bar{\Xi}}C_{1,1}-\frac{1}{4}\left(\sin\theta_{\bar{\Xi}} +5\sin3\theta_{\bar{\Xi}}\right)C_{1,3}\nonumber\\
&+\frac{45}{32}\left(2\sin\theta_{\bar{\Xi}}+7\sin3\theta_{\bar{\Xi}} +21\sin5\theta_{\bar{\Xi}}\right)C_{1,5},\nonumber\\
S_{1,3}= & P_{\Xi^{*}}\left[\frac{6}{35}\cos\theta_{\bar{\Xi}}C_{1,1}-\frac{1}{6}\left(3\cos\theta_{\bar{\Xi}} +5\cos3\theta_{\bar{\Xi}}\right)C_{1,3}\right.\nonumber\\
&\left.+\frac{15}{32}\left(30\cos\theta_{\bar{\Xi}}+35\cos3\theta_{\bar{\Xi}} +63\cos5\theta_{\bar{\Xi}}\right)C_{1,5}\right],\nonumber\\
S_{2,0}= & P_{\Xi^{*}}\left[C_{2,0}-\frac{3}{14}\left(1+3\cos\theta_{\bar{\Xi}}\right)C_{2,2}\right.\nonumber\\
&\left.+\frac{3}{32} \left(9+20\cos2\theta_{\bar{\Xi}}+35\cos4\theta_{\bar{\Xi}}\right)C_{2,4}\right],\nonumber\\
S_{3,1}= & P_{\Xi^{*}}\left[\frac{18}{35}\sin\theta_{\bar{\Xi}}C_{3,1}-\frac{1}{4}\left(\sin\theta_{\bar{\Xi}} +5\sin3\theta_{\bar{\Xi}}\right)C_{3,3}\right.\nonumber\\
&\left.+\frac{45}{32}\left(2\sin\theta_{\bar{\Xi}}+7\sin3\theta_{\bar{\Xi}} +21\sin5\theta_{\bar{\Xi}}\right)C_{3,5}\right],\nonumber\\
S_{3,3}= & \frac{6}{35}\cos\theta_{\bar{\Xi}}C_{3,1}-\frac{1}{6}\left(3\cos\theta_{\bar{\Xi}} +5\cos3\theta_{\bar{\Xi}}\right)C_{3,3}\nonumber\\
&+\frac{15}{32}\left(30\cos\theta_{\bar{\Xi}}+35\cos3\theta_{\bar{\Xi}} +63\cos5\theta_{\bar{\Xi}}\right) C_{3,5}.
\end{align}
These coefficients $C_{\mu\nu}$ are shown in Eq.~\eqref{eq:Cij_1h_5h_g}.
 
We observe that changes in the spin and parity of the $\bar{\Xi}^{*+}$ alter the specific formulation of the $S_{\mu\nu}^{\Xi\bar{\Xi}}$. As the spin of the $\bar{\Xi}^{*+}$ increases, the polarization correlation matrix exhibits more complex angular dependences. Additionally, the parity of the $\Xi^{*-}$($\bar{\Xi}^{*+}$) influences the signs of the polarization coefficients $S_{1,3}$, $S_{2,0}$, and $S_{3,3}$. These changes in the polarization correlation matrix are evident in the angular distribution of the products in $\Xi^-\bar{\Xi}^+$ decay processes.

We focus on their primary decay channels, $\Xi^{-}\rightarrow\Lambda\pi^{-}$ and $\bar{\Xi}^{+}\rightarrow\bar{\Lambda}\pi^{+}$. The joint angular distribution for the decay products is given by,
\begin{align}
\mathcal{W}\left(\omega,\eta\right)= & \sum_{\mu=0}^{3}\sum_{\nu=0}^{3}S_{\mu\nu}^{\Xi\bar{\Xi}}a_{\mu0}^{\Xi}a_{\nu0}^{\bar{\Xi}},\label{eq:joint_angle_distribution}
\end{align}
where $\vec{\omega}$ denotes the set of parameters associated with helicity amplitudes. For the $\bar{\Xi}^{*+}$ with spin 1/2, $\vec{\omega}=\left\{ \alpha_{\psi},\phi_{1},\alpha_{\Xi},\alpha_{\bar{\Xi}}\right\}$, and for the $\bar{\Xi}^{*+}$ with spins 3/2 or 5/2, $\vec{\omega}=\left\{ \alpha_{\psi},\alpha_{1},\phi_{1},\alpha_{\Xi},\alpha_{\bar{\Xi}}\right\}$. $\eta=\left\{ \theta_{\Xi},\theta_{\bar{\Xi}},\theta_{\Lambda},\phi_{\Lambda},\theta_{\bar{\Lambda}},\phi_{\bar{\Lambda}}\right\}$ represents the angles of all products. By analyzing the angular distributions of $\Lambda$ and $\bar{\Lambda}$, we can extract the coefficients of the polarization correlation matrix $S_{\mu\nu}^{\Xi\bar{\Xi}}$,
\begin{align}
\mu_{\mu\nu}= & \frac{1}{16\pi^{2}}\int\mathcal{W}\left(\omega,\eta\right)a_{\mu0}^{\Xi}a_{\nu0}^{\bar{\Xi}}d\Omega_{\Lambda}d\Omega_{\bar{\Lambda}}\nonumber \\
= & E_{\mu}\left(\alpha_{\Xi}\right)\bar{E}_{\nu}\left(\alpha_{\bar{\Xi}}\right)S_{\mu\nu}^{\Xi\bar{\Xi}},
\label{eq:moment1}
\end{align}
where $E_{\mu}\left(\alpha_{\Xi}\right)$ and $\bar{E}_{\nu}\left(\alpha_{\bar{\Xi}}\right)$ are functions of decay parameters $\alpha_{\Xi}$ and $\alpha_{\bar{\Xi}}$ respectively, given by,
\begin{align}
E_{\mu}\left(\alpha_{\Xi}\right)= & \begin{cases}
1 & \mu=0,\\
\frac{1}{3}\alpha_{\Xi}^{2} & \mu=1,2,3.
\end{cases}\label{eq:Emu}
\end{align}

Furthermore, we analyze the spin and parity of the $\bar{\Xi}^{*+}$ by examining the angular distribution of the polar angle $\theta_{\bar{\Xi}}$. We investigate the following moments of angular distributions,
\begin{align}
M_{\alpha\mu\nu}= & \frac{1}{64\pi^{3}}\int\mathcal{W}\left(\omega,\eta\right)a_{\mu0}^{\Xi}d^{\prime}_{\alpha\nu}a_{\nu0}^{\bar{\Xi}}d\Omega_{\Lambda}d\Omega_{\bar{\Lambda}}d\Omega_{\bar{\Xi}}\nonumber \\
= & \frac{1}{4\pi}\int d^{\prime}_{\alpha\nu}\mu_{\mu\nu}d\Omega_{\bar{\Xi}},\label{eq:Moment2}
\end{align}
where $d^{\prime}_{\alpha\nu}$ denotes a modified version of $d_{\mu\nu}$ in Appendix~\ref{sec:tranfer_matrix}, where we omit decay parameters and unnecessary coefficients. The non-zero coefficients are detailed as follows,
\begin{align}
d^{\prime}_{0,0}= & 1,\nonumber\\
d^{\prime}_{2,0}= & 1+3\cos2\theta_{\bar{\Xi}},\nonumber\\
d^{\prime}_{4,0}= & 9+20\cos2\theta_{\bar{\Xi}}+35\cos4\theta_{\bar{\Xi}},\nonumber\\
d^{\prime}_{1,1}= & \sin\theta_{\bar{\Xi}},\nonumber\\
d^{\prime}_{3,1}= & \sin\theta_{\bar{\Xi}}+5\sin3\theta_{\bar{\Xi}},\nonumber\\
d^{\prime}_{5,1}= & 2\sin\theta_{\bar{\Xi}}+7\sin3\theta_{\bar{\Xi}}+21\sin5\theta_{\bar{\Xi}},\nonumber\\
d^{\prime}_{1,3}= & \cos\theta,\nonumber\\
d^{\prime}_{3,3}= & 3\cos\theta_{\bar{\Xi}}+5\cos3\theta_{\bar{\Xi}},\nonumber\\
d^{\prime}_{5,3}= & 30\cos\theta_{\bar{\Xi}}+35\cos3\theta_{\bar{\Xi}}+63\cos5\theta_{\bar{\Xi}}.
\end{align}
\begin{table}[!htbp]
\caption{\label{tab:moments}The potential moments corresponding to angular distribution as defined in Eq.~\eqref{eq:Moment2}. When the spin state of the $\Xi^{*-}$($\bar{\Xi}^{*+}$) is 1/2, 3/2, or 5/2, the coefficients $C_{\mu\nu}$ are determined by Eq. ~\eqref{eq:Cij_1h_1h_g}, Eq.~\eqref{eq:Cij_1h_3h_g}, and Eq.~\eqref{eq:Cij_1h_5h_g} respectively.}
\begin{raggedright}
\[
\begin{array}{cccc}
\hline\hline J^{P}\text{ for the baryon }\Xi^{*-} & (1/2)^{P_{\Xi^{*}}} & (3/2)^{P_{\Xi^{*}}} & (5/2)^{P_{\Xi^{*}}}\\
\hline M_{0,0,0} & C_{0,0} & C_{0,0} & C_{0,0}\\
M_{2,0,0} & 0 & -\frac{4}{5}C_{0,4} & -\frac{24}{35}C_{0,2}\\
M_{4,0,0} & 0 & 0 & \frac{128C_{0,4}}{3}\\
\hline M_{1,1,1} & \frac{2}{27}\alpha_{\Xi}^{2}\alpha_{\bar{\Xi}}^{2}C_{1,3} & -\frac{8}{135}\alpha_{\Xi}^{2}\alpha_{\bar{\Xi}}^{2}C_{1,1} & \frac{4}{105}\alpha_{\Xi}^{2}\alpha_{\bar{\Xi}}^{2}C_{1,1}\\
M_{3,1,1} & 0 & \frac{64}{189}\alpha_{\Xi}^{2}\alpha_{\bar{\Xi}}^{2}C_{1,9} & -\frac{64}{189}\alpha_{\Xi}^{2}\alpha_{\bar{\Xi}}^{2}C_{1,3}\\
M_{5,1,1} & 0 & 0 & \frac{1024}{33}\alpha_{\Xi}^{2}\alpha_{\bar{\Xi}}^{2}C_{1,5}\\
\hline M_{1,1,3} & \frac{1}{27}P_{\Xi^{*}}\alpha_{\Xi}^{2}\alpha_{\bar{\Xi}}^{2}C_{1,3} & \frac{2}{135}P_{\Xi^{*}}\alpha_{\Xi}^{2}\alpha_{\bar{\Xi}}^{2}C_{1,1} & \frac{2}{315}P_{\Xi^{*}}\alpha_{\Xi}^{2}\alpha_{\bar{\Xi}}^{2}C_{1,1}\\
M_{3,1,3} & 0 & -\frac{16}{63}P_{\Xi^{*}}\alpha_{\Xi}^{2}\alpha_{\bar{\Xi}}^{2}C_{1,9} & -\frac{32}{189}P_{\Xi^{*}}\alpha_{\Xi}^{2}\alpha_{\bar{\Xi}}^{2}C_{1,3}\\
M_{5,1,3} & 0 & 0 & \frac{2560}{33}P_{\Xi^{*}}\alpha_{\Xi}^{2}\alpha_{\bar{\Xi}}^{2}C_{1,5}\\
\hline M_{0,2,0} & \frac{1}{3}P_{\Xi^{*}}\alpha_{\Xi}^{2}C_{2,0} & \frac{1}{3}P_{\Xi^{*}}\alpha_{\Xi}^{2}C_{2,0} & \frac{1}{3}P_{\Xi^{*}}\alpha_{\Xi}^{2}C_{2,0}\\
M_{2,2,0} & 0 & -\frac{4}{15}P_{\Xi^{*}}\alpha_{\Xi}^{2}C_{2,4} & -\frac{8}{35}P_{\Xi^{*}}\alpha_{\Xi}^{2}C_{2,2}\\
M_{4,2,0} & 0 & 0 & \frac{128}{9}P_{\Xi^{*}}\alpha_{\Xi}^{2}C_{2,4}\\
\hline M_{1,3,1} & \frac{2}{27}P_{\Xi^{*}}\alpha_{\Xi}^{2}\alpha_{\bar{\Xi}}^{2}C_{3,3} & -\frac{8}{135}P_{\Xi^{*}}\alpha_{\Xi}^{2}\alpha_{\bar{\Xi}}^{2}C_{3,1} & \frac{4}{105}P_{\Xi^{*}}\alpha_{\Xi}^{2}\alpha_{\bar{\Xi}}^{2}C_{3,1}\\
M_{3,3,1} & 0 & \frac{64}{189}P_{\Xi^{*}}\alpha_{\Xi}^{2}\alpha_{\bar{\Xi}}^{2}C_{3,9} & -\frac{64}{189}P_{\Xi^{*}}\alpha_{\Xi}^{2}\alpha_{\bar{\Xi}}^{2}C_{3,3}\\
M_{5,3,1} & 0 & 0 & \frac{1024}{33}P_{\Xi^{*}}\alpha_{\Xi}^{2}\alpha_{\bar{\Xi}}^{2}C_{3,5}\\
\hline M_{1,3,3} & \frac{1}{27}\alpha_{\Xi}^{2}\alpha_{\bar{\Xi}}^{2}C_{3,3} & \frac{2}{135}\alpha_{\Xi}^{2}\alpha_{\bar{\Xi}}^{2}C_{3,1} & \frac{2}{315}\alpha_{\Xi}^{2}\alpha_{\bar{\Xi}}^{2}C_{3,1}\\
M_{3,3,3} & 0 & -\frac{16}{63}\alpha_{\Xi}^{2}\alpha_{\bar{\Xi}}^{2}C_{3,9} & -\frac{32}{189}\alpha_{\Xi}^{2}\alpha_{\bar{\Xi}}^{2}C_{3,3}\\
M_{5,3,3} & 0 & 0 & \frac{2560}{33}\alpha_{\Xi}^{2}\alpha_{\bar{\Xi}}^{2}C_{3,5}
\\\hline\hline \end{array}
\]
\par\end{raggedright}
\end{table}

In Table~\ref{tab:moments}, we detail the potential moments for all products when $\bar{\Xi}^{+*}$($\Xi^{*-}$) possesses spins of 1/2, 3/2, or 5/2. As the spin of the $\bar{\Xi}^{*+}$($\Xi^{*-}$) increases, there is a noticeable expansion in the range of potential moments. This provides us with a method to verify the high spin of the $\bar\Xi^{*+}$($\Xi^{*-}$) by observing these additional moments. Additionally, the parity of the $\Xi^{*-}$ affects the sign correlation between moments $M_{\alpha,1,1}$ and $M_{\alpha,1,3}$, as well as between $M_{\alpha,3,1}$ and $M_{\alpha,3,3}$. For instance, when the $\Xi^{*-}$ has a spin of $1/2^{+}$, the moments $M_{\alpha,1,1}$ and $M_{\alpha,1,3}$ share the same sign. In contrast, when the spin is $1/2^{-}$, these moments exhibit opposite signs. Additionally, every increase in spin by 1, while parity remains unchanged, alters the sign correlation between these moments. These contrasts show the underlying reasons to determine the spin and parity of the $\Xi^{*-}$($\bar{\Xi}^{*+}$) effectively.

Considering the cascade decays of $\Xi^-\bar{\Xi}^+$,  analysis for the moments of angular distributions is similar. For instance, in the cascade decay of $\Xi^{-}$, $\Xi^{-}\rightarrow\Lambda\pi^{-}$, followed by $\Lambda\rightarrow p\pi^{-}$, we simply replace $a_{\mu0}^{\Xi}$ with $\sum\limits _{\mu^{\prime}=0}^{3}a_{\mu\mu^{\prime}}^{\Xi}a_{\mu^{\prime}0}^{\Lambda}$ in Eqs.~\eqref{eq:joint_angle_distribution},~\eqref{eq:moment1}, and~\eqref{eq:Moment2}, and include an additional phase space integral $d\Omega_{p}/(4\pi)$ in Eqs.~\eqref{eq:moment1} and~\eqref{eq:Moment2}. Correspondingly, in Eq.~\eqref{eq:moment1}, $E_{\mu}(\alpha_{\Xi})$ is substituted with $E_{\mu}(\alpha_{\Xi},\alpha_{\Lambda})$, which is given by,
\begin{align}
E_{\mu}\left(\alpha_{\Xi},\alpha_{\Lambda}\right)=  \begin{cases}
1+\frac{1}{3}\alpha_{\Xi}^{2}\alpha_{\Lambda}^{2} & \mu=0,\\
\frac{\alpha_{\Xi}^{2}}{3}+\frac{\alpha_{\Lambda}^{2}}{3}-\frac{2}{9}\alpha_{\Xi}^{2}\alpha_{\Lambda}^{2} & \mu=1,2,3.
\end{cases}\label{eq:Emu2}
\end{align}
This modification enhances the magnitude of the coefficients but preserves their sign. Consequently, resulting in Table~\ref{tab:moments}, this adjustment only affects the magnitude of the coefficients for the projected moments, without altering their sign. Thus, our prior analysis remains valid.

\subsection{Discussion}

In Sec.~\ref{sec:Production}, we present an analysis of the polarization correlations of two baryons with various spin combinations produced in the annihilation process. Specifically, we provide detailed formulas focusing on spin combinations of (1/2, 1/2), (1/2, 3/2), and (1/2, 5/2).  Following this, we introduce the method used to determine the spin and parity of excited baryons in Sec.~\ref{subsec:Xi_decay}. Several important aspects are highlighted for clarity.

For baryons with spins lower than 5/2, which constitute the majority of the predicted baryon spectrum, the absence of moments unique to spin 5/2 helps exclude the possibility of higher spins. In the experiment, this approach has been used in analysis of the spin of $\Xi(1530)$~\cite{BaBar:2008myc}. Combined with the measured mass and width, this method helps to identify new baryons. For example, a penta-quark $\Sigma^*$ state with $J^P=1/2^-$ near 1360--1380 MeV may exist, and it is around the decuplet $\Sigma^*(1385)$ state~\cite{Zou:2007mk, Wu:2009tu, Gao:2010hy}.

For baryons with spins of 5/2 or higher, the method effectively rules out their presence in lower spin states of 1/2 or 3/2, but does not exclude the possibility of them having higher spins beyond 5/2. This limitation arises because the potential moments of angular distribution for higher spin baryons can overlap those for lower spin particles, as shown in Table~\ref{tab:moments}. When considering the effect of parity, we can find that identifying a baryon with spin $5/2^{+}$ does not preclude the possibility of it being a particle with spins as $7/2^{-}$, $9/2^{+}$, and so on. Likewise, a baryon identified as having a spin of $5/2^{-}$ could still be a particle with spins as $7/2^{+}$, $9/2^{-}$, and so on.
To confirm a baryon with spin $J$, it is necessary to construct polarization coefficients for baryons with potential spins $J-1$, $J$, and $J+1$,  and ensure that the fit for the baryon with spin $J$ surpasses those for the other two spin cases.

In our analysis, we consider two baryons produced in the $e^+e^-$ annihilation process via spin-1 intermediate states, and one of the baryons is constrained to spin 1/2. In cases where the other baryon has a spin of 5/2 or higher, it is necessary to maintain sufficient angular momentum between the two baryons. This could potentially lead to a reduction in the production cross section~\cite{Zou:2002ar}. Therefore, exploring these high-spin excited baryons may require examining other spin combinations, such as (3/2, 5/2), (5/2, 5/2), and so on.

For simplicity, our analysis focuses solely on detailing the longitudinal polarization components for baryons with spin 5/2. To ascertain the spin and parity of the $\bar{\Xi}^{*+}$($\Xi^{*-}$), we average over the azimuthal angle $\phi_{\bar{\Xi}}$ in the decay $\bar{\Xi}^{*+}\rightarrow\bar{\Xi}^{+}\pi^{0}$. Once the spin and parity are determined, the entire polarization correlations should be used to analyze the transition form factors. We have already provided the complete expressions for spin combinations of (1/2, 1/2) and (1/2, 3/2) in Eqs.~\eqref{eq:Smunu_1h_1h} and \eqref{eq:Smunu_1h_3h}. Additionally, in Sec.~\ref{sec:sdm}, we provide a detailed methodology for constructing polarization projection matrices for baryons with spins 5/2 or higher, making the derivation of the associated polarization correlation matrices a straightforward process.

From the discussion above, it is evident that the study of the baryons with any spin can be conducted.  For particle identification, focusing solely on the longitudinal polarization components of the excited baryons is effective. However, for measuring transition form factors or other parameters, the complete polarization correlation matrix should be used.

\section{Transition Form factors and helicity amplitudes\label{sec:Transition_Form_factors}}

The polarization correlations of two baryons produced in the $e^+e^-$ annihilation process can be described using either helicity amplitudes $A_{i,j}$ or transition form factors. These approaches are essentially equivalent. Following the definition of the transition form factors in Refs.~\cite{Kramer:1973nv, Devenish:1975jd, Korner:1976hv, Perotti:2018wxm}, we clarify the direct relationship between the helicity amplitudes and transition form factors, bridging these two methods. The spinors used in this section, $u$ and $v$ for spin 1/2, $u^\alpha$ and $v^\alpha$ for spin 3/2, and $u^{\alpha\beta}$ and $v^{\alpha\beta}$ for spin 5/2, are detailed in Appendix~\ref{sec:spinnors}.

\subsection{Baryons with spin combination of (1/2, 1/2)\label{sec:transition_form_factor_1h_1h}}

When the baryon $B_{1}$ has $J^P=(\frac{1}{2})^{P_{B_{1}}}$ with mass $M_{1}$, and the antibaryon $\bar{B}_{2}$ has $J^P=(\frac{1}{2})^{P_{\bar{B}_{2}}}$ with mass $M_{2}$, the transition form factors are introduced by~\cite{Korner:1976hv}
\begin{align}
&\left\langle B_{1}\left(p_{1},s_{1}\right)\bar{B}_{2}\left(p_{2},s_{2}\right)\left|J_{\mu}\left(0\right)\right|0\right\rangle =\bar{u}\left(p_{1},s_{1}\right)\Gamma_{\mu}v\left(p_{2},s_{2}\right),
\end{align}
where $p_1, p_2$ and $s_1, s_2$ denote momenta and spins of $B_1$ and $\bar B_2$ respectively, $J_\mu(0)$ is the electromagnetic current. The transition form factors are contained in the vertex $\Gamma_\mu$. For $P_{B_1 B_2}=1$, the vertex can be decomposed as~\cite{Korner:1976hv}\footnote{In this paper, we focus on discussing the production processes of two baryons with different masses and $F_i$ is the transition form factor for these processes. For the baryon pair production processes, there are different conventions for the form factors. These form factors have been widely studied and can be found e.g., in \cite{Korner:1976hv, Perotti:2018wxm}.}
\begin{align}
\Gamma_{\mu} & =
F_{1}\left(q^{2}\gamma_{\mu}-\slashed{q}q_{\mu}\right)+F_{2}\left(P\cdot q\gamma_{\mu}-P_{\mu}\slashed{q}\right).
\end{align}
For $P_{B_1 B_2}=-1$, we have
\begin{align}
\Gamma_{\mu} & =G_{1}\left(q^{2}\gamma_{\mu}-\slashed{q}q_{\mu}\right)\gamma_{5}+G_{2}\left(P\cdot q\gamma_{\mu}-P_{\mu}\slashed{q}\right)\gamma_{5}.
\end{align}
Here, $P=\frac{1}{2}\left(p_{1}-p_{2}\right)$, $q=p_{1}+p_{2}$. We classify the transition form factors associated with $\gamma_{5}$ as $G_{i}$, and those unrelated to $\gamma_{5}$ as $F_{i}$. When the sign of $P_{B_{1}B_{2}}$ shifts, a swap of $F_{i}\leftrightarrow G_{i}$ is required, along with the multiplication of $\Gamma_{\mu}$ by $-\gamma_{5}$ from the left.

For the conjugate process, the transition form factors are defined similarly by,
\begin{align}
\left\langle \bar{B}_{1}\left(p_{1},s_{1}\right)B_{2}\left(p_{2},s_{2}\right)\left|J_{\mu}\left(0\right)\right|0\right\rangle  & =\bar{u}\left(p_{2},s_{2}\right)\Gamma_{\mu}v\left(p_{1},s_{1}\right).
\end{align}
For $P_{B_1 B_2}=1$, the vertex can be decomposed as,
\begin{align}
\Gamma_{\mu} & =
F_{1}\left(q^{2}\gamma_{\mu}-\slashed{q}q_{\mu}\right)+F_{2}\left(P\cdot q\gamma_{\mu}-P_{\mu}\slashed{q}\right).
\end{align}
For $P_{B_1 B_2}=-1$, we have
\begin{align}
\Gamma_{\mu} & =-G_{1}\left(q^{2}\gamma_{\mu}-\slashed{q}q_{\mu}\right)\gamma_{5}-G_{2}\left(P\cdot q\gamma_{\mu}-P_{\mu}\slashed{q}\right)\gamma_{5}.
\end{align}
When the sign of $P_{B_{1}B_{2}}$ changes, it requires a swap of $F_{i}\leftrightarrow G_{i}$, along with the multiplication of $\Gamma_{\mu}$ by $-\gamma_{5}$ from the right.

To transform the $B_1\bar{B}_2$ production process into its conjugate $\bar{B}_1B_2$ production process, the following substitutions for the transition form factors are necessary,
\begin{align}
  & F_1\rightarrow F_1, \quad \qquad F_2\rightarrow F_2 , \\
  & G_1\rightarrow -G_1, \quad \quad G_2\rightarrow -G_2.
\end{align}
The analysis of the conjugate process in later subsections mirrors the approach taken here. Thus, detailed expressions are not repeated. Instead, we detail how to perform substitutions for the transition form factors.

Since polarization correlations can be expressed using either helicity amplitudes or form factors, we link these two forms for both $B_1\bar{B}_2$ and $\bar{B}_1B_2$ production processes. For $P_{B_1 B_2} = 1$, we establish the following relationships,
\begin{align}
h_{1}=&\frac{1}{2}\sqrt{q^{2}}Q_{1}^{-}\left[2\left(M_{1}+M_{2}\right)F_{1}+\left(M_{1}-M_{2}\right)F_{2}\right],\nonumber\\
h_{2}=&\frac{\sqrt{2}}{2}Q_{1}^{-}\left[2q^{2}F_{1}+\left(M_{1}^{2}-M_{2}^{2}\right)F_{2}\right].
\end{align}
For $P_{B_1 B_2}=-1$, the relations are
\begin{align}
h_{1}=&\frac{1}{2}\sqrt{q^{2}}Q_{1}^{+}\left[2\left(M_{2}-M_{1}\right)G_{1}-\left(M_{1}+M_{2}\right)G_{2}\right],\nonumber\\
h_{2}=&\frac{\sqrt{2}}{2}Q_{1}^{+}\left[2q^{2}G_{1}+\left(M_{1}^{2}-M_{2}^{2}\right)G_{2}\right].
\end{align}
Here $Q_{1}^{\pm}=\sqrt{q^{2}-\left(M_{1}\pm M_{2}\right)^{2}}$. In relation to these equations, if the sign of $P_{B_{1}B_{2}}$ changes, a simple substitution is required: $M_{1}\rightarrow-M_{1}$.

\subsection{Baryons with spin combination of (1/2, 3/2)}
 
For the case where $B_{1}$ has $J^P=(\frac{1}{2})^{P_{B_{1}}}$ with mass $M_{1}$, and $\bar{B}_{2}$ has $J^P=(\frac{3}{2})^{P_{\bar{B}_{2}}}$ with mass $M_{2}$, the transition form factors are introduced by~\cite{Korner:1976hv}
\begin{align}
\left\langle B_{1}\left(p_{1},s_{1}\right)\bar{B}_{2}\left(p_{2},s_{2}\right)\left|J_{\mu}\left(0\right)\right|0\right\rangle  & =\bar{u}\left(p_{1},s_{1}\right)\Gamma_{\alpha\mu}v^{\alpha}\left(p_{2},s_{2}\right).
\end{align}
When $P_{B_{1}B_{2}}=1$, the vertex $\Gamma_{\alpha\mu}$ is decomposed as
\begin{align}
\Gamma_{\alpha\mu}=&
G_{1}\left(q_{\alpha}\gamma_{\mu}-\slashed{q}g_{\alpha\mu}\right)\gamma_{5}+G_{2}\left(q_{\alpha}p_{2\mu}-p_{2}\cdot qg_{\alpha\mu}\right)\gamma_{5}+G_{3}\left(q_{\alpha}q_{\mu}-q^{2}g_{\alpha\mu}\right)\gamma_{5}.
\end{align}
For $P_{B_{1}B_{2}}=-1$, we have
\begin{align}
\Gamma_{\alpha\mu}=&
F_{1}\left(q_{\alpha}\gamma_{\mu}-\slashed{q}g_{\alpha\mu}\right)-F_{2}\left(q_{\alpha}p_{2\mu}-p_{2}\cdot qg_{\alpha\mu}\right)-F_{3}\left(q_{\alpha}q_{\mu}-q^{2}g_{\alpha\mu}\right).
\end{align}
For the conjugate process, the substitutions for the transition form factors are as follows,
\begin{align}
  &G_1 \rightarrow G_1, \quad \, G_2 \rightarrow G_2, \quad \quad G_3 \rightarrow G_3, \\
  &F_1 \rightarrow -F_1,\quad F_2 \rightarrow F_2,\qquad F_3 \rightarrow F_3.
\end{align}
For $P_{B_{1}B_{2}}=1$, the relationships between the transition form factors and the helicity amplitudes are
\begin{align}
h_{1}=&\frac{\sqrt{q^{2}}Q_{1}^{-}}{\sqrt{6}M_{2}}\left[2M_{2}G_{1}+2M_{2}^{2}G_{2}+Q_{2} G_{3}\right],\nonumber\\
h_{2}=&\frac{Q_{1}^{-}}{2\sqrt{3}M_{2}}\left[2Q_{3}^{-}G_{1}+M_{2}Q_{2} G_{2}+2q^{2}M_{2}G_{3}\right],\nonumber\\
h_{3}=&\frac{Q_{1}^{-}}{2}\left[2\left(M_{1}+M_{2}\right)G_{1}+Q_{2} G_{2}+2q^{2}G_{3}\right].
\end{align}
For $P_{B_{1}B_{2}}=-1$, the relations are
\begin{align}
h_{1}=&\frac{\sqrt{q^{2}}Q_{1}^{+}}{\sqrt{6}M_{2}}\left[2M_{2}F_{1}+2M_{2}^{2}F_{2}+ Q_{2} F_{3}\right],\nonumber\\
h_{2}=&\frac{Q_{1}^{+}}{2\sqrt{3}M_{2}}\left[2Q_{3}^{+}F_{1}+M_{2}Q_{2} F_{2}+2q^{2}M_{2}F_{3}\right],\nonumber\\
h_{3}=&\frac{Q_{1}^{+}}{2}\left[2\left(M_{2}-M_{1}\right)F_{1}+Q_{2} F_{2}+2q^{2}F_{3}\right].
\end{align}
Here, $Q_{2}=q^{2}-M_{1}^{2}+M_{2}^{2}$ and $Q_{3}^{\pm}=q^{2}-M_{1}^{2}\pm M_{1}M_{2}$.

\subsection{Baryons with spin combination of (1/2, 5/2)}
 
For the case where $B_{1}$ has $J^P=(\frac{1}{2})^{P_{B_{1}}}$ with mass $M_{1}$, and $\bar{B}_{2}$ has $J^P=(\frac{5}{2})^{P_{\bar{B}_{2}}}$ with mass $M_{2}$, the transition form factors are introduced by~\cite{Korner:1976hv}
\begin{align}
\left\langle B_{1}\left(p_{1},s_{1}\right)\bar{B}_{2}\left(p_{2},s_{2}\right)\left|J_{\mu}\left(0\right)\right|0\right\rangle & =\bar{u}\left(p_{1},s_{1}\right)\Gamma_{\alpha\beta\mu}v^{\alpha\beta}\left(p_{2},s_{2}\right).
\end{align}
For $P_{B_{1}B_{2}}=1$, the vertex $\Gamma_{\alpha\beta\mu}$ is decomposed as
\begin{align}
\Gamma_{\alpha\beta\mu}= &
F_{1}q_{\beta}\left(q_{\alpha}\gamma_{\mu}-\slashed{q}g_{\alpha\mu}\right)-F_{2}q_{\beta}\left(q_{\alpha}p_{2\mu}-p_{2}\cdot qg_{\alpha\mu}\right)-F_{3}q_{\beta}\left(q_{\alpha}q_{\mu}-q^{2}g_{\alpha\mu}\right).
\end{align}
For $P_{B_{1}B_{2}}=-1$, we have
\begin{align}
\Gamma_{\alpha\beta\mu}=&
G_{1}q_{\beta}\left(q_{\alpha}\gamma_{\mu}-\slashed{q}g_{\alpha\mu}\right)\gamma_{5}+G_{2}q_{\beta}\left(q_{\alpha}p_{2\mu}-p_{2}\cdot qg_{\alpha\mu}\right)\gamma_{5}+G_{3}q_{\beta}\left(q_{\alpha}q_{\mu}-q^{2}g_{\alpha\mu}\right)\gamma_{5}.
\end{align}
For the conjugate process, the substitutions for the transition form factors are as follows,
\begin{align}
  &G_1 \rightarrow G_1, \quad \, G_2 \rightarrow G_2, \quad \quad G_3 \rightarrow G_3, \\
  &F_1 \rightarrow -F_1,\quad F_2 \rightarrow F_2,\qquad F_3 \rightarrow F_3.
\end{align}
For $P_{B_{1}B_{2}}=1$, the relationships between the transition form factors and helicity amplitudes are
\begin{align}
h_{1}=&\frac{\sqrt{q^{2}}Q_{1}^{-}(Q_{1}^{+})^{2}}{2\sqrt{10}M_{2}^{2}}\left[2M_{2}F_{1}+2M_{2}^{2}F_{2}+Q_{2} F_{3}\right],\nonumber\\
h_{2}=&\frac{Q_{1}^{-}(Q_{1}^{+})^{2}}{4\sqrt{5}M_{2}^{2}}\left[2Q_{3}^{+}F_{1}+M_{2}Q_{2}F_{2}+2M_{2}q^{2}F_{3}\right],\nonumber\\
h_{3}=&\frac{Q_{1}^{-}(Q_{1}^{+})^{2}}{2\sqrt{10}M_{2}}\left[2\left(M_{2}-M_{1}\right)F_{1}+Q_{2}F_{2}+2q^{2}F_{3}\right].
\end{align}
For $P_{B_{1}B_{2}}=-1$, the relations are
\begin{align}
h_{1}=&\frac{\sqrt{q^{2}}(Q_{1}^{-})^{2}Q_{1}^{+}}{2\sqrt{10}M_{2}^{2}}\left[2M_{2}G_{1}+2M_{2}^{2}G_{2}+Q_{2} G_{3}\right],\nonumber\\
h_{2}=&\frac{(Q_{1}^{-})^{2}Q_{1}^{+}}{4\sqrt{5}M_{2}^{2}}\left[2Q_{3}^{-}G_{1}+M_{2}Q_{2} G_{2}+2M_{2}q^{2}G_{3}\right],\nonumber\\
h_{3}=&\frac{(Q_{1}^{-})^{2}Q_{1}^{+}}{2\sqrt{10}M_{2}}\left[2\left(M_{1}+M_{2}\right)G_{1}+Q_{2}G_{2}+2q^{2}G_{3}\right].
\end{align}

\section{Summary\label{sec:Summary}}

This paper presents a polarization analysis of the process $e^{+}e^{-}\rightarrow B_{1}\bar{B}_{2}$, involving the baryon $B_{1}$ and the antibaryon $\bar{B}_{2}$ with a variety of potential spin combinations. This analysis aims to investigate baryon properties, including spin, parity, decay parameters, and form factors.

We present a universal methodology for decomposing spin density matrices for high-spin particles in both standard and Cartesian forms. The key step involves establishing complete sets of orthogonal spin projection matrices and spin components for each form. These two forms offer varied insights into the spin density matrix. The standard form clarifies the algebraic structure and rotation properties of spin projection matrices, while the Cartesian form offers a clearer physical interpretation of spin components. By linking spin projection matrices and spin components across these two forms, we unify them. Following this, we present the complete decomposition of the spin density matrix for spin-5/2 particles. Then we analyze the application of the spin density matrices within the helicity formalism, where the matrices are expressed with the polarization projected matrices and polarization expansion coefficients. We introduce a refined selection scheme for polarization projection matrices, linking them to spin projection matrices in the standard form through specific coefficients. This scheme ensures a direct match between the polarization expansion coefficients in the helicity formalism and the spin components in the Cartesian form. Subsequently, we present the spin density matrix for the two-particle system, where spin information is captured within the polarization correlation matrix $S_{\mu\nu}$.

Next, we outline the method to calculate the polarization correlations of two baryons, $B_{1}\bar{B}_{2}$, produced in electron-positron annihilation. Within the helicity formalism, we present the general expressions of the production density matrix of $B_{1}\bar{B}_{2}$. This matrix is formulated based on helicity amplitudes $A_{i,j}$ and corresponds to the spin density matrix of the two-baryon system. We analyze the selection rule for the possible correlation components based on their physical properties. Then, our analysis focuses on spin combinations of (1/2, 1/2), (1/2, 3/2), and (1/2, 5/2), for which we identify the non-zero helicity amplitudes and link them to transfer form factors $F_{i}$ or $G_{i}$.  We present parametrization schemes for these amplitudes, detail the polarization correlation coefficients, and establish boundaries for the normalized coefficients.
For the conjugate production process $\bar{B}_{1}B_{2}$, the polarization correlation coefficients have similar expressions, except for replacing $\theta_{B}$ with $\theta_{\bar{B}}$. These conjugate production processes offer new avenues to search for $CP$ violations.

Finally, we elaborate on the technique for the experimental measurement of polarization correlations via the decay processes. Within the helicity formalism, the decay expressions are represented by polarization transfer matrices. We outline the steps to calculate these matrices. Then we provide explicit expressions for $a_{\mu\nu}$, $b_{\mu\nu}$, and $d_{\mu\nu}$, corresponding to the transfer matrices for parent baryons with spins 1/2, 3/2, and 5/2, respectively. For baryons with established spins and parities, employing the relevant polarization correlation matrices and polarization transfer matrices allows for exploring associated parameters in specific processes. Our research further extends to the baryons with undetermined spins and parities, particularly the excited ${\Xi}^{*}$ baryons. Through the analysis of the $e^{+}e^{-}\rightarrow\Xi^{-}\bar{\Xi}^{*+}$ process, we present a method to ascertain the spin and parity of the $\bar{\Xi}^{*+}$($\Xi^{*-}$). The crucial part of this approach is identifying the potential moments of the angular distributions that are sensitive to the spin and parity of the $\bar{\Xi}^{*+}$($\Xi^{*-}$).

The electron-positron annihilation process, which is abundant in two-baryon production channels, offers excellent opportunities to investigate the spins and parities of excited baryons,  as well as to study the transfer form factors or the so-called effective transfer form factors between various baryons. This process enables a comprehensive understanding of the properties of baryons, consequently enhancing our knowledge of QCD properties at low-energy scales.

\acknowledgments{
The authors would like to thank  Zuo-tang Liang, Shuangshi Fang, Xiongfei Wang, Jianbin Jiao, Wenjing Zheng, and Minzhong Hu for useful discussions. This work is Supported by the National Natural Science Foundation of China (Approvals No. 12175117, No. 12175244, No. 12247121, No. 12305085, and No. 12321005) and the Shandong Province Natural Science Foundation (Approvals No. ZR2020MA098, No. ZR2021QA015, and No. ZFJH202303).}

\appendix

\section{THE DEFINITION AND PROBABILITY OF THE SPIN COMPONENTS FOR SPIN 5/2\label{sec:spin_5h}}

In Eq.~\eqref{eq:spin_5h_g}, we provide the general decomposition of the spin density matrix for spin 5/2. Here, we clarify the definitions and the physical interpretations of the 35 independent spin components.

The spin vector $S^{i}$ consists of three components
\begin{align}
S_{T}^{x}= & \left\langle \Sigma_{T}^{x}\right\rangle,\quad S_{T}^{y}=\left\langle \Sigma_{T}^{y}\right\rangle,\quad S_{L}=\left\langle \Sigma_{L}\right\rangle .
\end{align}
According to Eqs.~\eqref{eq:defination_SL}--\eqref{eq:coefficients_SLT_2}, the corresponding projection matrices are defined as
\begin{align}
\Sigma_{T}^{x}= & \Sigma^{x},\quad \Sigma_{T}^{y}=\Sigma^{y},\quad \Sigma_{L}=\Sigma^{z}.
\end{align}
Based on Eq.~\eqref{eq:Sigma_xyz}, the matrices $\Sigma^{i}$ in the $S_{z}$ representation are given by
\begin{align}
\Sigma^{x}= & \frac{1}{2}
\left(\begin{array}{cccccc}
0 & \sqrt{5} & 0 & 0 & 0 & 0\\
\sqrt{5} & 0 & 2\sqrt{2} & 0 & 0 & 0\\
0 & 2\sqrt{2} & 0 & 3 & 0 & 0\\
0 & 0 & 3 & 0 & 2\sqrt{2} & 0\\
0 & 0 & 0 & 2\sqrt{2} & 0 & \sqrt{5}\\
0 & 0 & 0 & 0 & \sqrt{5} & 0
\end{array}\right),\nonumber \\
\Sigma^{y}=& \frac{i}{2}
\left(\begin{array}{cccccc}
0 & -\sqrt{5} & 0 & 0 & 0 & 0\\
\sqrt{5} & 0 & -2\sqrt{2} & 0 & 0 & 0\\
0 & 2\sqrt{2} & 0 & -3 & 0 & 0\\
0 & 0 & 3 & 0 & -2\sqrt{2} & 0\\
0 & 0 & 0 & 2\sqrt{2} & 0 & -\sqrt{5}\\
0 & 0 & 0 & 0 & \sqrt{5} & 0
\end{array}\right),\nonumber \\
\Sigma^{z}= &  \frac{1}{2}
\left(\begin{array}{cccccc}
5 & 0 & 0 & 0 & 0 & 0\\
0 & 3 & 0 & 0 & 0 & 0\\
0 & 0 & 1 & 0 & 0 & 0\\
0 & 0 & 0 & -1 & 0 & 0\\
0 & 0 & 0 & 0 & -3 & 0\\
0 & 0 & 0 & 0 & 0 & -5
\end{array}\right).\label{eq:spin_5h_m1}
\end{align}
The relationship of these matrices to $Q_{M}^{5/2,1}$ is established in Eqs.~\eqref{eq:relation_QLM_SigmaLT} and~\eqref{eq:coefficients_QLM_SigmaLT}, where coefficients $t_{M}^{5/2,1}$ have values
\begin{align}
\left\{ t_{1}^{5/2,1},\, t_{0}^{5/2,1},\, t_{-1}^{5/2,1}\right\} =
\left\{ \frac{5}{6}\sqrt{21},\, \frac{5}{6}\sqrt{21},\, \frac{5}{6}\sqrt{21}\right\} .
\end{align}
The spin vector $S^{i}$ is determined by Eq.~\eqref{eq:Tijk}, and represented as
\begin{align}
S^{i}= & \left(S_{T}^{x},\,S_{T}^{y},\,S_{L}\right).
\end{align}
The rank-two spin tensor $T^{ij}$ comprises five spin components,
\begin{align}
&S_{LL}=  \left\langle \Sigma_{LL}\right\rangle ,\quad S_{LT}^{x}=\left\langle \Sigma_{LT}^{x}\right\rangle ,\quad S_{LT}^{y}=\left\langle \Sigma_{LT}^{y}\right\rangle ,\nonumber\\
&S_{TT}^{xx}=\left\langle \Sigma_{TT}^{xx}\right\rangle ,\quad S_{TT}^{xy}=\left\langle \Sigma_{TT}^{xy}\right\rangle.
\end{align}
The projection matrices corresponding to these components, as defined in Eqs.~\eqref{eq:defination_SL}--\eqref{eq:coefficients_SLT_2}, are:
\begin{align}
\Sigma_{LL}= & \Sigma^{zz},\quad \Sigma_{LT}^{x}=\Sigma^{xz},\quad \Sigma_{LT}^{y}=\Sigma^{yz},\nonumber\\
\Sigma_{TT}^{xx}= & \Sigma^{xx}-\Sigma^{yy},\quad \Sigma_{TT}^{xy}=\Sigma^{xy}.
\end{align}
The association of these matrices with $Q_{M}^{5/2,3}$ is specified in Eqs.~\eqref{eq:relation_QLM_SigmaLT} and~\eqref{eq:coefficients_QLM_SigmaLT}, where the coefficients $t_{M}^{5/2,2}$ are
\begin{align}
\left\{\begin{array}{c}
t_{2}^{5/2,2},\, t_{1}^{5/2,2},\, t_{0}^{5/2,2},\\
t_{-1}^{5/2,2},\, t_{-2}^{5/2,2}\end{array}\right\}
= \left\{ \begin{array}{c}
\frac{2}{3}\sqrt{210},\, \frac{1}{3}\sqrt{210},\, \frac{2}{3}\sqrt{70},\\
\frac{1}{3}\sqrt{210},\, \frac{1}{3}\sqrt{210}\end{array}\right\} .
\end{align}

The matrix representation of the rank-two spin tensor $T^{ij}$ is derived using Eq.~\eqref{eq:Tijk} and is expressed as
\begin{align}
T^{ij}=  \frac{1}{2}
\left(\begin{array}{ccc}
S_{TT}^{xx}-S_{LL} & 2S_{TT}^{xy} & 2S_{LT}^{x}\\
2S_{TT}^{xy} & -S_{TT}^{xx}-S_{LL} & 2S_{LT}^{y}\\
2S_{LT}^{x} & 2S_{LT}^{y} & 2S_{LL}
\end{array}\right).
\end{align}
The rank-three spin tensor $T^{ijk}$ includes seven spin components,
\begin{align}
S_{LLL}= & \left\langle \Sigma_{LLL}\right\rangle ,\quad S_{LLT}^{x}=\left\langle \Sigma_{LLT}^{x}\right\rangle ,\quad S_{LLT}^{y}=\left\langle \Sigma_{LLT}^{y}\right\rangle, \nonumber\\
S_{LTT}^{xx}= & \left\langle \Sigma_{LTT}^{xx}\right\rangle ,\quad S_{LTT}^{xy}=\left\langle \Sigma_{LTT}^{xy}\right\rangle ,\nonumber\\
S_{TTT}^{xxx}=&\left\langle \Sigma_{TTT}^{xxx}\right\rangle ,\quad S_{TTT}^{xxy}=\left\langle \Sigma_{TTT}^{xxy}\right\rangle.
\end{align}
The corresponding projection matrices are determined by Eqs.~\eqref{eq:defination_SL}--\eqref{eq:coefficients_SLT_2} and given by
\begin{align}
\Sigma_{LLL}= & \Sigma^{zzz},\quad \Sigma_{LLT}^{x}=\Sigma^{xzz},\quad \Sigma_{LLT}^{y}=\Sigma^{yzz},\nonumber\\
\Sigma_{LTT}^{xx}= & \Sigma^{xxz}-\Sigma^{yyz},\quad \Sigma_{LTT}^{xy}=\Sigma^{xyz},\nonumber\\
\Sigma_{TTT}^{xxx}= & \Sigma^{xxx}-3\Sigma^{yyx},\quad \Sigma_{TTT}^{xxy}=3\Sigma^{xxy}-\Sigma^{yyy}.
\end{align}
The relation of these matrices to $Q_{M}^{5/2,3}$ is defined in Eqs.~\eqref{eq:relation_QLM_SigmaLT} and~\eqref{eq:coefficients_QLM_SigmaLT}, where the coefficients $t_{M}^{5/2,3}$ are given by,
\begin{align}
\left\{\begin{array}{c}
t_{3}^{5/2,3},\, t_{2}^{5/2,3},\, t_{1}^{5/2,3},\\
t_{0}^{5/2,3},\, t_{-1}^{5/2,3},\\
t_{-2}^{5/2,3},\, t_{-3}^{5/2,3}\end{array}\right\} =
\left\{ \begin{array}{c}
6\sqrt{15},\, 3\sqrt{10},\, 6,\\
3\sqrt{6},\, 6,\\
\frac{3}{2}\sqrt{10},\, 6\sqrt{15}\end{array}\right\} .
\end{align}
The rank-three spin tensor $T^{ijk}$ is defined using Eq.~\eqref{eq:Tijk}, and the calculation process is similar to that for $S^{i}$ and $T^{ij}$. Although not difficult to calculate, the expressions for $T^{ijk}$ and higher orders like $T^{ijkl}$ and $T^{ijklm}$ are lengthy and thus are not presented with the details.

The rank-four spin tensor $T^{ijkl}$ comprises nine polarization components, represented as follows
\begin{align}
S_{LLLL}= & \left\langle \Sigma_{LLLL}\right\rangle,
\quad S_{LLLT}^{x}=\left\langle \Sigma_{LLLT}^{x}\right\rangle ,\nonumber\\
S_{LLLT}^{y}=&\left\langle \Sigma_{LLLT}^{y}\right\rangle,\quad
S_{LLTT}^{xx}= \left\langle \Sigma_{LLTT}^{xx}\right\rangle ,\nonumber\\
S_{LLTT}^{xy}=&\left\langle \Sigma_{LLTT}^{xy}\right\rangle ,
\quad S_{LTTT}^{xxx}=\left\langle \Sigma_{LTTT}^{xxx}\right\rangle,\nonumber\\
S_{LTTT}^{xxy}= & \left\langle \Sigma_{LTTT}^{xxy}\right\rangle ,
\quad S_{TTTT}^{xxxx}=\left\langle \Sigma_{TTTT}^{xxxx}\right\rangle ,\nonumber\\
S_{TTTT}^{xxxy}=&\left\langle \Sigma_{TTTT}^{xxxy}\right\rangle .
\end{align}
The corresponding projection matrices for these components are defined in Eqs.~\eqref{eq:defination_SL}--\eqref{eq:coefficients_SLT_2}, and their expressions are
\begin{align}
\Sigma_{LLLL}= & \Sigma^{zzzz},
\quad \Sigma_{LLLT}^{x}=\Sigma^{xzzz},\nonumber\\
\Sigma_{LLLT}^{y}=&\Sigma^{yzzz},\quad
\Sigma_{LLTT}^{xx}= \Sigma^{xxzz}-\Sigma^{yyzz},\nonumber\\
\Sigma_{LLTT}^{xy}=&\Sigma^{xyzz},\quad
\Sigma_{LTTT}^{xxx}= \Sigma^{xxxz}-3\Sigma^{xyyz},\nonumber\\
\Sigma_{LTTT}^{xxy}=& 3\Sigma^{xxyz}-\Sigma^{yyyz},\nonumber\\
\Sigma_{TTTT}^{xxxx}= & \Sigma^{xxxx}-6\Sigma^{xxyy}+\Sigma^{yyyy},\nonumber\\
\Sigma_{TTTT}^{xxxy}=&\Sigma^{xxxy}-\Sigma^{xyyy}.
\end{align}
The relationship between these matrices and $Q_{M}^{5/2,4}$ is detailed in Eqs.~\eqref{eq:relation_QLM_SigmaLT} and~\eqref{eq:coefficients_QLM_SigmaLT}, where the coefficients $t_{M}^{5/2,4}$ are specified as:
\begin{align}
\left\{ \begin{array}{c}
t_{4}^{5/2,4},\, t_{3}^{5/2,4},\, t_{2}^{5/2,4},\\
t_{1}^{5/2,4},\, t_{0}^{5/2,4},\, t_{-1}^{5/2,4},\\
t_{-2}^{5/2,4},\, t_{-3}^{5/2,4},\, t_{-4}^{5/2,4}
\end{array}\right\} =
\left\{\begin{array}{c}
20\sqrt{6},\, 10\sqrt{3},\, \frac{10}{7}\sqrt{42},\\
\frac{10}{7}\sqrt{21},\, \frac{4}{7}\sqrt{210},\, \frac{10}{7}\sqrt{21},\\
\frac{5}{7}\sqrt{42},\, 10\sqrt{3},\, 5\sqrt{6}
\end{array}\right\} .
\end{align}

Similarly, the rank-five spin tensor $T^{ijklm}$ consists of eleven spin components, expressed as
\begin{align}
S_{LLLLL}= & \left\langle \Sigma_{LLLLL}\right\rangle ,
\quad S_{LLLLT}^{x}=\left\langle \Sigma_{LLLLT}^{x}\right\rangle ,\nonumber\\
S_{LLLLT}^{y}=&\left\langle \Sigma_{LLLLT}^{y}\right\rangle ,\quad
S_{LLLTT}^{xx}= \left\langle \Sigma_{LLLTT}^{xx}\right\rangle ,\nonumber\\
S_{LLLTT}^{xy}=&\left\langle \Sigma_{LLLTT}^{xy}\right\rangle ,\quad
S_{LLTTT}^{xxx}=\left\langle \Sigma_{LLTTT}^{xxx}\right\rangle ,\nonumber\\
S_{LLTTT}^{xxy}= & \left\langle \Sigma_{LLTTT}^{xxy}\right\rangle ,\quad S_{LTTTT}^{xxxx}=\left\langle \Sigma_{LTTTT}^{xxxx}\right\rangle ,\nonumber\\
S_{LTTTT}^{xxxy}=&\left\langle \Sigma_{LTTTT}^{xxxy}\right\rangle,\quad
S_{TTTTT}^{xxxxx}= \left\langle \Sigma_{TTTTT}^{xxxxx}\right\rangle,\nonumber\\
S_{TTTTT}^{xxxxy}=&\left\langle \Sigma_{TTTTT}^{xxxxy}\right\rangle .
\end{align}
The relevant projection matrices for these components are outlined in Eqs.~\eqref{eq:defination_SL}--\eqref{eq:coefficients_SLT_2}, with their expressions being
\begin{align}
\Sigma_{LLLLL}= & \Sigma^{zzzzz},\quad
\Sigma_{LLLLT}^{x}=\Sigma^{xzzzz},\nonumber\\
\Sigma_{LLLLT}^{y}=&\Sigma^{yzzzz},\quad
\Sigma_{LLLTT}^{xx}= \Sigma^{xxzzz}-\Sigma^{yyzzz},\nonumber\\
\Sigma_{LLLTT}^{xy}=&\Sigma^{xyzzz},\quad
\Sigma_{LLTTT}^{xxx}= \Sigma^{xxxzz}-3\Sigma^{xyyzz},\nonumber\\
\Sigma_{LLTTT}^{xxy}=&3\Sigma^{xxyzz}-\Sigma^{yyyzz},\nonumber\\
\Sigma_{LTTTT}^{xxxx}= & \Sigma^{xxxxz}-6\Sigma^{xxyyz}+\Sigma^{yyyyz},\nonumber\\
\Sigma_{LTTTT}^{xxxy}=&\Sigma^{xxxyz}-\Sigma^{xyyyz},\nonumber\\
\Sigma_{TTTTT}^{xxxxx}= & \Sigma^{xxxxx}-10\Sigma^{xxxyy}+5\Sigma^{xyyyy},\nonumber\\
\Sigma_{TTTTT}^{xxxxy}=& 5\Sigma^{xxxxy}-10\Sigma^{xxyyy}+\Sigma^{yyyyy}.
\end{align}
The association of these matrices with $Q_{M}^{5/2,5}$ is determined in Eqs.~\eqref{eq:relation_QLM_SigmaLT} and~\eqref{eq:coefficients_QLM_SigmaLT}, and the coefficients $t_{M}^{5/2,5}$ are
\begin{align}
\left\{ \begin{array}{c}
t_{5}^{5/2,5},\, t_{4}^{5/2,5},\, t_{3}^{5/2,5},\\
t_{2}^{5/2,5},\, t_{1}^{5/2,5},\, t_{0}^{5/2,5},\\
t_{-1}^{5/2,5},\, t_{-2}^{5/2,5},\, t_{-3}^{5/2,5},\\
t_{-4}^{5/2,5},\quad t_{-5}^{5/2,5}
\end{array}\right\} =
\left\{ \begin{array}{c}
20\sqrt{15},\, 10\sqrt{6},\, \frac{20}{3}\sqrt{3},\\
5\sqrt{2},\, \frac{10}{7}\sqrt{14},\, \frac{10}{21}\sqrt{210},\\
\frac{10}{7}\sqrt{14},\, \frac{5}{2}\sqrt{2},\, \frac{20}{3}\sqrt{3},\\
\frac{5}{2}\sqrt{6},\quad 20\sqrt{15}
\end{array}\right\}.
\end{align}
The total degree of polarization is determined by Eq.~\eqref{eq:degree}, given by
\begin{align}
d=&\frac{1}{\sqrt{2s}}\sqrt{(2s+1)\text{Tr}[\rho_s^2]-1}\nonumber\\
=&\frac{\sqrt{6}}{\sqrt{5}}\left(\frac{2}{35}S^{i}S^{i}+\frac{1}{56}T^{ij}T^{ij}+\frac{1}{162}T^{ijk}T^{ijk}+\frac{1}{360}T^{ijkl}T^{ijkl}+\frac{1}{450}T^{ijklm}T^{ijklm}\right)^{1/2}.
\end{align}

Following the approach presented in Ref.~\cite{Bacchetta:2000jk}, we offer the probabilistic interpretation for these spin components. We introduce the spin projection operator along the $(\theta,\phi)$ direction as
\begin{align}
\Sigma^{i}\hat{n}_{i}= & \Sigma^{x}\sin\theta\cos\phi+\Sigma^{y}\sin\theta\sin\phi+\Sigma^{z}\cos\theta.
\end{align}
The probability of finding this specific eigenstate from the spin density matrix can be expressed as,
\begin{align}
P\left(m_{\left(\theta,\phi\right)}\right)= & \text{Tr}\left[\rho\left|m_{\left(\theta,\phi\right)}\right\rangle \left\langle m_{\left(\theta,\phi\right)}\right|\right].
\end{align}
We focus on providing the physical interpretations for the following longitudinal polarization components,
\begin{align}
&S_{L}  =\left\langle \Sigma^{z}\right\rangle ,\quad S_{LL}=\left\langle \Sigma^{zz}\right\rangle ,\quad S_{LLL}=\left\langle \Sigma^{zzz}\right\rangle, \nonumber \\
&S_{LLLL}  =\left\langle \Sigma^{zzzz}\right\rangle ,\quad S_{LLLLL}=\left\langle \Sigma^{zzzzz}\right\rangle. \label{eq:probability_definition}
\end{align}
Similar methods can be employed to determine other spin components. While the calculations for these components are direct, their expressions can be lengthy. To maintain focus and conciseness, we have chosen not to present the specifics of these additional components. Using the shorthand notation, $\left|m_{z}\right\rangle =\left|m_{(0,0)}\right\rangle $, the probabilistic interpretation of longitudinal spin components can be represented as follows,
\begin{align}
S_{L}= & \frac{5}{2}\left[P_{z}\left(\frac{5}{2}\right)-P_{z}\left(-\frac{5}{2}\right)\right]+\frac{3}{2}\left[P_{z}\left(\frac{3}{2}\right)-P_{z}\left(-\frac{3}{2}\right)\right]+\frac{1}{2}\left[P_{z}\left(\frac{1}{2}\right)-P_{z}\left(-\frac{1}{2}\right)\right],\\
S_{LL}= & \frac{10}{3}\left[P_{z}\left(\frac{5}{2}\right)+P_{z}\left(-\frac{5}{2}\right)\right]-\frac{2}{3}\left[P_{z}\left(\frac{3}{2}\right)+P_{z}\left(-\frac{3}{2}\right)\right]-\frac{8}{3}\left[P_{z}\left(\frac{1}{2}\right)+P_{z}\left(-\frac{1}{2}\right)\right],\\
S_{LLL}= & 3\left[P_{z}\left(\frac{5}{2}\right)-P_{z}\left(-\frac{5}{2}\right)\right]-\frac{21}{5}\left[P_{z}\left(\frac{3}{2}\right)-P_{z}\left(-\frac{3}{2}\right)\right]-\frac{12}{5}\left[P_{z}\left(\frac{1}{2}\right)-P_{z}\left(-\frac{1}{2}\right)\right],\\
S_{LLLL}= & \frac{12}{7}\left[P_{z}\left(\frac{5}{2}\right)+P_{z}\left(-\frac{5}{2}\right)\right]-\frac{36}{7}\left[P_{z}\left(\frac{3}{2}\right)+P_{z}\left(-\frac{3}{2}\right)\right]+\frac{24}{7}\left[P_{z}\left(\frac{1}{2}\right)+P_{z}\left(-\frac{1}{2}\right)\right],\\
S_{LLLLL}= & \frac{10}{21}\left[P_{z}\left(\frac{5}{2}\right)-P_{z}\left(-\frac{5}{2}\right)\right]-\frac{50}{21}\left[P_{z}\left(\frac{3}{2}\right)-P_{z}\left(-\frac{3}{2}\right)\right]+\frac{100}{21}\left[P_{z}\left(\frac{1}{2}\right)-P_{z}\left(-\frac{1}{2}\right)\right].
\end{align}
The domains for these polarization components can be obtained from these probability interpretations,
\begin{align}
& S_{L} \in\left[-\frac{5}{2},\frac{5}{2}\right],S_{LL}\in\left[-\frac{8}{3},\frac{10}{3}\right],S_{LLL}\in\left[-\frac{21}{5},\frac{21}{5}\right],\nonumber \\
&S_{LLLL} \in\left[-\frac{36}{7},\frac{24}{7}\right],S_{LLLLL}\in\left[-\frac{100}{21},\frac{100}{21}\right].
\end{align}

\section{SPIN 3/2 AND SPIN 5/2 POLARIZATION PROJECT MATRICES\label{sec:basis matrices}}

In this section, we present explicit expressions of the polarization projection matrices for spin-3/2 and spin-5/2 particles in the helicity formalism.

For spin 3/2 particles, we follow the matrix basis set selection in Refs.~\cite{Zhang:2023wmd,Zhang:2023box},
\begin{align}
&\Sigma_{0}=  \frac{1}{4}\left(\begin{array}{cccc}
1 & 0 & 0 & 0\\
0 & 1 & 0 & 0\\
0 & 0 & 1 & 0\\
0 & 0 & 0 & 1
\end{array}\right),\quad \Sigma_{1}=\frac{1}{10}\left(\begin{array}{cccc}
3 & 0 & 0 & 0\\
0 & 1 & 0 & 0\\
0 & 0 & -1 & 0\\
0 & 0 & 0 & -3
\end{array}\right), \Sigma_{2}=\frac{1}{10}\left(\begin{array}{cccc}
0 & \sqrt{3} & 0 & 0\\
\sqrt{3} & 0 & 2 & 0\\
0 & 2 & 0 & \sqrt{3}\\
0 & 0 & \sqrt{3} & 0
\end{array}\right),\nonumber\\
&\Sigma_{3}=  \frac{i}{10}\left(\begin{array}{cccc}
0 & -\sqrt{3} & 0 & 0\\
\sqrt{3} & 0 & -2 & 0\\
0 & 2 & 0 & -\sqrt{3}\\
0 & 0 & \sqrt{3} & 0
\end{array}\right),\quad \Sigma_{4}=\frac{1}{4}\left(\begin{array}{cccc}
1 & 0 & 0 & 0\\
0 & -1 & 0 & 0\\
0 & 0 & -1 & 0\\
0 & 0 & 0 & 1
\end{array}\right),\quad\Sigma_{5}=\frac{\sqrt{3}}{12}\left(\begin{array}{cccc}
0 & 1 & 0 & 0\\
1 & 0 & 0 & 0\\
0 & 0 & 0 & -1\\
0 & 0 & -1 & 0
\end{array}\right),\nonumber\\
&\Sigma_{6}=  \frac{i\sqrt{3}}{12}\left(\begin{array}{cccc}
0 & -1 & 0 & 0\\
1 & 0 & 0 & 0\\
0 & 0 & 0 & 1\\
0 & 0 & -1 & 0
\end{array}\right),\quad\Sigma_{7}=\frac{\sqrt{3}}{12}\left(\begin{array}{cccc}
0 & 0 & 1 & 0\\
0 & 0 & 0 & 1\\
1 & 0 & 0 & 0\\
0 & 1 & 0 & 0
\end{array}\right),\quad\Sigma_{8}=\frac{i\sqrt{3}}{12}\left(\begin{array}{cccc}
0 & 0 & -1 & 0\\
0 & 0 & 0 & -1\\
1 & 0 & 0 & 0\\
0 & 1 & 0 & 0
\end{array}\right),\nonumber\\
&\Sigma_{9}=  \frac{1}{6}\left(\begin{array}{cccc}
1 & 0 & 0 & 0\\
0 & -3 & 0 & 0\\
0 & 0 & 3 & 0\\
0 & 0 & 0 & -1
\end{array}\right),\quad\Sigma_{10}=\frac{\sqrt{3}}{6}\left(\begin{array}{cccc}
0 & 1 & 0 & 0\\
1 & 0 & -\sqrt{3} & 0\\
0 & -\sqrt{3} & 0 & 1\\
0 & 0 & 1 & 0
\end{array}\right),\quad\Sigma_{11}=\frac{i\sqrt{3}}{6}\left(\begin{array}{cccc}
0 & -1 & 0 & 0\\
1 & 0 & \sqrt{3} & 0\\
0 & -\sqrt{3} & 0 & -1\\
0 & 0 & 1 & 0
\end{array}\right),\nonumber\\
&\Sigma_{12}=  \frac{\sqrt{3}}{12}\left(\begin{array}{cccc}
0 & 0 & 1 & 0\\
0 & 0 & 0 & -1\\
1 & 0 & 0 & 0\\
0 & -1 & 0 & 0
\end{array}\right),\quad\Sigma_{13}=\frac{i\sqrt{3}}{12}\left(\begin{array}{cccc}
0 & 0 & -1 & 0\\
0 & 0 & 0 & 1\\
1 & 0 & 0 & 0\\
0 & -1 & 0 & 0
\end{array}\right),\quad\Sigma_{14}=\frac{1}{6}\left(\begin{array}{cccc}
0 & 0 & 0 & 1\\
0 & 0 & 0 & 0\\
0 & 0 & 0 & 0\\
1 & 0 & 0 & 0
\end{array}\right),\nonumber\\
&\Sigma_{15}=  \frac{i}{6}\left(\begin{array}{cccc}
0 & 0 & 0 & -1\\
0 & 0 & 0 & 0\\
0 & 0 & 0 & 0\\
1 & 0 & 0 & 0
\end{array}\right).
\end{align}

For spin 5/2 particles, the matrix basis subset associated with longitudinal polarization components is given by,
\begin{align}
&\Sigma_{0}=\frac{1}{6} \left(\begin{array}{cccccc}
1 & 0 & 0 & 0 & 0 & 0\\
0 & 1 & 0 & 0 & 0 & 0\\
0 & 0 & 1 & 0 & 0 & 0\\
0 & 0 & 0 & 1 & 0 & 0\\
0 & 0 & 0 & 0 & 1 & 0\\
0 & 0 & 0 & 0 & 0 & 1
\end{array}\right),\quad\Sigma_{1}=\frac{1}{35}\left(\begin{array}{cccccc}
5 & 0 & 0 & 0 & 0 & 0\\
0 & 3 & 0 & 0 & 0 & 0\\
0 & 0 & 1 & 0 & 0 & 0\\
0 & 0 & 0 & -1 & 0 & 0\\
0 & 0 & 0 & 0 & -3 & 0\\
0 & 0 & 0 & 0 & 0 & -5
\end{array}\right),\nonumber \\
&\Sigma_{2}=\frac{1}{56}\left(\begin{array}{cccccc}
5 & 0 & 0 & 0 & 0 & 0\\
0 & -1 & 0 & 0 & 0 & 0\\
0 & 0 & -4 & 0 & 0 & 0\\
0 & 0 & 0 & -4 & 0 & 0\\
0 & 0 & 0 & 0 & -1 & 0\\
0 & 0 & 0 & 0 & 0 & 5
\end{array}\right),\quad\Sigma_{3}= \frac{1}{108}\left(\begin{array}{cccccc}
5 & 0 & 0 & 0 & 0 & 0\\
0 & -7 & 0 & 0 & 0 & 0\\
0 & 0 & -4 & 0 & 0 & 0\\
0 & 0 & 0 & 4 & 0 & 0\\
0 & 0 & 0 & 0 & 7 & 0\\
0 & 0 & 0 & 0 & 0 & -5
\end{array}\right),\nonumber \\
&\Sigma_{4}=\frac{1}{48}\left(\begin{array}{cccccc}
1 & 0 & 0 & 0 & 0 & 0\\
0 & -3 & 0 & 0 & 0 & 0\\
0 & 0 & 2 & 0 & 0 & 0\\
0 & 0 & 0 & 2 & 0 & 0\\
0 & 0 & 0 & 0 & -3 & 0\\
0 & 0 & 0 & 0 & 0 & 1
\end{array}\right),\quad\Sigma_{5}=\frac{1}{120}\left(\begin{array}{cccccc}
1 & 0 & 0 & 0 & 0 & 0\\
0 & -5 & 0 & 0 & 0 & 0\\
0 & 0 & 10 & 0 & 0 & 0\\
0 & 0 & 0 & -10 & 0 & 0\\
0 & 0 & 0 & 0 & 5 & 0\\
0 & 0 & 0 & 0 & 0 & -1
\end{array}\right).
\end{align}

\section{POLARIZATION TRANSFER MATRIX\label{sec:tranfer_matrix}}

In Sec.~\ref{sec:Decay_chains}, we describe the decay processes of baryons using transition matrices $a_{\mu\nu}$, $b_{\mu\nu}$, and $d_{\mu\nu}$. These matrices are crucial for understanding the polarization dynamics in baryon decays, and we provide detailed expressions for each matrix.

For the decay process $1/2\rightarrow1/2+0$, the polarization transition matrix $a_{\mu\nu}$ contains 14 non-zero elements. Among these, 10 terms do not involve $\alpha_{D}$, given by~\cite{Perotti:2018wxm}
\begin{align}
a_{0,0}= & 1,\nonumber\\
a_{1,1}= & \gamma_{D}\cos\theta\cos\phi-\beta_{D}\sin\phi,\nonumber\\
a_{1,2}= & -\beta_{D}\cos\theta\cos\phi-\gamma_{D}\sin\phi,\nonumber\\
a_{1,3}= & \sin\theta\cos\phi,\nonumber\\
a_{2,1}= & \beta_{D}\cos\phi+\gamma_{D}\cos\theta\sin\phi,\nonumber\\
a_{2,2}= & \gamma_{D}\cos\phi-\beta_{D}\cos\theta\sin\phi,\nonumber\\
a_{2,3}= &\sin\theta \sin\phi,\nonumber\\
a_{3,1}= & -\gamma_{D}\sin\theta,\nonumber\\
a_{3,2}= & \beta_{D}\sin\theta,\nonumber\\
a_{3,3}= & \cos\theta.
\end{align}
The remaining 4 terms, which are dependent on $\alpha_{D}$, are expressed as
\begin{align}
\left\{ a_{0,3},\, a_{1,0},\, a_{2,0},\, a_{3,0}\right\} = & \alpha_{D}\left\{ a_{0,0},\, a_{1,3},\, a_{2,3},\, a_{3,3}\right\},
\end{align}
where the items in the left-hand list are equal to the corresponding items in the right-hand list multiplied by $\alpha_{D}$, for example, $a_{0,3}=\alpha_{D} \, a_{0,0}$.

For the decay process $3/2\rightarrow1/2+0$ the polarization transition matrix $b_{\mu\nu}$ includes 52 non-zero elements. Of these, 36 terms do not involve $\alpha_{D}$, given by~\cite{Zhang:2023box}
\begin{align}
b_{0,0}= & 1,\nonumber\\
b_{1,1}= & -\frac{4}{5}\gamma_{D}\sin\theta,\nonumber\\
b_{1,2}= & \frac{4}{5}\beta_{D}\sin\theta,\nonumber\\
b_{1,3}= & \frac{2}{5}\cos\theta,\nonumber\\
b_{2,1}= & -\frac{4}{5}\left(-\gamma_{D}\cos\theta\cos\phi+\beta_{D}\sin\phi\right),\nonumber\\
b_{2,2}= & -\frac{4}{5}\left(\beta_{D}\cos\theta\cos\phi+\gamma_{D}\sin\phi\right),\nonumber\\
b_{2,3}= & \frac{2}{5}\sin\theta\cos\phi,\nonumber\\
b_{3,1}= & \frac{4}{5}\left(\beta_{D}\cos\phi+\gamma_{D}\cos\theta\sin\phi\right),\nonumber\\
b_{3,2}= & \frac{4}{5}\left(\gamma_{D}\cos\phi-\beta_{D}\cos\theta\sin\phi\right),\nonumber\\
b_{3,3}= & \frac{2}{5}\sin\theta\sin\phi,\nonumber\\
b_{4,0}= & -\frac{1}{4}\left(1+3\cos2\theta\right),\nonumber\\
b_{5,0}= & -\sin\theta\cos\theta\cos\phi,\nonumber\\
b_{6,0}= & -\sin\theta\cos\theta\sin\phi,\nonumber\\
b_{7,0}= & -\frac{1}{2}\sin^{2}\theta\cos2\phi,\nonumber\\
b_{8,0}= & -\sin^{2}\theta\sin\phi\cos\phi,\nonumber\\
b_{9,1}= & \frac{1}{4}\gamma_{D}\left(\sin\theta+5\sin3\theta\right),\nonumber\\
b_{9,2}= & -\frac{1}{4}\beta_{D}\left(\sin\theta+5\sin3\theta\right),\nonumber\\
b_{9,3}= & -\frac{1}{4}\left(3\cos\theta+5\cos3\theta\right),\nonumber\\
b_{10,1}= & \frac{1}{8}\left[2\beta_{D}\left(3+5\cos2\theta\right)\sin\phi-\gamma_{D}\left(\cos\theta+15\cos3\theta\right)\cos\phi\right],\nonumber\\
b_{10,2}= & \frac{1}{8}\left[2\gamma_{D}\left(3+5\cos2\theta\right)\sin\phi+\beta_{D}\left(\cos\theta+15\cos3\theta\right)\cos\phi\right],\nonumber\\
b_{10,3}= & -\frac{3}{8}\left(\sin\theta+5\sin3\theta\right)\cos\phi,\nonumber\\
b_{11,1}= & -\frac{1}{8}\left[2\beta_{D}\left(3+5\cos2\theta\right)\cos\phi+\gamma_{D}\left(\cos\theta+15\cos3\theta\right)\sin\phi\right],\nonumber\\
b_{11,2}= & -\frac{1}{8}\left[2\gamma_{D}\left(3+5\cos2\theta\right)\cos\phi-\beta_{D}\left(\cos\theta+15\cos3\theta\right)\sin\phi\right],\nonumber\\
b_{11,3}= & -\frac{3}{8}\left(\sin\theta+5\sin3\theta\right)\sin\phi,\nonumber\\
b_{12,1}= & \frac{1}{4}\sin\theta\left[4\beta_{D}\cos\theta\sin2\phi-\gamma_{D}\left(1+3\cos2\theta\right)\cos2\phi\right],\nonumber\\
b_{12,2}= & \frac{1}{4}\sin\theta\left[4\gamma_{D}\cos\theta\sin2\phi+\beta_{D}\left(1+3\cos2\theta\right)\cos2\phi\right],\nonumber\\
b_{12,3}= & -\frac{3}{2}\sin^{2}\theta\cos\theta\cos2\phi,\nonumber\\
b_{13,1}= & -\frac{1}{4}\sin\theta\left[4\beta_{D}\cos\theta\cos2\phi+\gamma_{D}\left(1+3\cos2\theta\right)\sin2\phi\right],\nonumber\\
b_{13,2}= & -\frac{1}{4}\sin\theta\left[4\gamma_{D}\cos\theta\cos2\phi-\beta_{D}\left(1+3\cos2\theta\right)\sin2\phi\right],\nonumber\\
b_{13,3}= & -3\sin^{2}\theta\cos\theta\sin\phi\cos\phi,\nonumber\\
b_{14,1}= & \frac{1}{2}\sin^{2}\theta\left(\beta_{D}\sin3\phi-\gamma_{D}\cos\theta\cos3\phi\right),\nonumber\\
b_{14,2}= & \frac{1}{2}\sin^{2}\theta\left(\gamma_{D}\sin3\phi+\beta_{D}\cos\theta\cos3\phi\right),\nonumber\\
b_{14,3}= & -\frac{1}{2}\sin^{3}\theta\cos3\phi,\nonumber\\
b_{15,1}= & -\frac{1}{2}\sin^{2}\theta\left(\beta_{D}\cos3\phi+\gamma_{D}\cos\theta\sin3\phi\right),\nonumber\\
b_{15,2}= & -\frac{1}{2}\sin^{2}\theta\left(\gamma_{D}\cos3\phi-\beta_{D}\cos\theta\sin3\phi\right),\nonumber\\
b_{15,3}= & -\frac{1}{2}\sin^{3}\theta\sin3\phi.
\end{align}
The remaining 16 terms, which depend on $\alpha_{D}$, are expressed as
\begin{align}
\left\{ \begin{array}{ccc}
b_{0,3},&b_{1,0},&b_{2,0},\\
b_{3,0},&b_{4,3},&b_{5,3},\\
b_{6,3},&b_{7,3},&b_{8,3},\\
b_{9,0},&b_{10,0},&b_{11,0},\\
b_{12,0},&b_{13,0},&b_{14,0},\\
&b_{15,0}
\end{array}\right\} = &
\alpha_{D}\left\{\begin{array}{ccc}
b_{0,0},&b_{1,3},&b_{2,3},\\
b_{3,3},&b_{4,0},&b_{5,0},\\
b_{6,0},&b_{7,0},&b_{8,0},\\
b_{9,3},&b_{10,3},&b_{11,3}\\
b_{12,3},&b_{13,3},&b_{14,3},\\
&b_{15,3}
\end{array}\right\}.
\end{align}

For the decay process $5/2\rightarrow1/2+0$,  when focusing only on the longitudinal polarization components of the parent particle, the polarization transition matrix $d_{\mu\nu}$ contains 18 non-zero elements. Among these, 12 terms do not involve  $\alpha_{D}$, given by
\begin{align}
d_{0,0}= & 1,\nonumber\\
d_{2,0}= & -\frac{3}{14}\left(1+3\cos2\theta\right),\nonumber\\
d_{4,0}= & \frac{3}{32}\left(9+20\cos2\theta+35\cos4\theta\right),\nonumber\\
d_{1,3}= & \frac{6}{35}\cos\theta,\nonumber\\
d_{3,3}= & -\frac{1}{6}\left(3\cos\theta+5\cos3\theta\right),\nonumber\\
d_{5,3}= & \frac{15}{32}\left(30\cos\theta+35\cos3\theta+63\cos5\theta\right),\nonumber\\
d_{1,1}= & -\frac{18}{35}\gamma_{D}\sin\theta,\nonumber\\
d_{3,1}= & \frac{1}{4}\gamma_{D}\left(\sin\theta+5\sin3\theta\right),\nonumber\\
d_{5,1}= & -\frac{45}{32}\gamma_{D}\left(2\sin\theta+7\sin3\theta+21\sin5\theta\right),\nonumber\\
d_{1,2}= & \frac{18}{35}\beta_{D}\sin\theta,\nonumber\\
d_{3,2}= & -\frac{1}{4}\beta_{D}\left(\sin\theta+5\sin3\theta\right),\nonumber\\
d_{5,2}= & \frac{45}{32}\beta_{D}\left(2\sin\theta+7\sin3\theta+21\sin5\theta\right).
\end{align}
The remaining 6 terms, which depend on $\alpha_{D}$, are described as
\begin{align}
\left\{ \begin{array}{ccc}
d_{0,3},&d_{1,0},&d_{2,3},\\
d_{3,0},&d_{4,3},&d_{5,0}
\end{array}\right\} = &
\alpha_{D}\left\{ \begin{array}{ccc}
d_{0,0},&d_{1,3},&d_{2,0},\\
d_{3,3},&d_{4,0},&d_{5,3}
\end{array}\right\}.
\end{align}

\section{CONVENTION FOR THE SPIN-1/2 SPIN-1 SPIN-3/2 SPIN-5/2 SPINNORS\label{sec:spinnors}}

In this section,  we present the field spinors for particles with spins 1/2, 1, 3/2, and 5/2 in the Pauli-Dirac representation. We denote the mass of particle as $m$ and the four-momentum as
\begin{align}
p^\mu=(E,|\vec{p}| \sin \theta \cos \phi,|\vec{p}| \sin \theta \sin \phi,|\vec{p}| \cos \theta).
\end{align}

For spin 1/2 fields, which satisfy the Dirac equation~\cite{Dirac:1928hu}, the polarization vectors with helicity $\lambda=\pm 1$ are given by
\begin{align}
&u\left(p,\pm\right) =\left(\begin{array}{c}
\sqrt{E+m}\chi_{\pm}\\
\pm\sqrt{E-m}\chi_{\pm}
\end{array}\right),\nonumber \\
&v\left(p,\pm\right)=\left(\begin{array}{c}
\sqrt{E-m}\chi_{\mp}\\
\mp\sqrt{E+m}\chi_{\mp}
\end{array}\right),
\end{align}
where $\chi_{\pm}$ are the two-component spinors defined as
\begin{eqnarray}
\chi_{+} & = & \left(\begin{array}{c}
\cos\frac{\theta}{2}\\
\sin\frac{\theta}{2}e^{i\phi}
\end{array}\right),\chi_{-}=\left(\begin{array}{c}
-\sin\frac{\theta}{2}\\
\cos\frac{\theta}{2}e^{i\phi}
\end{array}\right).
\end{eqnarray}

For spin 1 fields, which are solutions to the Proca equation~\cite{Proca:1936fbw}, the polarization vectors for helicities $\lambda=\pm1,0$ are given by
\begin{align}
&\varepsilon^{\mu}\left(p,\pm\right) = \frac{1}{\sqrt{2}}\left(\begin{array}{c}
0\\
\mp\cos\theta\cos\phi+i\sin\phi\\
\mp\cos\theta\sin\phi-i\cos\phi\\
\pm\sin\theta
\end{array}\right),\nonumber\\
&\varepsilon^{\mu}\left(p,0\right)=\frac{1}{m}\left(\begin{array}{c}
\left|\vec{p}\right|\\
E\sin\theta\cos\phi\\
E\sin\theta\sin\phi\\
E\cos\theta
\end{array}\right).
\end{align}

For higher-spin spinors, solutions are obtained using Klein-Gordon equations for integer spins, and Rarita-Schwinger equations~\cite{Rarita:1941mf} for half-integer spins. We follow conventions in Ref.~\cite{Huang:2003ym}.

For spin 3/2, the polarization vectors with helicities $\lambda=\pm3/2,\,\pm1/2$ are given by
\begin{align}
u^{\mu}\left(p,\pm\frac{3}{2}\right) =&\varepsilon_{\pm}^{\mu}\left(p\right)u_{\pm}\left(p\right),\nonumber\\
u^{\mu}\left(p,\pm\frac{1}{2}\right) =& \sqrt{1/3}\varepsilon_{\pm}^{\mu}\left(p\right)u_{\mp}\left(p\right)+\sqrt{2/3}\varepsilon_{0}^{\mu}\left(p\right)u_{\pm}\left(p\right),\nonumber\\
v^{\mu}\left(p,\pm\frac{3}{2}\right) =&\varepsilon_{\pm}^{\mu*}\left(p\right)v_{\pm}\left(p\right),\nonumber\\
v^{\mu}\left(p,\pm\frac{1}{2}\right) = &\sqrt{1/3}\varepsilon_{\pm}^{\mu*}\left(p\right)v_{\mp}\left(p\right)+\sqrt{2/3}\varepsilon_{0}^{\mu*}\left(p\right)v_{\pm}\left(p\right),
\end{align}

For spin 5/2, the polarization vectors with helicities $\lambda=\pm5/2,\,\pm3/2,\,\pm1/2$ are given by
\begin{align}
u^{\mu\nu}\left(p,\pm\frac{5}{2}\right)= & \varepsilon_{\pm}^{\mu}\left(p\right)\varepsilon_{\pm}^{\nu}\left(p\right)u_{\pm}\left(p\right),\nonumber\\
u^{\mu\nu}\left(p,\pm\frac{3}{2}\right)= & \sqrt{1/5}\varepsilon_{\pm}^{\mu}\left(p\right)\varepsilon_{\pm}^{\nu}\left(p\right)u_{\mp}\left(p\right)+\sqrt{2/5}\varepsilon_{0}^{\mu}\left(p\right)\varepsilon_{\pm}^{\nu}\left(p\right)u_{\pm}\left(p\right)\nonumber \\
&+\sqrt{2/5}\varepsilon_{\pm}^{\mu}\left(p\right)\varepsilon_{0}^{\nu}\left(p\right)u_{\pm}\left(p\right),\nonumber\\
u^{\mu\nu}\left(p,\pm\frac{1}{2}\right)= & \sqrt{1/10}\varepsilon_{\pm}^{\mu}\left(p\right)\varepsilon_{\mp}^{\nu}\left(p\right)u_{\pm}\left(p\right)+\sqrt{1/10}\varepsilon_{\mp}^{\mu}\left(p\right)\varepsilon_{\pm}^{\nu}\left(p\right)u_{\pm}\left(p\right)\nonumber \\
&+\sqrt{1/5}\varepsilon_{0}^{\mu}\left(p\right)\varepsilon_{\pm}^{\nu}\left(p\right)u_{\mp}\left(p\right)+\sqrt{1/5}\varepsilon_{\pm}^{\mu}\left(p\right)\varepsilon_{0}^{\nu}\left(p\right)u_{\mp}\left(p\right)\nonumber \\
&+\sqrt{2/5}\varepsilon_{0}^{\mu}\left(p\right)\varepsilon_{0}^{\nu}\left(p\right)u_{\pm}\left(p\right),\nonumber\\
v^{\mu\nu}\left(p,\pm\frac{5}{2}\right)= & \varepsilon_{\pm}^{\mu*}\left(p\right)\varepsilon_{\pm}^{\nu*}\left(p\right)u_{\pm}\left(p\right),\nonumber\\
v^{\mu\nu}\left(p,\pm\frac{3}{2}\right)= & \sqrt{1/5}\varepsilon_{\pm}^{\mu*}\left(p\right)\varepsilon_{\pm}^{\nu*}\left(p\right)u_{\mp}\left(p\right)+\sqrt{2/5}\varepsilon_{0}^{\mu*}\left(p\right)\varepsilon_{\pm}^{\nu*}\left(p\right)u_{\pm}\left(p\right)\nonumber \\
&+\sqrt{2/5}\varepsilon_{\pm}^{\mu*}\left(p\right)\varepsilon_{0}^{\nu*}\left(p\right)u_{\pm}\left(p\right),\nonumber\\
v^{\mu\nu}\left(p,\pm\frac{1}{2}\right)= & \sqrt{1/10}\varepsilon_{\pm}^{\mu*}\left(p\right)\varepsilon_{\mp}^{\nu*}\left(p\right)u_{\pm}\left(p\right)+\sqrt{1/10}\varepsilon_{\mp}^{\mu*}\left(p\right)\varepsilon_{\pm}^{\nu*}\left(p\right)u_{\pm}\left(p\right)\nonumber \\
&+\sqrt{1/5}\varepsilon_{0}^{\mu*}\left(p\right)\varepsilon_{\pm}^{\nu*}\left(p\right)u_{\mp}\left(p\right)+\sqrt{1/5}\varepsilon_{\pm}^{\mu*}\left(p\right)\varepsilon_{0}^{\nu*}\left(p\right)u_{\mp}\left(p\right)\nonumber \\
&+\sqrt{2/5}\varepsilon_{0}^{\mu*}\left(p\right)\varepsilon_{0}^{\nu*}\left(p\right)u_{\pm}\left(p\right).
\end{align}

\end{document}